\documentclass[aps,prc,twocolumn,nofootinbib,superscriptaddress,amsmath,amssymb]{revtex4-1}

\usepackage{epsfig}
\usepackage{graphicx}
\usepackage{dcolumn}
\usepackage{bm}
\usepackage{subfigure}

\begin{document}

\title{Final Analysis of Proton Form Factor Ratio Data at
  $\mathbf{Q^2 = 4.0}$, 4.8 and 5.6 GeV$\mathbf{^2}$}
\author{A. J. R. Puckett} \email[Corresponding author: ]{puckett@jlab.org}
\affiliation{Thomas Jefferson National Accelerator Facility, Newport
  News, VA 23606}
\affiliation{Los Alamos National Laboratory, Los Alamos, NM 87545}
\author{E. J. Brash}
\affiliation{Christopher Newport University, Newport News, VA 23606}
\affiliation{Thomas Jefferson National Accelerator Facility, Newport
  News, VA 23606}
\author{O. Gayou}
\affiliation{College of William and Mary, Williamsburg, VA 23187}
\affiliation{Universit\'{e} Blaise Pascal/CNRS-IN2P3, F-63177 Aubi\`{e}re, France}
\author{M. K. Jones}
\affiliation{Thomas Jefferson National Accelerator Facility, Newport News, VA 23606}
\author{L. Pentchev}
\author{C. F. Perdrisat}
\affiliation{College of William and Mary, Williamsburg, VA 23187}
\author{V. Punjabi}
\affiliation{Norfolk State University, Norfolk, VA 23504}
\author{K. A. Aniol}
\affiliation{California State University, Los Angeles, Los Angeles,
  CA 90032}
\author{T. Averett}
\affiliation{College of William and Mary, Williamsburg, VA 23187}
\author{F. Benmokhtar}
\affiliation{Rutgers, The State University of New Jersey,  Piscataway,
  NJ 08855}
\author{W. Bertozzi}
\affiliation{Massachusetts Institute of Technology, Cambridge, MA
  02139}
\author{L. Bimbot}
\affiliation{Institut de Physique Nucl\'{e}aire, F-91406 Orsay,
  France}
\author{J. R. Calarco}
\affiliation{University of New Hampshire, Durham, NH 03824}
\author{C. Cavata}
\affiliation{CEA Saclay, F-91191 Gif-sur-Yvette, France}
\author{Z. Chai}
\affiliation{Massachusetts Institute of Technology, Cambridge, MA
  02139}
\author{C.-C. Chang}
\affiliation{University of Maryland, College Park, MD 20742}
\author{T. Chang}
\affiliation{University of Illinois, Urbana-Champaign, IL 61801}
\author{J. P. Chen}
\affiliation{Thomas Jefferson National Accelerator Facility, Newport
  News, VA 23606}
\author{E. Chudakov}
\affiliation{Thomas Jefferson National Accelerator Facility, Newport
  News, VA 23606}
\author{R. De Leo}
\affiliation{INFN, Sezione di Bari and University of Bari, 70126 Bari,
  Italy}
\author{S. Dieterich}
\affiliation{Rutgers, The State University of New Jersey,  Piscataway,
  NJ 08855}
\author{R. Endres}
\affiliation{Rutgers, The State University of New Jersey,  Piscataway,
  NJ 08855}
\author{M. B. Epstein}
\affiliation{California State University, Los Angeles, Los Angeles,
  CA 90032}
\author{S. Escoffier}
\affiliation{CEA Saclay, F-91191 Gif-sur-Yvette, France}
\author{K. G. Fissum}
\affiliation{University of Lund, PO Box 118, S-221 00 Lund, Sweden}
\author{H. Fonvieille}
\affiliation{Universit\'{e} Blaise Pascal/CNRS-IN2P3, F-63177
  Aubi\`{e}re, France}
\author{S. Frullani}
\affiliation{INFN, Sezione Sanit\`{a} and Istituto Superiore di Sanit\`{a}, 
00161 Rome, Italy}
\author{J. Gao}
\affiliation{Massachusetts Institute of Technology, Cambridge, MA
  02139}
\author{F. Garibaldi}
\affiliation{INFN, Sezione Sanit\`{a} and Istituto Superiore di Sanit\`{a}, 
00161 Rome, Italy}
\author{S. Gilad}
\affiliation{Massachusetts Institute of Technology, Cambridge, MA
  02139}
\author{R. Gilman}
\affiliation{Rutgers, The State University of New Jersey,  Piscataway,
  NJ 08855}
\affiliation{Thomas Jefferson National Accelerator Facility, Newport
  News, VA 23606}
\author{A. Glamazdin}
\affiliation{Kharkov Institute of Physics and Technology, Kharkov
  310108, Ukraine}
\author{C. Glashausser}
\affiliation{Rutgers, The State University of New Jersey,  Piscataway,
  NJ 08855}
\author{J. Gomez}
\affiliation{Thomas Jefferson National Accelerator Facility, Newport
  News, VA 23606}
\author{J.-O. Hansen}
\affiliation{Thomas Jefferson National Accelerator Facility, Newport
  News, VA 23606}
\author{D. Higinbotham}
\affiliation{Massachusetts Institute of Technology, Cambridge, MA
  02139}
\author{G. M. Huber}
\affiliation{University of Regina, Regina, SK S4S OA2, Canada}
\author{M. Iodice}
\affiliation{INFN, Sezione Sanit\`{a} and Istituto Superiore di Sanit\`{a}, 
00161 Rome, Italy}
\author{C. W. de Jager}
\affiliation{Thomas Jefferson National Accelerator Facility, Newport
  News, VA 23606}
\affiliation{University of Virginia, Charlottesville, VA 22901}
\author{X. Jiang}
\affiliation{Rutgers, The State University of New Jersey,  Piscataway,
  NJ 08855}
\author{M. Khandaker}
\affiliation{Norfolk State University, Norfolk, VA 23504}
\author{S. Kozlov}
\affiliation{University of Regina, Regina, SK S4S OA2, Canada}
\author{K. M. Kramer}
\affiliation{College of William and Mary, Williamsburg, VA 23187}
\author{G. Kumbartzki}
\affiliation{Rutgers, The State University of New Jersey,  Piscataway,
  NJ 08855}
\author{J. J. LeRose}
\affiliation{Thomas Jefferson National Accelerator Facility, Newport
  News, VA 23606}
\author{D. Lhuillier}
\affiliation{CEA Saclay, F-91191 Gif-sur-Yvette, France}
\author{R. A. Lindgren}
\affiliation{University of Virginia, Charlottesville, VA 22901}
\author{N. Liyanage}
\affiliation{Thomas Jefferson National Accelerator Facility, Newport
  News, VA 23606}
\author{G. J. Lolos}
\affiliation{University of Regina, Regina, SK S4S OA2, Canada}
\author{D. J. Margaziotis}
\affiliation{California State University, Los Angeles, Los Angeles,
  CA 90032}
\author{F. Marie}
\affiliation{CEA Saclay, F-91191 Gif-sur-Yvette, France}
\author{P. Markowitz}
\affiliation{Florida International University, Miami, FL 33199}
\author{K. McCormick}
\affiliation{Kent State University, Kent, OH 44242}
\author{R. Michaels}
\affiliation{Thomas Jefferson National Accelerator Facility, Newport
  News, VA 23606}
\author{B. D. Milbrath}
\affiliation{Eastern Kentucky University, Richmond, KY 40475}
\author{S. K. Nanda}
\affiliation{Thomas Jefferson National Accelerator Facility, Newport
  News, VA 23606}
\author{D. Neyret}
\affiliation{CEA Saclay, F-91191 Gif-sur-Yvette, France}
\author{N. M. Piskunov}
\affiliation{JINR-LHE, 141980 Dubna, Moscow Region, Russian
  Federation}
\author{R. D. Ransome}
\affiliation{Rutgers, The State University of New Jersey,  Piscataway,
  NJ 08855}
\author{B. A. Raue}
\affiliation{Florida International University, Miami, FL 33199}
\author{R. Roch\'e}
\affiliation{Florida State University, Tallahassee, FL 32306}
\author{M. Rvachev}
\affiliation{Massachusetts Institute of Technology, Cambridge, MA
  02139}
\author{C. Salgado}
\affiliation{Norfolk State University, Norfolk, VA 23504}
\author{S. Sirca}
\affiliation{Institut Jozef Stefan, University of Ljubljana, SI-1001
  Ljubljana}
\author{I. Sitnik}
\affiliation{JINR-LHE, 141980 Dubna, Moscow Region, Russian
  Federation}
\author{S. Strauch}
\affiliation{Rutgers, The State University of New Jersey,  Piscataway,
  NJ 08855}
\author{L. Todor}
\affiliation{Carnegie-Mellon University, Pittsburgh, PA 15213}
\author{E. Tomasi-Gustafsson}
\affiliation{CEA Saclay, F-91191 Gif-sur-Yvette, France}
\author{G. M. Urciuoli}
\affiliation{INFN, Sezione Sanit\`{a} and Istituto Superiore di Sanit\`{a}, 
00161 Rome, Italy}
\author{H. Voskanyan}
\affiliation{Yerevan Physics Institute, Yerevan 375036, Armenia}
\author{K. Wijesooriya}
\affiliation{University of Illinois, Urbana-Champaign, IL 61801}
\author{B. B. Wojtsekhowski}
\affiliation{Thomas Jefferson National Accelerator Facility, Newport
  News, VA 23606}
\author{X. Zheng}
\affiliation{Massachusetts Institute of Technology, Cambridge, MA
  02139}
\author{L. Zhu}
\affiliation{Massachusetts Institute of Technology, Cambridge, MA 02139}
\collaboration{The Jefferson Lab Hall A Collaboration}
\noaffiliation
\date{\today}

\begin{abstract}
  Precise measurements of the proton electromagnetic form factor ratio
  $R = \mu_p G_E^p/G_M^p$ using the polarization transfer method at
  Jefferson Lab have
  revolutionized the understanding of nucleon structure by revealing the
  strong decrease of $R$ with momentum transfer $Q^2$ for $Q^2 \gtrsim
  1$ GeV$^2$, in strong disagreement with
  previous extractions of $R$ from cross section measurements. In
  particular, the polarization transfer results have exposed the
  limits of applicability of the one-photon-exchange approximation and
  highlighted the role of quark orbital angular momentum in the
  nucleon structure. The
  GEp-II experiment in Jefferson Lab's Hall A measured $R$ at four $Q^2$
  values in the range $3.5$ GeV$^2 \le Q^2 \le 5.6$ GeV$^2$. A possible discrepancy
  between the originally published GEp-II results and more recent
  measurements at higher $Q^2$ motivated a new analysis of the GEp-II
  data. This article presents the final results of the GEp-II
  experiment, including details of the new analysis, an expanded description of the apparatus and an overview
  of theoretical progress since the original publication. The key
  result of the final analysis is a systematic increase in the
  results for $R$, improving the consistency of the
  polarization transfer data in the high-$Q^2$ region. This increase
  is the result of an improved selection of elastic events which
  largely removes the systematic effect of the inelastic
  contamination, underestimated by the original analysis. 
\end{abstract}
\maketitle
\section{Introduction}
The electromagnetic form factors (FFs) of the nucleon have been
revived as a subject of high interest in hadronic physics since a
series of 
precise recoil polarization measurements of the ratio of the proton's
electric ($G_E^p$) and
magnetic ($G_M^p$) FFs
\cite{Jones00,*Punjabi05,*Punjabi05erratum,Gayou02} in Jefferson Lab's Hall A
established the rapid decrease with momentum transfer $Q^2$
of $R = \mu_p G_E^p/G_M^p$, where $\mu_p$ is the proton's magnetic
moment, for 0.5 GeV$^2 \le Q^2 \le 5.6$ GeV$^2$. These measurements
disagreed strongly with previous extractions of $G_E^p$ from cross
section data \cite{PerdrisatPunjabiVanderhaegen2007} using
the Rosenbluth method \cite{Rosenbluth1950}, which found $\mu_p G_E^p/G_M^p
\approx 1$. Subsequent investigations of both experimental techniques,
including a novel ``Super-Rosenbluth'' measurement using $^1H(e,p)e'$
cross section measurements to reduce systematic uncertainties \cite{Qattan05}, found
no neglected sources of error in either data set, pointing to
incompletely understood physics as the source of the discrepancy. 

Theoretical investigations of the discrepancy have focused on
higher-order QED corrections to the cross section and
polarization observables in elastic $ep$ scattering
\cite{AfanasevRadCorr,TPEXreview}, including radiative corrections and
two-photon-exchange (TPEX) effects. The amplitude for
elastic electron-proton scattering involving the exchange of two or more
hard\footnote{``Hard'' in this context means that both exchanged
  photons carry an appreciable fraction of the total momentum
  transfer.} photons cannot presently be calculated model-independently. In the $Q^2$ region of the discrepancy, model
calculations of TPEX \cite{BlundenMelnitTjon,GPDTPEX} predict relative
corrections to both the cross section and polarization observables that are typically at the
few-percent level. At large $Q^2$, the sensitivity of the Born
(one-photon-exchange) cross
section to $G_E^p$ becomes similar to or smaller than the sensitivity of
the measured cross section to poorly-known TPEX
corrections, obscuring the extraction of $G_E^p$. On the other hand, the polarization transfer ratio $R$ defined in equations
\eqref{recoilformulas} is directly proportional to $G_E^p/G_M^p$, such that the extraction of $G_E^p$ is much less sensitive to corrections beyond the Born approximation. For this reason, a general consensus
has emerged that the polarization transfer data most reliably
determine $G_E^p$ at large $Q^2$. Nonetheless, active experimental and
theoretical investigation of the discrepancy and the role of TPEX
continues \cite{Arrington2011782}. Owing to the lack of a model-independent theoretical prescription for TPEX corrections, precise measurements of elastic $ep$
scattering observables sensitive
to TPEX effects continue to play an important role in the resolution of the discrepancy.


The revised experimental understanding of the proton form factors led to
an onslaught of theoretical work. The constancy of the
Rosenbluth data for $G_E^p/G_M^p$ was consistent with a ``precocious''
onset
of pQCD dimensional scaling laws
\cite{BrodskyFarrar1975,*BrodskyLepage1979}, valid for 
asymptotically large $Q^2$, an interpretation which had to be
abandoned in light of the polarization data. The decrease of $R$ with
$Q^2$ was later interpreted in a pQCD-scaling framework including higher-twist
corrections \cite{BelitskyJiYuan2003}, demonstrating the importance of
quark orbital angular momentum in the interpretation of nucleon
structure. The relations between nucleon form factors and Generalized
Parton Distributions (GPDs) have placed this connection on a more
quantitative footing
\cite{JiGPDPRL1997,GuidalGPD2005,DiehlGPDtomography2005}. Furthermore,
the GPD-form factor sum rules have been used to derive
model-independent representations of the nucleon transverse charge and
magnetization densities as two-dimensional Fourier transforms
of the Dirac ($F_1$) and Pauli ($F_2$) form
factors
\cite{MillerGPDchargedensity2007,*MillerGPDmagnetdensity2008}. In the
context of calculations based on QCD's Dyson-Schwinger Equations (DSEs)
\cite{Cloet09,EichmannFaddeev}, the form factor data are instrumental in elucidating
the dynamical interplay between the nucleon's dressed-quark core,
diquark correlations, and the pseudoscalar meson cloud~\cite{CD200850}. Recent measurements of
the neutron form factors at large $Q^2$ \cite{Riordan,Lachniet} have
enabled for the first time a detailed flavor-decomposition
\cite{FlavorDecomposition} of the form factor data, leading to new
insights. In addition, the form
factor data have been interpreted within a large number of
phenomenological models; a recent review of the large body of 
theoretical work relevant to the nucleon FFs is given in
\cite{PerdrisatPunjabiVanderhaegen2007}, and a current overview is given in section \ref{subsec:theory} of this work.

The recoil polarization method exploits the relation between
the transferred polarization in elastic $\vec{e}p$ scattering and the ratio
$G_E^p/G_M^p$. In the one-photon-exchange approximation, the polarization transferred to recoiling protons in
the elastic scattering of longitudinally polarized electrons by
unpolarized protons has longitudinal ($P_\ell$) and
transverse ($P_t$) components in the reaction plane given by \cite{AkhiezerRekalo3,*AkhiezerRekalo1,*AkhiezerRekalo2,ArnoldCarlsonGross}
\begin{eqnarray}
  \label{recoilformulas} P_t &=& -h P_e \sqrt{\frac{2\epsilon(1-\epsilon)}{\tau}}
  \frac{r}{1+\frac{\epsilon}{\tau}r^2} \nonumber \\
  P_\ell &=& h P_e \frac{\sqrt{1-\epsilon^2}}{1+\frac{\epsilon}{\tau}r^2} \\
  r \equiv \frac{G_E^p}{G_M^p} &=& -\frac{P_t}{P_\ell}\sqrt{\frac{\tau(1+\epsilon)}{2\epsilon}} = \frac{R}{\mu_p}
  \nonumber, 
\end{eqnarray}
where $h = \pm 1$ is the electron beam helicity, $P_e$ is the beam polarization, $\tau \equiv
Q^2/4M_p^2$, $M_p$ is the proton mass, and $\epsilon \equiv
\left[1+2(1+\tau)\tan^2\left(\theta_e/2\right)\right]^{-1}$, with $\theta_e$ the electron
scattering angle in the proton rest (lab) frame,
corresponds to the
longitudinal polarization of the virtual photon in the
one-photon-exchange approximation.

Recent measurements
from Jefferson Lab's Hall C
\cite{PuckettPRL10} extended the $Q^2$ reach of the polarization transfer method to 8.5
GeV$^2$. The published data from Hall A are well described by a
linear $Q^2$ dependence \cite{PerdrisatPunjabiVanderhaegen2007}, 
\begin{eqnarray}
  R = 1.0587 - 0.14265 Q^2, \label{gayoufit}
\end{eqnarray}
with $Q^2$ in GeV$^2$, valid for $Q^2 \ge 0.4$ GeV$^2$. On the other
hand, all three of
the recent Hall C data points are at least 1.5 standard
deviations above this line, including the measurement at overlapping $Q^2 = 5.17$
GeV$^2$, which lies 1.8$\sigma$ above equation \eqref{gayoufit}.
Due to the strong, incompletely understood discrepancy between the
Rosenbluth and polarization transfer methods of extracting $G_E^p/G_M^p$ and
the fact that the new Hall C measurements are the first to
check the reproducibility of the Hall A data using a completely
different apparatus in the $Q^2$ region where the discrepancy is
strongest, understanding possible systematic differences between the
experiments is important. 

This article reports an updated,
final data analysis of the three higher-$Q^2$ measurements of $R$ from Hall
A, originally published in \cite{Gayou02}, along with expanded details
of the experiment. To avoid confusion, a naming
convention is adopted throughout the remainder of this article for the
most frequently cited experiments: GEp-I for Ref.~\cite{Punjabi05}, GEp-II for
Ref.~\cite{Gayou02}, the subject of this article, GEp-III for Ref.~\cite{PuckettPRL10} and
GEp-2$\gamma$ for Ref.~\cite{MezianePRL11}. Section
\ref{sec:expsetup} presents the kinematics of the measurements, an
expanded description of the experimental apparatus, and a comparison
of the GEp-II and GEp-III/GEp-2$\gamma$ experiments. Section \ref{sec:analysis}
presents the data analysis method, including the selection of elastic
events, the extraction of polarization observables, and the estimation
and subtraction of the non-elastic background contribution. Section
\ref{sec:results} presents the final results of the experiment and discusses the impact of the revised data on the world database of
proton electromagnetic form factor measurements, in the context of the considerable advances in theory
since the original publication. The conclusions and summary are given in section \ref{sec:conclusion}.

\section{Experiment Setup}
\label{sec:expsetup}
Table \ref{kintable} shows the central
kinematics of the measurements
from the GEp-II experiment. The kinematic variables given in Table
\ref{kintable} are the beam energy $E_e$, the scattered electron energy $E'_e$, the
electron scattering angle $\theta_e$, the scattered proton momentum
$p_p$, and the proton scattering angle $\theta_p$.  
\begin{table*}
  \caption{\label{kintable} Central kinematics of
    the GEp-II experiment. The central $Q^2$ value is defined by the
    central momentum of the left HRS in which the proton was
    detected. $\epsilon$ is the parameter appearing in equation \eqref{recoilformulas}, $E_e$ is the
    beam energy, $E'_e$ is the scattered electron energy, $\theta_e$ is
    the electron scattering angle, $p_p$ is the proton momentum, $\theta_p$ is
    the proton scattering angle, $\chi$ is the central spin precession angle, $P_e$ is the beam polarization, and
    $R_{cal}$ is the distance from the target to the
    calorimeter surface. At the lowest $Q^2$ of 3.5 GeV$^2$, the
    scattered electron was detected in the right HRS.}
  \begin{ruledtabular}
  \begin{tabular}{cccccccccc}
    Nominal $Q^2$ (GeV$^2$) & $\epsilon$ & 
    $E_e$ (GeV) & $E'_e$ (GeV) & $\theta_e$ ($^\circ$) & $p_p$ (GeV) &
    $\theta_p$ ($^\circ$) & $\chi$ ($^\circ$) & $P_e$ (\%) & $R_{cal}$ (m)
    \\ \hline
    3.5 & 0.77 & 4.61 & 2.74 & 30.6 & 2.64 & 31.8 & 241 & 70 & HRSR \\
    4.0 & 0.71 & 4.61 & 2.47 & 34.5 & 2.92 & 28.6 & 264 & 70 & 17.0 \\
    4.8 & 0.59 & 4.59 & 2.04 & 42.1 & 3.36 & 23.8 & 301 & 73 & 12.5 \\ 
    5.6 & 0.45 & 4.60 & 1.61 & 51.4 & 3.81 & 19.3 & 337 & 71 & 9.0 \\
  \end{tabular}
  \end{ruledtabular}
\end{table*}

\subsection{Experimental Apparatus}

The GEp-II experiment ran in Hall A at Jefferson Lab during November and December of 2000.
A polarized electron beam was scattered off a liquid hydrogen 
target. Hall A is equipped with two High Resolution Spectrometers
(HRS)~\cite{HallANIM}, which are identical in design. In this
experiment, the left HRS (HRSL) was used to detect the 
recoil proton, while the right HRS (HRSR) was used to detect the scattered electron at the 
lowest $Q^2$ of 3.5 GeV$^2$. For the three 
highest $Q^2$ points at 4.0, 4.8 and 5.6 GeV$^2$, electrons were
detected by a lead-glass calorimeter. The focal plane of the HRSL was equipped with a focal plane polarimeter (FPP) to 
measure the polarization of the recoil proton. 

The Continuous Electron Beam Accelerator at the Thomas Jefferson National
Accelerator Facility (JLab) delivers a high quality, longitudinally
polarized electron beam with $\sim$100\% duty factor. The beam energy was 
measured using the Arc and ep methods. The ep method determines the energy 
by measuring the opening angle between the scattered electron and the recoil proton 
in ep elastic scattering, while the Arc method uses the standard technique of measuring 
a bend angle in a series of dipole magnets. The combined absolute accuracy of both
methods is $\Delta E/E \sim 10^{-4}$, while the beam energy spread is
at the $10^{-5}$ level. The nominal beam energy in this experiment 
was 4.6 GeV. The beam polarization was measured by Compton and
M\"{o}ller polarimeters. Details of the standard Hall A equipment
can be found in~\cite{HallANIM} and references therein.

The hydrogen target cell used in this experiment was 15 cm long along the beam direction.
The target was operated at a constant temperature of 19~K and pressure of 
25~psi, resulting in a density of about 0.072~g/cm$^3$. To minimize
the target density fluctuations due to localized heat deposition by
the intense electron beam, a fast raster system consisting of a pair
of dipole magnets was used to increase the transverse size of the beam in the horizontal and vertical directions. The raster
shape was square or 
circular in the plane transverse to the beam axis. In 
this experiment, the raster size was approximately $4\times 4$~mm$^2$.

Recoil protons were detected in the high resolution spectrometer
located  on the beam left (HRSL)~\cite{HallANIM}. The HRSL has a central bend angle of 
$45^\circ$, and subtends a $6.5$~msr
solid-angle for charged particles with momenta up to 4 GeV with $\pm
5\%$ momentum acceptance. Two vertical drift chambers
measure the particle's position and trajectory at the focal
plane. With knowledge of the optics of the HRSL magnets and precise
beam position monitoring, the proton scattering angles, momentum and vertex
coordinates were reconstructed with FWHM resolutions of $\sim$2.6 (4.0) mrad for the
in-plane (out-of-plane) angle, $\Delta p/p
\sim 2.6 \times 10^{-4}$ for the momentum, and $\sim$3.1 mm for the
vertex coordinate in the plane transverse to
the HRSL optical axis.


\subsubsection{ Focal Plane Polarimeter}
\begin{figure}
\includegraphics[width=.48\textwidth]{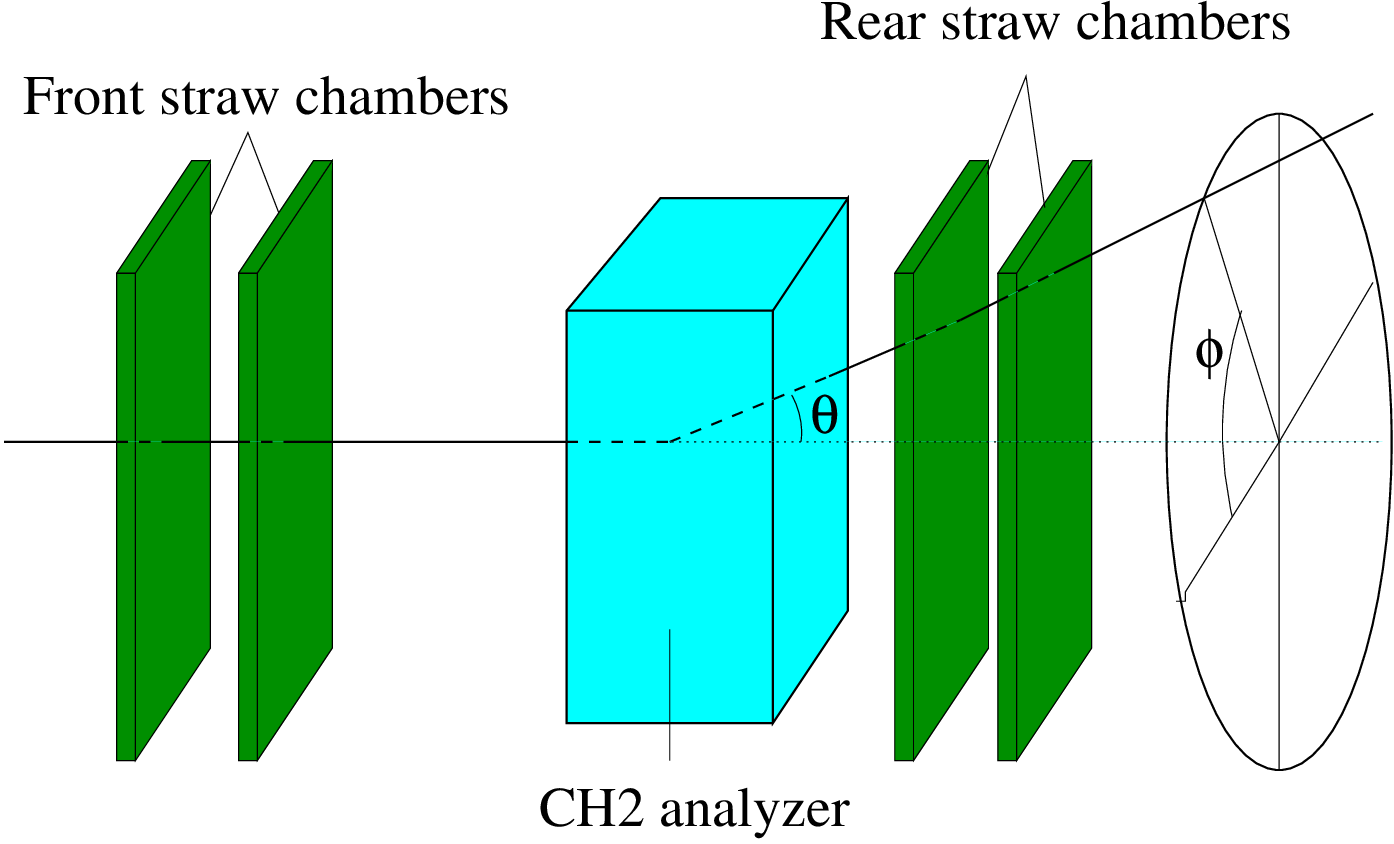}
\caption{(color online) Layout of the Focal Plane Polarimeter.}
\label{fig:fpp}
\end{figure}

The central instrument of this experiment was the Focal Plane
Polarimeter (FPP)~\cite{Punjabi05}, installed in the focal plane of HRSL.
The FPP measures the transverse polarization of the recoil proton. 
The protons
are scattered in the focal plane region by an analyzer, as shown in 
Figure~\ref{fig:fpp}. If the protons are polarized transverse to their momentum 
direction, an azimuthal asymmetry results from the spin-orbit interaction with 
the analyzing nuclei. 

\begin{figure}
\includegraphics[width=.48\textwidth]{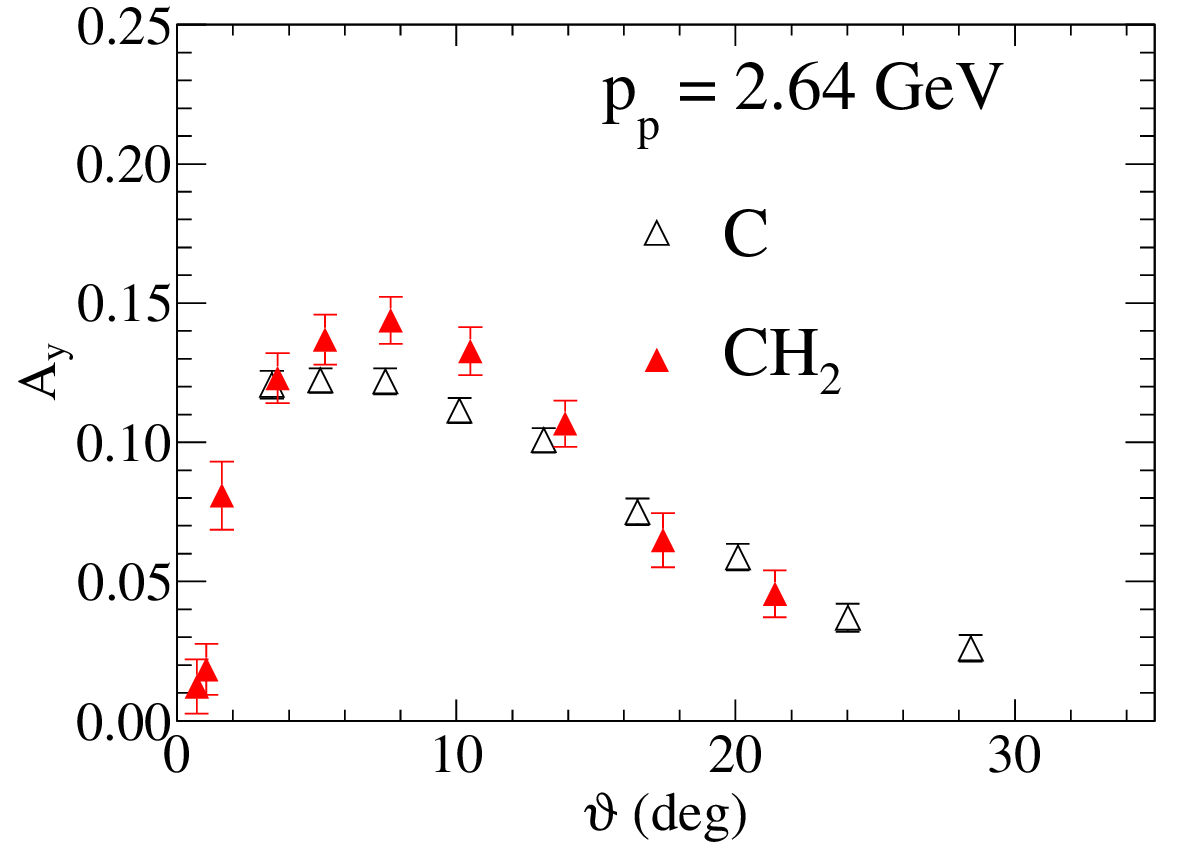}
\caption{(color online) Analyzing power versus scattering angle in the
  analyzer for graphite (49.5 cm-thick) and polyethylene (58 cm-thick) at
  a proton momentum of 2.64 GeV, corresponding to $Q^2 = 3.5$ GeV$^2$.}
\label{fig:Ay_CvsCH2_theta}
\end{figure}

\begin{figure}
\includegraphics[width=.48\textwidth]{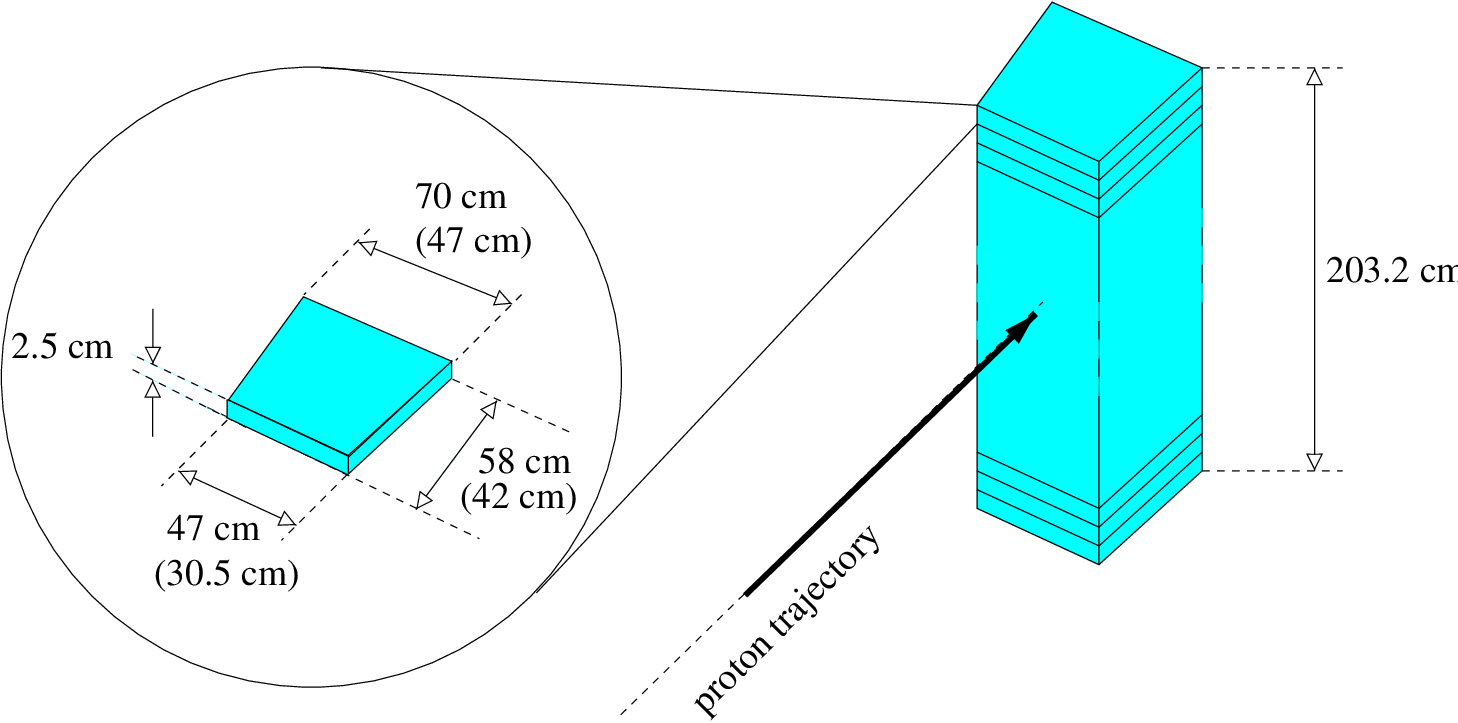}
\caption{(color online) Stack of polyethylene plates for the
  analyzer. The dimensions shown on the plate are for the 58~cm (42~cm) stack, and were chosen to match
the envelope of elastically scattered protons in HRSL.}
\label{fig:ch2}
\end{figure}

The FPP has been described in detail in~\cite{Punjabi05}, so
only a brief summary of its characteristics will be given here. 
The only significant difference in the configuration of the FPP between the GEp-I and
GEp-II experiments was a change of 
the analyzer material
from carbon to polyethylene. During GEp-I, the analyzer consisted
of four doors of carbon which could be combined to produce
a maximum thickness of 51~cm. For cost, safety and efficiency reasons, carbon is ideal 
for measuring proton polarization with a momentum up to 
2.4~GeV, which was sufficient for GEp-I. For GEp-II, 
the maximum proton momentum  was 3.8~GeV. At 
this momentum, the analyzing power of carbon, which contributes to the size
of the asymmetry, and therefore to the size of the error bar, decreases
significantly. An experiment was carried out at the Laboratory for High Energy
(LHE) at the Joint Institute for Nuclear Research (JINR) in Dubna, Russia to find an optimal analyzing
material and its thickness for protons at 3.8~GeV \cite{Ay_pCH2}.
Polyethylene, a compound of carbon and hydrogen, was found to increase
the analyzing power relative to carbon as shown in Figure~\ref{fig:Ay_CvsCH2_theta}. 
A stack of eighty 2.5~cm-thick plates, each 58~cm deep along the
direction of incident protons, was installed between the unused, 
opened, doors of the carbon analyzer, as shown in Figure~\ref{fig:ch2}. This 58~cm thickness was used
for the $Q^2$ = 3.5~GeV$^2$ kinematics. For the three higher $Q^2$
kinematics, an additional stack of polyethylene with a thickness of 42~cm was installed on a rail just upstream of 
the 58 cm stack to give a total thickness of 100~cm. 

\subsubsection{Electron detection at $Q^2=3.5$~GeV$^2$}
\label{subsubsec:e35}
For the measurement at $Q^2=3.5$~GeV$^2$, the electron was detected in
the high resolution spectrometer
located on the beam right (HRSR). The trigger was defined by a coincidence between an 
electron in HRSR and a proton in HRSL. The precise measurement of the scattered
electron kinematics using a high-resolution magnetic spectrometer
provides for an extremely clean selection of elastic $ep$ events with
cuts on the reconstructed missing energy and momentum, as shown in Fig. 4.8 of Ref.~\cite{GayouThesis}.

\subsubsection{Electron detection at $Q^2 \geq 4.0$~GeV$^2$}

\begin{figure}
\includegraphics[width=.48\textwidth]{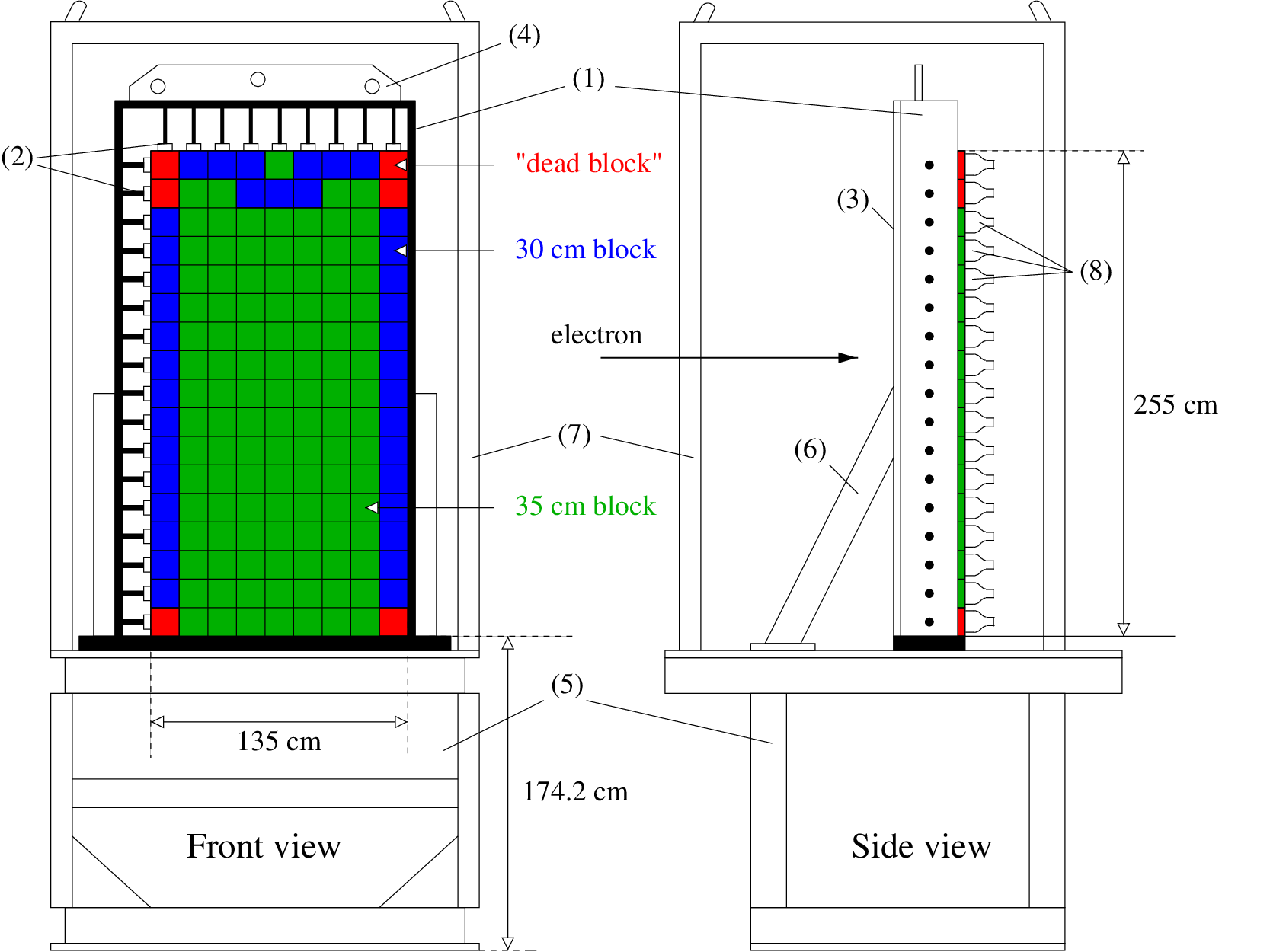}
\caption{(color) Design of the calorimeter used to detect the scattered electron. 
In the front view, the 2.54~cm-thick aluminum plate in front of the blocks
is not shown. See text for details.}
\label{fig:calo}
\end{figure}

For the measurements at $Q^2 \ge 4.0$ GeV$^2$, a lead-glass calorimeter was used to detect
electrons due to the larger electron solid angle compared to the
proton solid angle defined by HRSL. The lead-glass blocks from the standard HRSR calorimeter were used to
assemble this calorimeter along with some additional spare blocks.
Figure~\ref{fig:calo} shows a front and a side view of the calorimeter on its
platform. The blocks of lead-glass, of cross-sectional area 
$15\times 15$~cm$^2$, were individually wrapped in one foil of
aluminized mylar, and one foil of black paper, to avoid light leaks. Each
block was then tested, and the current drawn in the phototube due to
noise was found to be less than 100 nA for all blocks.
The blocks were assembled in a rectangular array of 9 columns and 17
rows, requiring a total of 153 blocks. Most of the 
blocks, in green in Figure~\ref{fig:calo}, were 35~cm long, corresponding to 
13.7~radiation lengths. 37 blocks positioned on the edges of the calorimeter
were only 30~cm long, corresponding to 11.8~radiation lengths (in blue in
Figure~\ref{fig:calo}). The highest electron energy, at $Q^2$ = 4.0~GeV$^2$, was 
2.5 GeV. At this energy, the shower stops after 7.7 radiation 
lengths, so that the entire shower is
contained in the block. Since only 147 blocks were available, 6 blocks were missing to form a complete 
rectangle.
These were replaced by wood placeholder blocks at the corners of the 
detector (in red in Figure~\ref{fig:calo}). The active area of the calorimeter was 3.31~m$^2$. 

The blocks were placed in the steel support frame (1), and held together using
wood plates (2). The front of the support was covered with a
one inch thick aluminum plate
(3) to absorb very low energy particles. The ensemble was lifted by the top
steel plate (4) onto the platform (5) using the Hall A crane. Balance
on the platform was maintained by the steel support legs (6). The
ensemble was placed on wheels and moved with the help of the Hall A crane attached to
the steel lifting frame (7). The calorimeter was placed at the
distance from the target required to match the electron solid angle to
that of the proton at each kinematic setting. The acceptance matching
was only approximate, due to the complicated shape of the spectrometer acceptance. Overall, 
about 5\% of elastic events with a proton detected in HRSL were lost due to acceptance
mismatching. 
The Cherenkov light
emitted by primary electrons and shower secondaries in the lead-glass
was collected by Photonis XP2050 photomultipliers (8), and the signals
were digitized by
LeCroy 1881 integrating ADCs and LeCroy 1877 TDCs. 
The trigger for the measurements at $Q^2 \ge 4.0$ GeV$^2$ was defined by a single proton in the HRS, signaled by a coincidence of 
two planes of scintillators in the focal plane. For each single proton event in the left HRS, the ADC and TDC 
information from the calorimeter was read out for all blocks, and elastic 
events were selected offline by applying software cuts
to the calorimeter data.

\subsection{Comparison to Hall C Experiments}
\label{subsec:hallccomp}
The GEp-II experiment shares several important features with
GEp-III. Both experiments used magnetic
spectrometers instrumented with Focal Plane
Polarimeters (FPPs) to detect protons and measure their polarization,
and large acceptance electromagnetic
calorimeters to detect electrons in coincidence. The use of calorimeters in both
experiments was driven by the
requirement of 
acceptance matching; at large $Q^2$ and $\theta_e$, the
Jacobian of the reaction magnifies the electron solid angle
compared to the proton solid angle fixed by the spectrometer acceptance. The drawbacks of this
choice compared to electron detection using a magnetic spectrometer are
twofold. First, the energy resolution of lead-glass calorimeters is
relatively poor, so that elastic and inelastic reactions are not
well separated in reconstructed electron energy. Second, the signals in lead-glass from
electrons and photons of similar energies are indistinguishable,
leaving one vulnerable to photon backgrounds from the decay of
$\pi^0$, which played an important role in the analysis of both
experiments.

The high-$Q^2$ measurements of the GEp-III experiment \cite{PuckettPRL10} were
carried out consecutively with the GEp-2$\gamma$ experiment, a series
of precise measurements of
$R$ at $Q^2 = 2.5$ GeV$^2$ \cite{MezianePRL11} designed to search for effects beyond the
Born approximation, thought to explain the disagreement between
Rosenbluth and polarization data \cite{TPEXreview}. Using the same
apparatus and analysis procedure as GEp-III, the results of
GEp-$2\gamma$ \cite{MezianePRL11} are in excellent agreement with the
GEp-I data from Hall A \cite{Punjabi05} at nearly identical
$Q^2$, as shown in Figure \ref{ratiofig}. The background
corrections to the GEp-2$\gamma$ data were negligible after applying the cuts
described in \cite{PuckettPRL10,MezianePRL11}. Similarly, electrons
were detected in the HRSL in the GEp-I experiment,
so that the selection of elastic events was practically
background-free~\cite{Punjabi05}. In the absence of major
background corrections, the agreement between precise measurements at
the same $Q^2$ using different polarimeters and magnetic
spectrometers limits the size of 
any potentially neglected systematic errors arising from sources other than
background.


The liquid hydrogen targets used in Halls A and C had radiation lengths
of $\sim$2\%, leading to a significant Bremsstrahlung flux
across the target length, in addition to the virtual photon flux due
to the presence of the electron beam. The kinematics of $\pi^0$
photoproduction ($\gamma + p \rightarrow \pi^0 + p$) near end point
($E_\gamma \rightarrow E_e$) are very similar to elastic $ep$
scattering at high energies ($E_\gamma \gg m_\pi$), such that protons
from $\gamma + p \rightarrow \pi^0 + p$ overlap with elastically scattered protons within
experimental resolution. In the lab frame, asymmetric $\pi^0$ decays with one
photon emitted at a forward angle relative to the $\pi^0$ momentum, carrying
most of the $\pi^0$ energy, are detected with a high
probability. At high energies and momentum transfers, the $\pi^0$
photoproduction cross section is observed to scale as $s^{-7}$ for fixed
$\Theta_{CM}$ \cite{Anderson1976}, where $s$ is the center-of-momentum
(CM) energy squared and $\Theta_{CM}$
is the CM $\pi^0$ production angle. In addition, the CM angular distribution is
peaked at forward and backward angles.
The goal of the GEp-III experiment was to
measure to the highest possible $Q^2$, given the maximum available beam
energy of 5.71 GeV. At $Q^2 = 8.5$ GeV$^2$, the relatively high
$Q^2/s$ ratio, with $\Theta_{CM} \in$ 129-143$^\circ$, led to a $\pi^0p$:$ep$ ratio of $\sim$40:1. 
The severity of the background conditions required
maximal exploitation of elastic kinematics to suppress the $\pi^0$ background. Even after all cuts described in
\cite{PuckettPRL10}, the remaining background was estimated at $\sim$6\% of accepted
events. Given the large difference between the signal and background
polarizations, this level of contamination required a substantial
\emph{positive} correction to $R$. 

In light of the improved understanding of the importance of the
$\pi^0$ background gained during the analysis of the GEp-III data, an underestimation of its
effect in the GEp-II analysis was considered as a potential source of
disagreement between the two experiments. Therefore, the GEp-II data
for $Q^2 = $ 4.0, 4.8, and 5.6 GeV$^2$ were reanalyzed to investigate the systematics
of the $\pi^0$ background. The data from GEp-II at $Q^2=3.5$ GeV$^2$ were not
reanalyzed, since electrons were detected in the HRSR and the $\pi^0$
background was absent. The systematics of this configuration were thus irrelevant to the
comparison between GEp-II and GEp-III at higher $Q^2$. 

\section{Data Analysis}
\label{sec:analysis}
\subsection{Elastic Event Selection}
\label{section:elasticeventselection}
\begin{figure*} 
  \includegraphics[width=.325\textwidth]{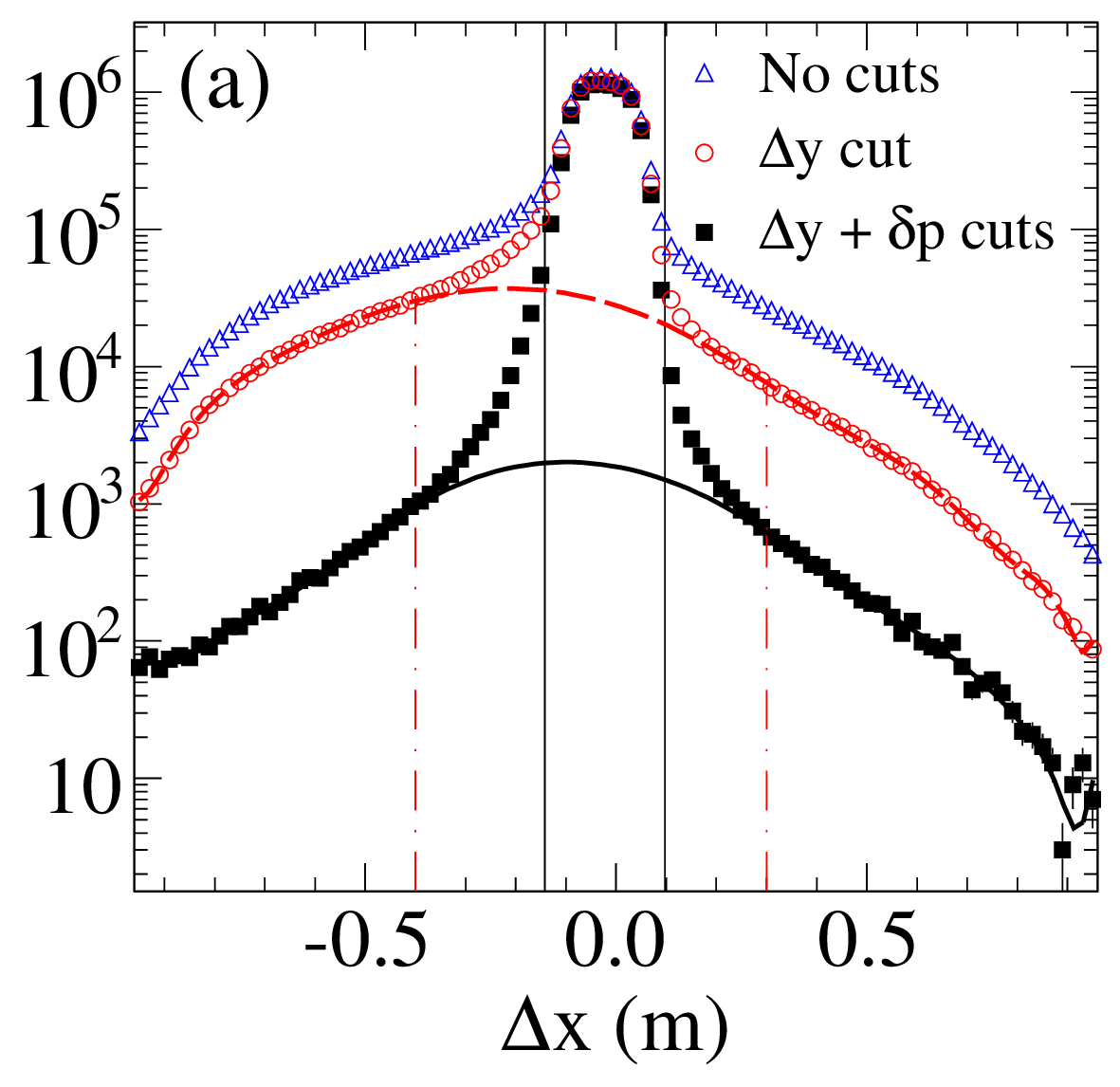}
  \includegraphics[width=.325\textwidth]{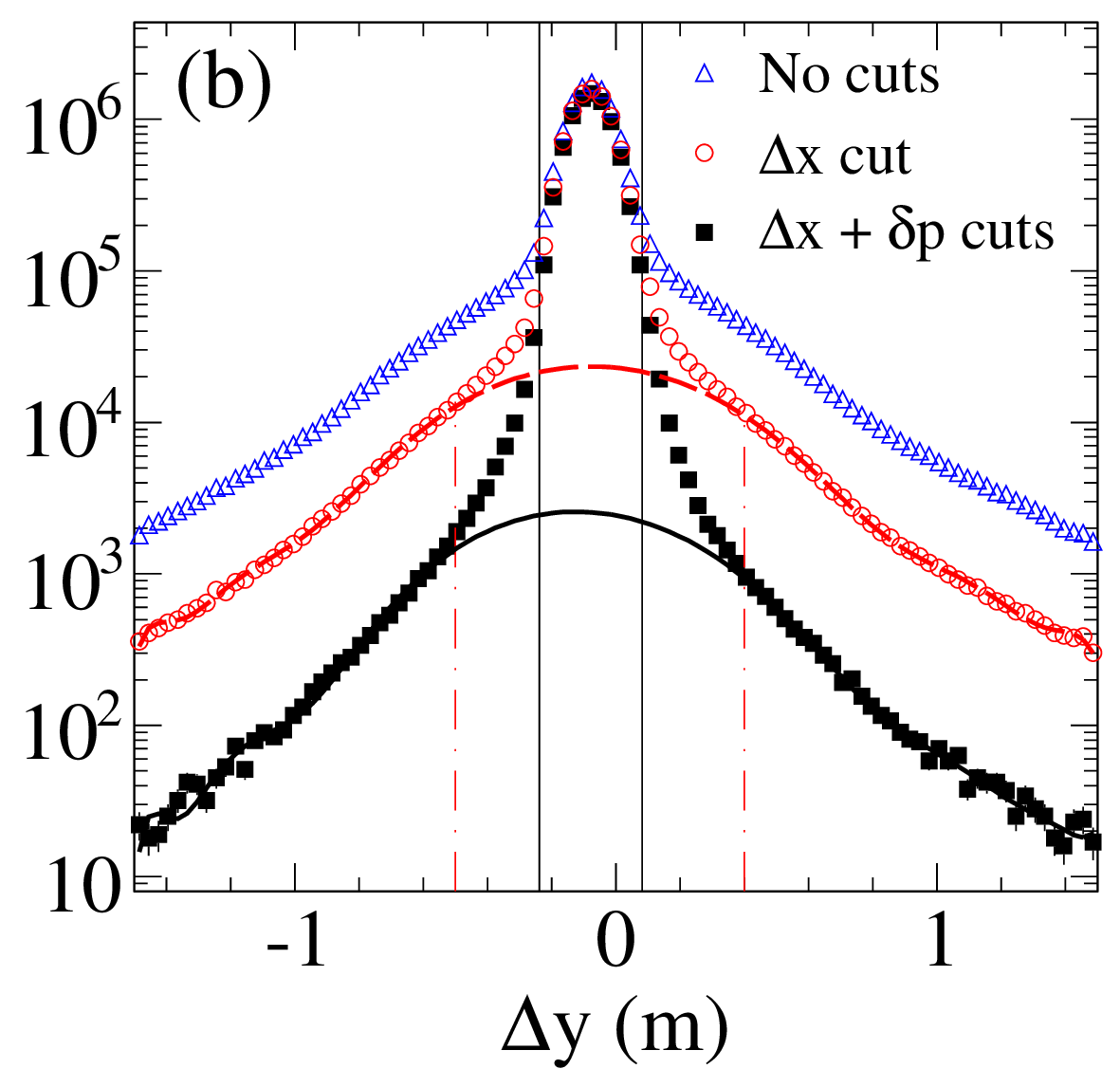}
  \includegraphics[width=.325\textwidth]{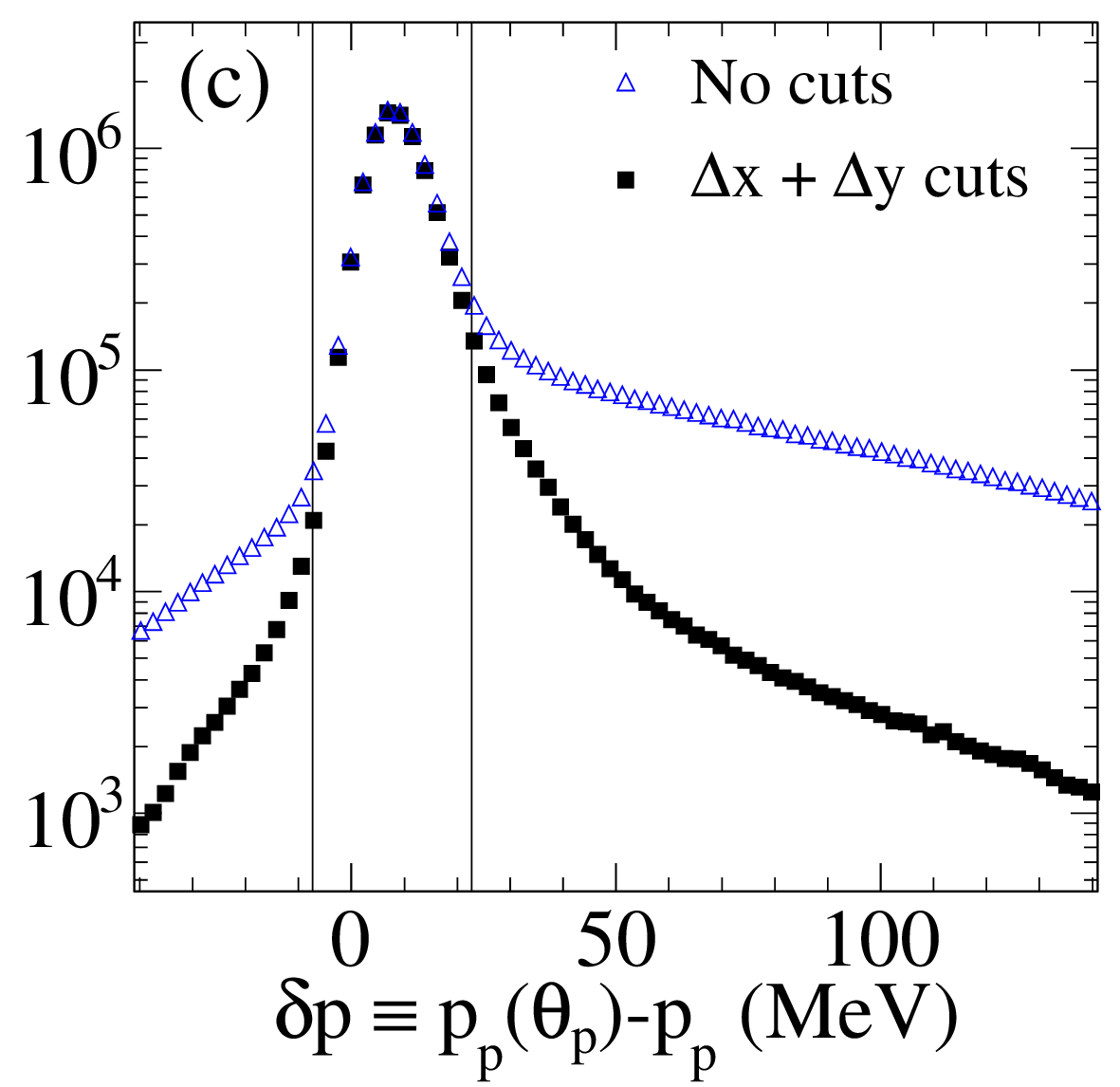}
  \caption{\label{elasticplots} (color online) Elastic event selection at $Q^2 = 4.8$ GeV$^2$. The effects of cuts are shown for the
    horizontal calorimeter coordinate difference $\Delta x$ in panel (a), the
    vertical difference $\Delta y$ in panel (b), and the proton
    momentum difference $\delta p \equiv p_p(\theta_p) - p_p$ in
    panel (c). Solid vertical
    lines indicate the cut applied. Empty triangles show the
    distribution of events before applying any cuts. Filled squares show events passing the cuts on \emph{both} of the
    other two variables. Empty circles in panels (a) and (b) show the $\Delta x$
    ($\Delta y$) distribution of events passing the $\Delta y$
    ($\Delta x$) cut, regardless of $\delta p$. In panels (a) and (b), the dashed and
    solid curves show the estimated background before and after the
    $\delta p$ cut. Dot-dashed vertical lines indicate the range of
    the elastic peak excluded from the fit to the background.}
\end{figure*}
Figure \ref{elasticplots} shows a representative example of the procedure for isolating
elastic events in the GEp-II data, at $Q^2 = 4.8$ GeV$^2$. As
described in \cite{Gayou02} and \cite{GayouThesis}, cuts were
applied to the difference between the HRS and calorimeter time
signals ($\pm 4$ ns at $Q^2 = 4.0$ and 4.8 GeV$^2$, and $\pm 5$ ns at
$Q^2 = 5.6$ GeV$^2$) and the missing
energy $\left(\left|E_{miss} \equiv E_e + M_p - \sqrt{p_p^2+M_p^2} - E_{calo}
\right| \le 1000 \mbox{ MeV}\right)$ to suppress random coincidences and low-energy
inelastic backgrounds, respectively\footnote{The loose missing energy cut reflects the relatively poor energy resolution of lead-glass.}. The remaining backgrounds from
$^1$H$(\gamma, \pi^0 p)$ and quasielastic Al$(e,e'p)$ reactions in the
target cell windows were rejected using
the kinematic correlations between the electron and proton arms. The
measured proton kinematics were used to predict the scattered
electron's trajectory assuming elastic scattering, and then the
predicted electron trajectory, defined by the polar scattering angle
$\theta_e^{(p)}$ and the azimuthal scattering angle $\phi_e^{(p)}$ (where
$^{(p)}$ denotes the value predicted from the measured proton
kinematics), was projected from the measured interaction
vertex\footnote{The interaction vertex is defined as the
  intersection of the beamline with the projection of the reconstructed
  proton trajectory on the horizontal plane.} to the surface of the
calorimeter. 

Figures \ref{elasticplots}(a) and \ref{elasticplots}(b) show
the horizontal ($\Delta x$) and vertical ($\Delta y$) differences
between the measured shower coordinates at the calorimeter and the
coordinates calculated from the measured proton kinematics
assuming elastic scattering. Figure \ref{elasticplots}(c) shows the difference $\delta p \equiv p_p(\theta_p)
- p_p$ between the measured $p_p$ and the momentum required by elastic
kinematics at the measured $\theta_p$, given by
\begin{eqnarray}
     p_p(\theta_p) &=& \frac{2M_pE_e(M_p+E_e)\cos \theta_p}{M_p^2 + 2M_pE_e +E_e^2\sin^2\theta_p}. \label{pthetap}
\end{eqnarray}
In each panel of Figure~\ref{elasticplots}, the distribution of the
plotted variable is shown
before and after applying cuts (illustrated by vertical lines) to
\emph{both} of the other two variables, which most nearly corresponds
to the GEp-III analysis. In addition, the
$\Delta x$ ($\Delta y$) distribution is shown after applying the cut
to $\Delta y$ ($\Delta x$), \emph{regardless of
  $\delta p$}, which most nearly corresponds to the selection of the
original GEp-II analysis, in which no cut was applied to $\delta p$. Each
spectrum exhibits a clear elastic peak near zero on top of  a smooth
background distribution. The background in the $\Delta x$ and $\Delta
y$ spectra is dominated by $\pi^0$ photoproduction events. The estimated background curves shown in
panels (a) and (b) of Figure \ref{elasticplots} were obtained using
the polynomial sideband fitting method described in section
\ref{subsubsec:sideband}. The $\delta p$ cut clearly has significant
additional background suppression power relative to $\Delta x$ and
$\Delta y$ cuts alone.
In the $\delta p$
spectrum, the background distribution is highly asymmetric about the
peak, reflecting the fact that elastically scattered protons carry the
highest kinematically allowed momenta at a given
$\theta_{p}$. 

Since the two-body reaction
kinematics are overdetermined, the method used to calculate $\Delta
x$ and $\Delta y$ is not unique. In combination with the precisely
known beam energy, the expected electron polar scattering angle
$\theta_e^{(p)}$ can
be calculated from either the measured proton momentum $p_p$, the
measured proton scattering angle $\theta_p$, or a combination of
both. Different methods were used by the GEp-II and
GEp-III data analyses to calculate $\Delta x$ and $\Delta y$. In the original
GEp-II analysis, the calculation was formulated in terms of Cartesian
components of the outgoing particle momenta rather than polar and
azimuthal scattering angles. The effective $\theta_e^{(p)}$ in the
GEp-II approach depends on both $\theta_p$ and $p_p$. The exact
equations used can be found in Appendix D of
Ref.~\cite{GayouThesis}. In the GEp-III analysis, $\theta_e^{(p)}$ was calculated from $p_p$, as described in
\cite{PuckettPRL10}. Both methods were tested in the present
reanalysis. The $\Delta x$ and $\Delta y$ distributions in
Figures \ref{elasticplots}(a) and \ref{elasticplots}(b) were calculated
using the GEp-II method, in order to demonstrate the background suppression power of the added $\delta p$ cut of
Fig.~\ref{elasticplots}(c) \emph{relative to
the original analysis}. For events selected by this cut, $p_p
\approx p_p(\theta_p)$, such that the $\Delta x$ values obtained from
the GEp-II and GEp-III methods are equal up to detector resolution.

A key difference between the GEp-III and
GEp-II experiments is the dominant source of resolution in the
variables used to select elastic events. The cell size of the
GEp-II calorimeter was 15 $\times$ 15 cm$^2$, compared to the 4
$\times$ 4 cm$^2$ cell size of the GEp-III calorimeter. In GEp-II, the
resolution of $\Delta x$ and $\Delta y$ is dominated by the
calorimeter coordinate measurement, and is therefore largely insensitive to the choice of proton variables used
to calculate the expected electron angles. In GEp-III, on the other hand, the scattered electron
angles were measured with excellent resolution by the highly-segmented
BigCal, such that the proton arm resolution was dominant. Given
the kinematics of GEp-III and the angular and momentum resolution of the High
Momentum Spectrometer (HMS) in
Hall C \cite{HornLongPaper}, the best $\Delta x$ resolution was
obtained by using $p_p$ to
calculate $\theta_e^{(p)}$. In the GEp-II analysis, the main practical difference between the two methods
is the resulting background shape. In kinematics for which the reaction
Jacobian necessitates the use of a calorimeter for electron detection,
the GEp-III method generally results in a wider and more asymmetric
$\Delta x$ distribution of the background, with inelastic events
assuming predominantly negative $\Delta x$ values. In the GEp-III analysis, using $\theta_e^{(p)}(p_p)$
provided the best possible $\Delta x$ resolution \emph{and} a wider $\Delta x$
distribution of the background. In the GEp-II case, calculating
$\Delta x$ using the GEp-III method spreads out the background without
affecting the width of the elastic peak, thus reducing the background in
the $\Delta x$ spectrum with no $\delta p$ cut. After applying the $\delta
p$ cut, however, the $\Delta x$ distributions obtained from the two
calculations are practically identical, and the choice becomes
arbitrary. As discussed in section~\ref{sec:systematics}, the results
for $R$ obtained with the $\delta p$ cut included do not depend on the method
used to calculate $\Delta x$. The final reanalysis results were
obtained with $\Delta x$ and $\Delta y$ calculated using the GEp-II method.

The original GEp-II analysis applied a two-dimensional polygon cut
to the correlated $\Delta y$ versus $\Delta x$ distribution. Using
identical cuts to the original analysis, the published results
\cite{Gayou02} were successfully reproduced. In the final analysis,
however, one-dimensional (rectangular) cuts were applied to $\Delta x$ and $\Delta
y$, which simplifies the background estimation
procedure. For all three $Q^2$ points, a cut of $\pm 12( \pm 16)$ cm was
applied to $\Delta x$($\Delta y$), centered at the midpoint between
half-maxima on either side of the elastic
peak, as in Figure \ref{elasticplots}. The width of the cuts was
chosen to be similar to the effective width
of the polygon cut applied by the original analysis, and reflects the
dominant contribution of the calorimeter cell size to the resolution
of $\Delta x$ and $\Delta y$. In addition, a cut of $\pm 15$ MeV, also
centered at the midpoint between half-maxima of the elastic peak, was
applied to $\delta p$, as in Figure \ref{elasticplots}(c). The width
of the $\delta p$ cut was chosen to be $\pm 3\sigma$, where $\sigma
\approx 5$ MeV
is the $\delta p$ resolution, which was roughly independent of the proton momentum in this experiment. 
While the difference
in the selection of events from using a different shape of the
$\Delta x$ and $\Delta y$ cuts is small, the $\delta p$ cut removes a
rather substantial 6.0\%, 7.3\%, and 10.7\% of events relative to the original analysis
for $Q^2 = 4.0$, 4.8 and 5.6 GeV$^2$, respectively. 

While a fraction of the events rejected by the $\delta p$ cut are elastic, including
events in the $ep$ radiative tail and elastic events with $\delta p$
smeared by non-Gaussian tails of the HRS resolution, most
of the rejected events are part of the background, and contribute very
little to the statistical precision of the data. Moreover, even real elastic
events reconstructed outside the peak region of $\delta p$ do not meaningfully
contribute to the accurate determination of the form factor ratio, because such
events are either (a) part of the radiative tail and therefore subject to
radiative corrections that are in principle
calculable \cite{AfanasevRadCorr} but practically difficult due to
large backgrounds in the radiative tail region, or (b) have unreliable
angle or momentum reconstruction, which distorts the spin transport
matrix of the HRS (see Ref.~\cite{Ron11} and section \ref{subsubsec:precession} below)
in an uncontrolled fashion. Therefore, the application of the
$\delta p$ cut has benefits beyond mere background suppression, as it
also suppresses radiative corrections and the (potential) systematic
effects of large angle or momentum reconstruction errors. The
estimation of the background contamination and the background-related corrections to the polarization
transfer observables are discussed in section
\ref{subsec:background}. The next section discusses the procedure for the extraction of polarization observables from the
``raw'' asymmetries measured by the FPP.

\subsection{Extraction of Polarization Observables}
\label{section:polextract}
As detailed in \cite{Punjabi05,GayouThesis}, useful scattering events
in the FPP were selected by requiring a good reconstructed track in
both the front and
rear straw chambers and requiring the scattering vertex $z_{close}$,
defined by the point of closest approach between incident and
scattered tracks, to lie within the physical extent of the CH$_2$
analyzer. Events with polar scattering angles $\vartheta < 0.5^\circ$
were rejected, since at small angles comparable to the angular
resolution of the FPP, the azimuthal angle resolution diverges. Moreover, the small-angle region is dominated by multiple Coulomb
scattering, which has zero analyzing power. 
\subsubsection{Focal Plane Asymmetry}
Spin-orbit coupling causes a
left-right asymmetry in the angular distribution of protons scattered
by carbon and hydrogen nuclei in the CH$_2$ analyzer of the FPP
with respect to the transverse polarization of the incident
proton\footnote{In this context, ``transverse'' means orthogonal to the
  incident proton's momentum direction.}. The
measured angular distribution for incident protons with momentum $p$
and transverse polarization components $P_x^{FPP}$ and
$P_y^{FPP}$ for a beam helicity of $\pm 1$ can be expressed as\footnote{In the assumed coordinate system, the $z$ axis is
along the incident proton momentum, while the $x$ and $y$ axes describe
the transverse coordinates in relation to the proton trajectory and the
detector coordinate system, as described in the text.}
\begin{eqnarray}
  N^\pm (p,\vartheta,\varphi) &=& N_0^\pm \frac{\varepsilon(p,\vartheta)}{2\pi} \left[1
    + (\pm A_y P_x^{FPP} + c_1) \cos \varphi + \right. \nonumber \\ 
  & & \left. (\mp A_y P_y^{FPP} + s_1) \sin \varphi + \right. \nonumber \\ 
  & & \left. c_2 \cos (2\varphi) + s_2 \sin (2\varphi) + \ldots \right], \label{fpasym}
\end{eqnarray}
where $N_0^\pm$ is the total number of incident protons for beam
helicity $\pm 1$, $\varepsilon(p,\vartheta)$ is the polarimeter \emph{efficiency}
defined as the fraction of protons of momentum $p$ scattered
at an angle $\vartheta$, $A_y(p,\vartheta)$ is the analyzing power of
the $\vec{p}+\mbox{CH}_2$ reaction, and $\varphi$ is the azimuthal scattering angle. The
additional terms $c_1, s_1, c_2, s_2, \ldots$ represent false or
instrumental asymmetries caused by non-uniform acceptance or
efficiency, and possible $\varphi$-dependent reconstruction errors. These
terms depend on $p$, $\vartheta$, and the incident proton
trajectory, on which the geometric acceptance depends. Normalized angular distributions $n_\pm \equiv N^\pm(\varphi)/N_0^\pm$
can be defined for each helicity state. The helicity-sum distribution
$n_+ + n_-$ cancels the helicity-dependent asymmetries corresponding to the
transferred polarization, providing
access to the false asymmetries, while the helicity-difference
distribution $n_+ - n_-$ cancels the helicity-independent false asymmetries, providing
access to the physical asymmetries.

False asymmetry effects are strongly suppressed in
the extraction of the transferred polarization components by the rapid
(30 Hz) beam helicity reversal, which cancels the false
asymmetry contribution (to first order) and also cancels slow variations of luminosity and
detection efficiency, resulting in the same effective integrated
luminosity for each beam helicity state. Since the elastic scattering cross section on an
unpolarized proton target is independent of electron helicity, equal numbers of protons incident on CH$_2$ are detected for positive and
negative beam helicities. In the GEp-II experiment, the numbers of
events in each helicity state were always found to be equal within statistical
uncertainties at the 10$^{-4}$ level. In a polarization transfer measurement, equal integrated luminosities for each beam helicity
are \emph{not} strictly required to robustly separate the physical from the
instrumental asymmetries, since the angular distribution can be
normalized to the number of incident protons for each
helicity state. Nonetheless, having equal numbers of events in
each helicity state maximizes the statistical precision of the measured
asymmetry while minimizing the systematic uncertainty in its
extraction. The false asymmetry coefficients determined from Fourier
analysis of the helicity-sum distribution can be used to correct the
residual second-order effect of the false asymmetry, which is small
compared to other uncertainties in the data of this experiment and therefore neglected (see
section \ref{subsubsec:falseasym}).

\begin{figure}
  \begin{center}
    \includegraphics[width=.48\textwidth]{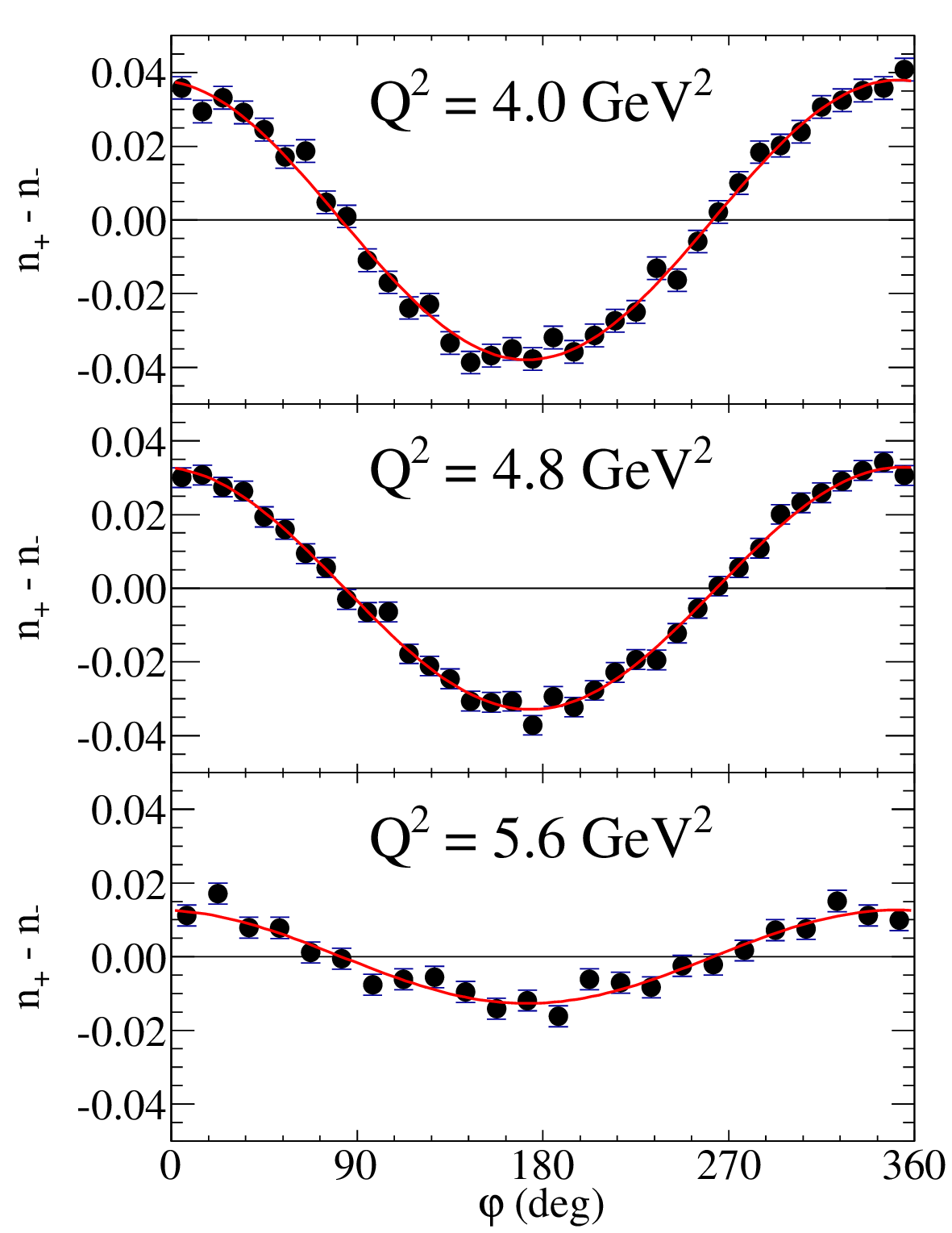}
  \end{center}
  \caption{\label{fig:fpasym} (color online) Focal-plane helicity-difference
    asymmetry $n_+ - n_- \equiv
    \left(N_{bins}/2\right)\left[N^+(\varphi)/N_0^+ -
      N^-(\varphi)/N_0^- \right]$, where $N_{bins}$ is the
    number of $\varphi$ bins and $N^\pm(\varphi), N_0^\pm$
    are defined as in equation \eqref{fpasym}, for the three highest $Q^2$
    points from GEp-II. Curves are fits to the data. See text for details.}
\end{figure}
Figure \ref{fig:fpasym} shows the helicity-difference asymmetry $n_+ -
n_-$ for the three highest $Q^2$ points from GEp-II, integrated over
the range of polar angles
$\vartheta$ with non-zero analyzing power. The data were fitted with
$n_+ - n_- = a\cos \varphi + b \sin \varphi$, with a resulting
$\chi^2/\mbox{ndf}$ of 0.90, 0.53 and 0.92 for $Q^2 = 4.0$, 4.8 and 5.6
GeV$^2$, respectively. At each $Q^2$, the asymmetry exhibits a
clear sinusoidal behavior, with a large $\cos \varphi$ amplitude
proportional to $P_x^{FPP}$ and a smaller $\sin \varphi$ amplitude
proportional to $P_y^{FPP}$. There is no evidence in the
data for a constant offset or the presence of higher harmonics,
judging from the good $\chi^2$ of the fit with only $\cos
\varphi$ and $\sin \varphi$ terms\footnote{Fits with Fourier
  modes up to $4\varphi$ and a constant term found that the
  coefficients of all terms other than $\cos \varphi$ and $\sin
  \varphi$ were zero within statistical uncertainties.}. The amplitude of the asymmetry is
proportional to the product of the weighted-average analyzing power
and the magnitude of the proton polarization, while the phase of the
asymmetry is determined by the ratio $P_y^{FPP}/P_x^{FPP}$ of the
proton's transverse polarization components at the focal plane.

\subsubsection{Spin Precession}
\label{subsubsec:precession}
The asymmetry measured by the FPP is determined by the proton's
transverse polarization after
undergoing spin precession in the magnets of the HRS. To
extract the transferred polarization components at the target
corresponding to equations \eqref{recoilformulas} requires accurate
knowledge of the spin transport properties of the HRS. It is worth
noting that without spin precession in magnetic spectrometers, a
common feature of the GEp-I, GEp-II, GEp-III and GEp-2$\gamma$ experiments, proton polarimetry based on
nuclear scattering would not work, since the spin-orbit coupling
responsible for the azimuthal asymmetry is
insensitive to the proton's longitudinal polarization, which can only
be measured by rotating the longitudinal component into a transverse component. 

The precession of the spin of particles moving
relativistically in a magnetic field is governed by the
Thomas-B.M.T. equation \cite{BMTequation}. The dominant precession
effect in all of the aforementioned experiments is caused by the large
vertical bend of the proton trajectory in the dipoles
of the magnetic spectrometers. In first approximation, the proton
spin precesses in the dispersive (vertical) plane by an angle $\chi = \gamma
\kappa_p \theta_{bend}$ relative to the proton trajectory, where
$\gamma^2 = 1+p_p^2/M_p^2$ is the proton's relativistic boost
factor, $\kappa_p$ is the proton's anomalous magnetic moment, and
$\theta_{bend}$ is the vertical trajectory bend angle. In this
idealized approximation, the proton spin does not precess in the
horizontal plane. The sensitivity of the FPP
asymmetry to $P_\ell$ is maximized when $\left|\sin \chi\right| =
1$. The central values of $\chi$ for the four kinematic settings of
GEp-II are given in Table \ref{kintable}. 

Because the
central value of $\chi$ is close to 360$^\circ$ at $Q^2 = 5.6$
GeV$^2$ and the range of $\chi$ accepted by the HRSL is roughly $285^\circ \le \chi \le
390^\circ$, the dominant $\cos \varphi$ amplitude of the focal plane
asymmetry, which is roughly
proportional to $P_\ell \sin \chi$, is reduced when averaged over the full
$\chi$ acceptance, as in the bottom panel of Figure
\ref{fig:fpasym}. However, the adverse impact of the unfavorable precession
angle on the precision of the data is
mitigated by the large $\chi$ acceptance of the HRS and the
fact that $P_\ell$ is quite large for the kinematics in
question. The $\chi$-dependence of the asymmetry is accounted for by
the weighting of events in the unbinned maximum-likelihood analysis
described below, which optimizes the statistical precision of the
extraction without explicitly removing events near $\chi =
360^\circ$. Moreover, the $\chi$ and $Q^2$ acceptances of the HRSL are
only weakly correlated, so that the range of $Q^2$ contributing to the
determination of $R$ is not strongly affected.  

The presence of quadrupole magnets complicates the spin transport
calculation by introducing precession in the horizontal
(non-dispersive) plane, which
mixes $P_t$ and $P_\ell$. The trajectory bend angle in the
non-dispersive plane is zero for the spectrometer central ray, but 
non-zero for trajectories with angular and/or spatial deviations from the
HRS optical axis. Because of the strong in-plane angle ($\theta_p$)
dependence of the cross section, the acceptance-averaged horizontal
precession angle is generally significantly non-zero. The quadrupole
effects are qualitatively characterized by the non-dispersive precession angle $\chi_\phi \equiv
\gamma \kappa_p \phi_{bend}$, where $\phi_{bend}$ is the total
trajectory bend angle in the non-dispersive plane. 

The spin transport calculation for the final analysis was performed
using COSY \cite{COSY}, a differential algebra-based software library
for charged particle optics and other applications. Since each proton
trajectory through the HRS magnets is unique, the spin transport
matrix must
be calculated for each event. Rather than perform a computationally
expensive numerical integration of the BMT equation for each proton
trajectory, a polynomial expansion of the forward spin transport
matrix up to fifth order in the proton trajectory angles, vertex
coordinates and momentum was fitted to a sample of random test
trajectories that were propagated through a detailed layout of the HRS
magnetic elements including fringe fields. The coefficients of this
polynomial expansion were then used to calculate the spin rotation
matrix for each event. Unlike the optics matrices used for particle
transport, which are independent of the HRS central momentum setting
due to the fixed central bend angle, the spin transport matrix depends
on the central momentum setting because the precession frequency relative to the proton trajectory is
proportional to $\gamma$. Therefore, the fitting
procedure for the COSY matrices had to be carried out separately for
each $Q^2$. The Taylor expansion of
the matrix elements in powers of the small deviations from the central
ray within the acceptance of the HRSL converges quite rapidly to an accuracy better than the
spectrometer resolution.

Several coordinate rotations are involved in the calculation of the
spin transport matrix elements for each event. First, the reaction plane
coordinate system defines $P_t$ and $P_\ell$: $P_\ell$ is directed along the
recoiling proton's momentum and $P_t$ is transverse to the proton
momentum but parallel to the scattering plane, in the direction
of decreasing $\theta_p$. A rotation is applied
from the reaction plane to the fixed transport coordinate system in
which the $z$-axis is along the HRS optical axis, the $x$-axis points
along the dispersive plane in the direction of increasing particle
momentum (vertically downward), and the $y$-axis is chosen as $\hat{y}
= \hat{z} \times \hat{x}$ so that
$(\hat{x},\hat{y},\hat{z})$ forms a right-handed Cartesian coordinate
system. The COSY
calculations are performed in this fixed coordinate system. After
applying the COSY rotation, which transports the spin from the target
to the focal plane in transport coordinates, a final rotation is
applied to express the rotated spin vector in the comoving coordinates of the proton
trajectory at the focal plane, in which the $z$ axis is along the proton
momentum, the $y$ axis is chosen perpendicular to the proton momentum
and parallel to the $yz$ plane of the transport coordinate
system, and the $x$ axis is chosen as $\hat{x} = \hat{y} \times
\hat{z}$. The definition of the $x$ and $y$ axes of the comoving
coordinate system at the focal plane is arbitrary as long as
it is applied consistently with the other coordinate systems
involved. In the original analysis of GEp-II
\cite{Gayou02,GayouThesis}, the azimuthal FPP scattering angle $\varphi$
was measured counterclockwise from the $y$ axis toward the $x$ axis, as
viewed along the $z$ axis. This convention is also used in the
present work, but it is worth noting that a different convention was
used in the analysis of the
GEp-III \cite{PuckettPRL10} and GEp-2$\gamma$ \cite{MezianePRL11}
experiments, in which $\varphi$ was measured clockwise from the $x$
axis toward the $y$ axis. 

The observables $P_t$, $P_\ell$, and $R$ were extracted from the data
using an unbinned maximum-likelihood method. Up to an overall
normalization constant independent of $P_t$ and $P_\ell$, the
likelihood function is given by
\begin{eqnarray}
  \mathcal{L}(P_t,P_\ell) &=& \prod_{i=1}^{N_{event}}
  \frac{1}{2\pi}\bigg[ 1+ \lambda_0(\varphi_i) +
  \nonumber \\ 
  & &  h_i P_e A_y^{(i)} \left( (S_{xt}^{(i)} P_t + S_{x\ell}^{(i)} P_\ell)\cos
      \varphi_i - \right. \nonumber \\ 
  & & \left. (S_{yt}^{(i)} P_t + S_{y\ell}^{(i)} P_\ell) \sin
    \varphi_i \right) \bigg], \label{likelihoodfunction}
\end{eqnarray} 
where $\lambda_0$ represents the sum of all false asymmetry terms,
$h_i$ and $P_e$ are the beam helicity and polarization, respectively,
$A_y^{(i)}$ is the analyzing power, and the $S_{jk}^{(i)}$ with $j=x,y$ and
$k=t,\ell$ are the spin transport matrix elements. The values of $P_t$
and $P_\ell$ extracted by maximizing the likelihood function
\eqref{likelihoodfunction} correspond to those of equations
\eqref{recoilformulas} in the case $P_e = 1$, i.e., the beam is 100\%
polarized. Converting the product over all events into a sum by
taking the logarithm and keeping only terms up to second order in the
Taylor-expansion\footnote{The maximum truncation error in the expansion of the logarithm for $x
= 0.1$, an upper limit corresponding to the largest
$\vartheta$-dependent asymmetries observed
in the GEp-II data, is approximately 0.3\% (relative).} of the logarithm ($\ln(1+x) = x - x^2/2 +
\mathcal{O}x^3$, where $x$ corresponds to the asymmetry) reduces the coupled, nonlinear system of partial differential equations to a
linear system of algebraic equations for the polarization transfer components:
\begin{eqnarray}
  \left(\begin{array}{cc} \left(\lambda_t^{(i)}\right)^2
      & \lambda_t^{(i)} \lambda_\ell^{(i)} \\ \lambda_t^{(i)}
      \lambda_\ell^{(i)} &
      \left(\lambda_{\ell}^{(i)}\right)^2 \end{array}\right)\left(\begin{array}{c}
      P_t \\ P_\ell \end{array}\right)
  &=& \left(\begin{array}{c} \lambda_t^{(i)}(1
      - \lambda_0^{(i)})
      \\ \lambda_\ell^{(i)}(1-\lambda_0^{(i)}) \end{array}\right), \label{weightedsum} 
\end{eqnarray}
in which a sum over all events ($\sum_{i=1}^{N_{event}}$) is implied, and the coefficients $\lambda_t$ and $\lambda_\ell$ are defined
for the $i^{th}$ event as 
\begin{eqnarray}
  \lambda_t^{(i)} &\equiv& h_i P_e A_y^{(i)} \left(S_{xt}^{(i)} \cos
    \varphi_i - S_{yt}^{(i)} \sin \varphi_i \right) \nonumber \\
  \lambda_\ell^{(i)} &\equiv& h_i P_e A_y^{(i)} \left(S_{x\ell}^{(i)}
    \cos \varphi_i - S_{y\ell}^{(i)} \sin \varphi_i\right).
\end{eqnarray}

Equation~\eqref{weightedsum} can be written as a matrix
equation $M \mathbf{P} = \mathbf{b}$, where $M$ is the $2\times 2$
matrix of sums multiplying the vector $\mathbf{P}$ of
polarization transfer components, and $\mathbf{b}$ is the vector of
sums on the right-hand-side of
\eqref{weightedsum}. The solution of this equation is $\mathbf{P} =
M^{-1} \mathbf{b}$, and the standard statistical variances in $P_t$
and $P_\ell$ are
obtained from the diagonal elements of the covariance matrix
$M^{-1}$. The corresponding statistical error in $R = \mu_p
G_E^p/G_M^p$ is obtained by appropriate error propagation through
equation~\eqref{recoilformulas}. The kinematic factor in equation~\eqref{recoilformulas} is calculated for each event from the
reconstructed kinematics, and is averaged over all events in the
calculation of $R$. Since the reconstruction of the kinematics
is not unique and can be fixed by choosing any two of $E_e$, $E'_e$,
$\theta_e$, $p_p$ and $\theta_p$, the choice was made to use the
quantities measured with the highest precision, namely $p_p$ and
$E_e$, to calculate $Q^2$
and $\epsilon$ for each event. The kinematic factor
$\sqrt{\tau(1+\epsilon)/2\epsilon}$ is known to a much better accuracy
than the statistical and systematic accuracy of $P_t/P_\ell$
and therefore makes a negligible contribution to the total
uncertainty. 

It is worth remarking that ``bin centering''
effects due to the finite $Q^2$ and $\epsilon$ acceptance within each
data point are essentially negligible, since the $Q^2$ acceptance is small compared
to the magnitude of $Q^2$. The difference between the
average value of the kinematic factor
$\sqrt{\tau(1+\epsilon)/2\epsilon}$ and its value calculated at the
average $Q^2$ is negligible compared to the uncertainty in the ratio
$P_t/P_\ell$. Furthermore, both the observed and expected\footnote{Expected
  variations are based on
  the best current knowledge
  of the $Q^2$ dependence of $G_E^p/G_M^p$.} variations of $P_t$, $P_\ell$ and $R$ within
the acceptance of each data point are small compared to their
statistical uncertainties. Therefore, all data from each $Q^2$ point are combined into
a single result quoted at the average $Q^2$.

The forward spin transport matrix depends on all parameters of the scattered
proton trajectory before it enters the HRSL. Since the expected variation
of $R$ within the acceptance of
each data point is small, any anomalous dependence of the
extracted $R$ on the reconstructed proton trajectory parameters is a
signature of problems with the spin transport calculation. Conversely,
the absence of anomalous dependence serves as
a powerful data quality check.
\begin{figure}
  \begin{center}
    \includegraphics[width=.49\textwidth]{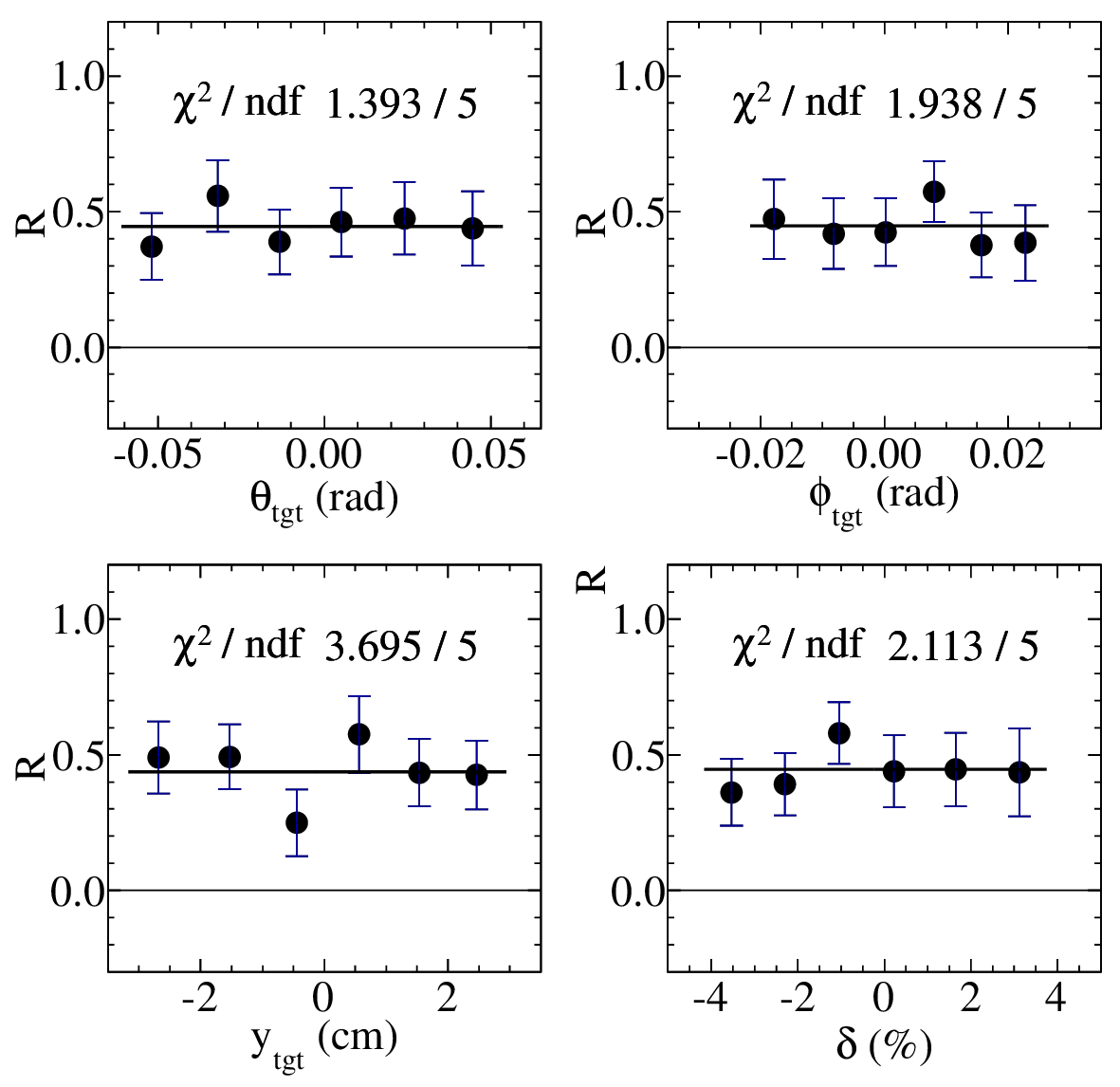}
  \end{center}
  \caption{\label{fig:precessioncheck} Dependence of extracted $R =
    \mu_p G_E^p/G_M^p$ values on the reconstructed proton trajectory parameters
    at $Q^2 = 4.8$ GeV$^2$: $\theta_{tgt}$ ($\phi_{tgt}$) is the
    proton trajectory angle relative to the HRS optical axis in the
    dispersive (non-dispersive) plane, $y_{tgt}$ is the position of
    the interaction vertex in spectrometer coordinates (see text for
    details), and $\delta$ is the percentage deviation
    of the proton momentum from the central HRS momentum setting. In
    each panel, the data are integrated over the
    other three variables.}
\end{figure}
Figure \ref{fig:precessioncheck} shows the dependence of $R$ at $Q^2 =
4.8$ GeV$^2$, extracted using equation \eqref{weightedsum}, on all
four proton trajectory parameters that enter the spin transport
calculation. These include the trajectory angles $\theta_{tgt} =
\tan^{-1} (dx/dz)$ and $\phi_{tgt} = \tan^{-1} (dy/dz)$ relative to
the HRS optical axis, the vertex coordinate $y_{tgt}$, defined as the
horizontal position of the intersection of the proton trajectory
with the plane normal to the HRS optical axis containing the
origin\footnote{Assuming that the HRSL points at
  the origin of Hall A, $y_{tgt}$ is related to the position $z_{vtx}$ of the
  interaction point along the beamline by $y_{tgt} = -z_{vtx} \left( \sin
    \Theta_p + \cos \Theta_p \tan \phi_{tgt} \right)$, where $\Theta_p$ is the
  HRS central angle (given as $\theta_p$ in Table \ref{kintable}).},
and $\delta \equiv 100 \times
(p-p_0)/p_0$, the percentage deviation of the measured proton momentum
from the HRS central momentum setting. There is no evidence for a
dependence of $R$ on any of the variables involved in the precession
calculation, indicating the excellent quality of the COSY
model. Linear and quadratic fits to the individual dependencies were also performed, and all
non-constant terms included in the fits were found to be consistent
with zero.

\subsubsection{Analyzing Power Calibration}
\label{subsubsec:Ay}
The $\vec{p}+\mbox{CH}_2$ analyzing power relating the size of the
measured asymmetry to the proton polarization depends on the initial
proton momentum and the scattering angle
$\vartheta$. Given the relatively small momentum acceptance of the
HRS, the $p$-dependence of $A_y$ within the acceptance of each $Q^2$
point is much weaker than the very strong $\vartheta$ dependence, and
can be neglected as a first approximation. Dedicated measurements of $A_y$
\cite{Ay_pCH2} at and above the momentum range of the GEp-II
experiment were performed prior to the GEp-III experiment. However,
precise independent knowledge of $A_y$ is not required in the analysis
because of the self-calibrating
nature of elastic $ep$ scattering, explained below.

\begin{figure}
  \begin{center}
    \includegraphics[width=.44\textwidth]{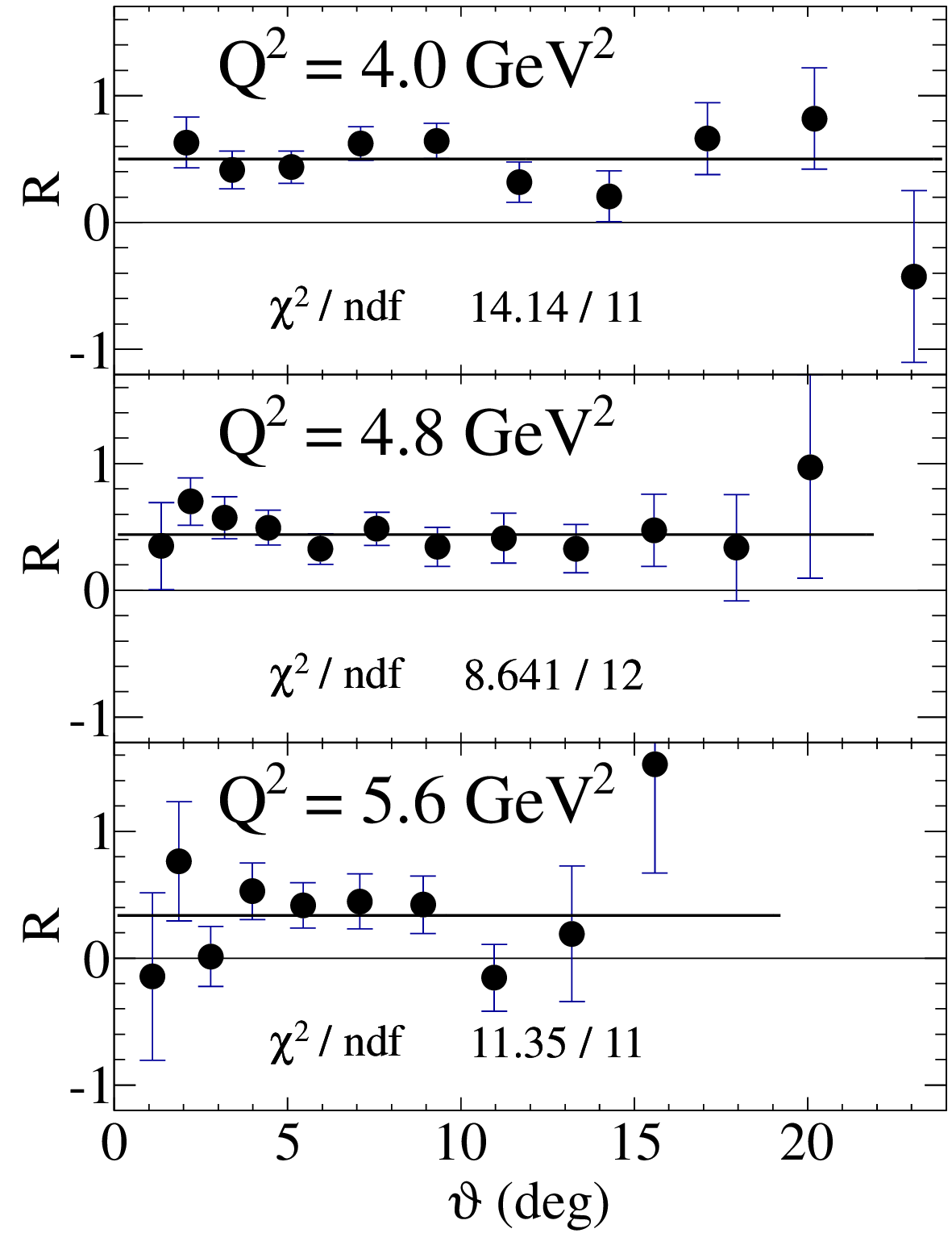}
  \end{center}
  \caption{\label{fig:rthfpp} Dependence of the form factor ratio $R$ on the FPP polar
    scattering angle $\vartheta$ for the three highest
    $Q^2$ values of GEp-II. The constant behavior of $R$
    confirms the cancellation of $A_y$ in the ratio $P_t/P_\ell$. }
\end{figure}
Provided that the effective
$\vartheta$ acceptance is $\varphi$-independent, the analyzing
power cancels in the ratio $P_t/P_\ell$ from which the form factor
ratio $R$ is extracted, implying that the result for $R$ is
independent of $A_y$. Uniform $\vartheta$ acceptance is guaranteed
by applying a ``cone test'' in the selection of FPP events, which
requires that the projection to the rearmost FPP detector plane of a
track originating at the reconstructed $\vec{p}+$CH$_2$ scattering
vertex $z_{close}$ at a polar angle $\vartheta$ falls within
the active detector area for all azimuthal
angles $\varphi$. Moreover, the cancellation can be verified by binning
the results in $\vartheta$ and checking the constancy of $R
\propto P_t/P_\ell$ as a function of
$\vartheta$. Figure \ref{fig:rthfpp} shows the $\vartheta$ dependence
of $R$ for the three highest $Q^2$ points of GEp-II. At each $Q^2$, a
constant fit to the data gives a good $\chi^2$ and no systematic
trends are observed. 

\begin{figure}
  \begin{center}
    \includegraphics[width=.48\textwidth]{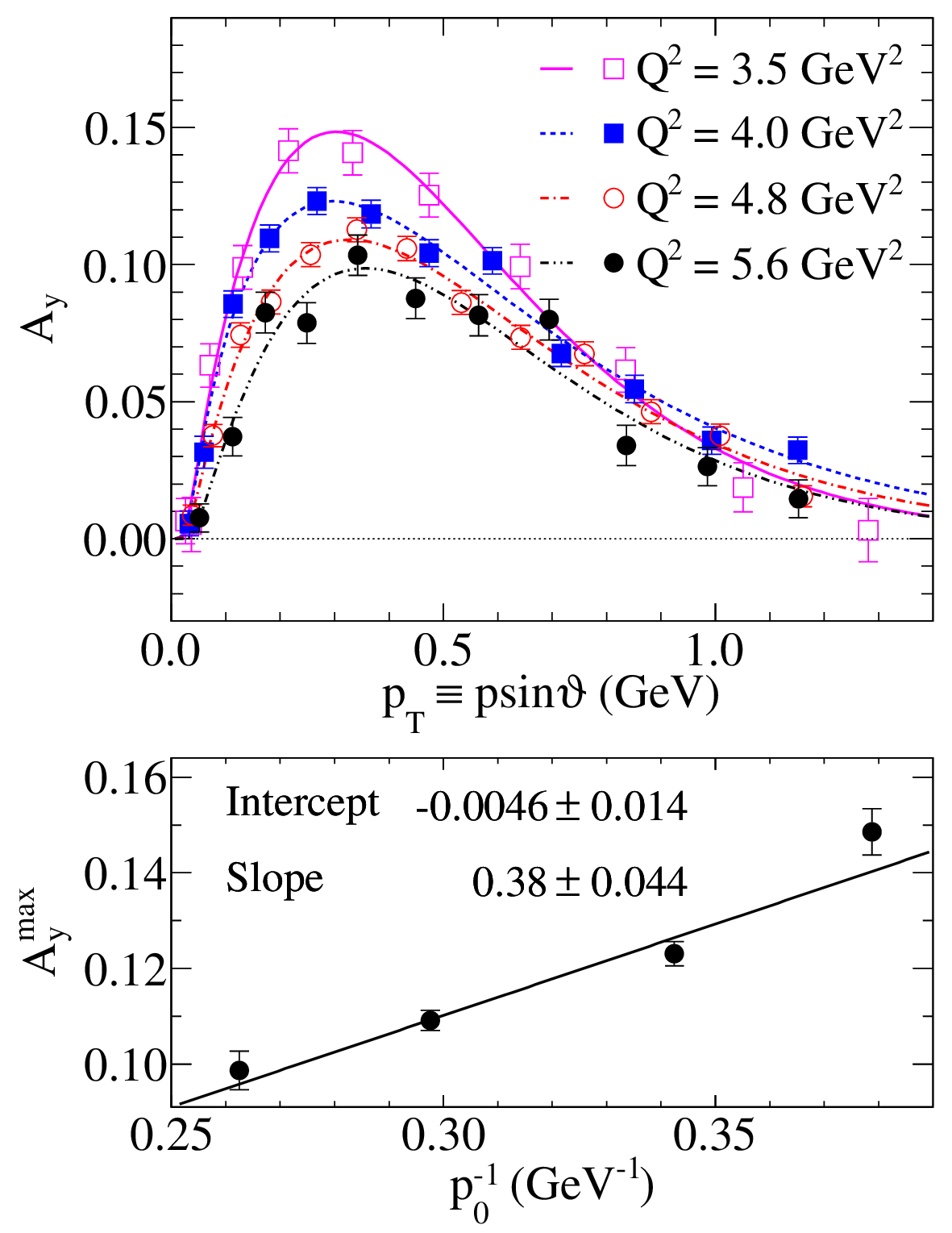}
    \caption{\label{fig:Ay} (color online) Top panel: extracted analyzing power as a function of $p
      \sin \vartheta$, where $p$ is the proton momentum incident on
      CH$_2$ (corrected event-by-event for energy loss in CH$_2$ up to
      the reconstructed scattering vertex), using the $P_e$
      values of Table \ref{kintable}, for all four $Q^2$ values of the
      GEp-II experiment. Curves are fits to the data (see text
      for details). Bottom panel: maximum analyzing power vs. $1/p_0$ in
      GeV$^{-1}$, where $p_0$ is the central proton momentum, for the
      four $Q^2$ points. Error bars in $A_y$ and $A_y^{max}$ values are statistical only. See Supplemental Material at \cite{SupplementalOnlineMaterial} for data tables with numerical $A_y$ and $A_y^{max}$ results.}
  \end{center}
\end{figure} 
The fact that $P_t$ and $P_\ell$ depend only on
$R$ and kinematic factors implies that the product $P_eA_y$ can be extracted by
comparing the measured asymmetries $P_e A_y P_t$ and $P_e A_y P_\ell$
to the values of $P_t$ and $P_\ell$ obtained from equation
\eqref{recoilformulas}. Combined with the measurements of $P_e$ to within an overall accuracy of $\pm 3\%$ by M\"{o}ller and
Compton polarimetery, $A_y$ was directly extracted from the data of
this experiment. The $p$ and $\vartheta$ dependences of $A_y$ thus
obtained were then used in equation \eqref{weightedsum} to improve the
statistical precision of the form factor ratio extraction by
weighting events according to their analyzing power.

Figure \ref{fig:Ay} shows the measured
$A_y$ as a function of the ``transverse momentum'' $p_T \equiv p \sin \vartheta$ for each $Q^2$ point, where $p$ is the incident proton
momentum corrected for energy loss in CH$_2$ up to the
reconstructed scattering vertex, illustrating the approximate scaling of the angular distribution of
$A_y$ with momentum. The results
shown in Figure~\ref{fig:Ay} are in fairly good agreement with the
unpublished results from the original analysis in \cite{GayouThesis},
despite using the more restrictive elastic event selection cuts of the
present work. This is due in part to the fact that the sensitivity of $P_\ell$, from which $A_y$ is primarily determined, to
$r = G_E^p/G_M^p$ is rather weak (see
equation~\eqref{recoilformulas}). Nonetheless, for the three highest $Q^2$
points, the improved suppression of the background in this analysis leads to a slight
systematic increase in $A_y$, since the asymmetry of the background included in the original analysis partially
cancels that of the signal. $A_y$ rises rapidly from zero in the
region dominated by Coulomb scattering to a maximum at $p_T \approx 0.3$ GeV and then tapers off to nearly zero beyond
about 1.5 GeV. The measured angular distribution at each $Q^2$ was
fitted using a simple parametrization $A_y(p_T) = (p_T -
p_T^0)^\alpha e^{-b(p_T-p_T^0)^\beta}$, where $p_T^0$, $\alpha$, $b$
and $\beta$ are adjustable parameters. This parametrization
incorporates the main features of the angular distribution with
sensible limiting behavior and is
sufficiently flexible to give a good description of the data. The
fit results for each $Q^2$ are given in Table \ref{ayfittable}.
\begin{table}
  \begin{ruledtabular}
    \begin{tabular}{cccc}
      $Q^2$ (GeV$^2$) & $p_T^0$ (GeV) & $\alpha$ & $\beta$ \\ \hline 
      3.5 & $0.030 \pm 0.008$ & $0.89 \pm 0.06$ & $1.19 \pm 0.10$ \\
      4.0 & $0.031 \pm 0.003$ & $0.88 \pm 0.03$ & $0.95 \pm 0.04$ \\
      4.8 & $0.029 \pm 0.005$ & $1.02 \pm 0.04$ & $1.03 \pm 0.05$ \\
      5.6 & $0.038 \pm 0.011$ & $1.14 \pm 0.09$ & $1.12 \pm 0.11$ \\ \hline 
      $Q^2$ (GeV$^2$) & $b$ & $\chi^2/\mbox{n.d.f.}$ & $A_y^{max}$ \\ \hline 
      3.5 & $3.51 \pm 0.15$ & 1.23 & $0.149 \pm 0.005$ \\
      4.0 & $3.28 \pm 0.06$ & 1.22 & $0.123 \pm 0.003$ \\
      4.8 & $3.44 \pm 0.06$ & 1.30 & $0.109 \pm 0.002$ \\
      5.6 & $3.67 \pm 0.15$ & 2.02 & $0.099 \pm 0.004$
    \end{tabular}
  \end{ruledtabular}
  \caption{\label{ayfittable} $A_y$ fit results. Parametrization is
    $A_y(p_T) = \left(p_T - p_T^0\right)^\alpha e^{-b\left(p_T -
        p_T^0\right)^\beta}$. The uncertainty in $A_y^{max}$ was
    calculated from the full covariance matrix of the fit result.}
\end{table}
The quality of the fit was improved by including the zero offset $p_T^0$, as
the data seem to prefer a vanishing $A_y$ at finite $p_T^0 \approx
0.03$ GeV, independent of $Q^2$. For $p_T < p_T^0$, $A_y = 0$ was assumed. The results for
the exponents $\alpha$ and $\beta$ are essentially compatible with the
product of a
linear rise and an exponential decay. An alternate parametrization
which fixes $\alpha =1$ and $\beta = 1$ and adds an overall normalization
constant as a free parameter in addition to the slope parameter $b$ does not describe the
data as well as the chosen parametrization in which $\alpha$ and $\beta$ are
free parameters but the overall normalization is fixed. The amplitude of the measured $A_y$ distribution, as measured by its
maximum value, scales approximately with $1/p$, as shown in the
bottom panel of Figure~\ref{fig:Ay}. Notably, the intercept of the
linear fit to the $1/p$ dependence of $A_y^{max}$ is compatible with
zero, suggesting that the analyzing power for $\vec{p}+\mbox{CH}_2$ scattering
vanishes for asymptotically large proton momenta, rather than crossing
zero at a finite momentum. The fitted curves shown in
Figure~\ref{fig:Ay} were used to describe $A_y(p_T)$ in the
analysis. 

The observed proportionality of $A_y$ to $1/p$ allows the momentum
dependence of $A_y$ to be accounted for in the analysis by simply
scaling its value for each event by a factor $p_0/p$, where
$p_0$ is the central proton momentum and $p$ is the proton momentum
for the event in question\footnote{For this purpose, the central momentum $p_0$ was
  corrected for energy loss in half the thickness of CH$_2$, while the
  momentum $p$ for the event in question was
  corrected for energy loss up to the
  reconstructed scattering vertex.}. This is because the fitted $A_y(p_T)$
curve, which is averaged over the $\pm 5\%$ momentum bite of the HRS
at each $Q^2$, essentially gives $A_y(p_0,p_T)$, where $p_0$ is the central
momentum. Assuming that the $1/p$ slope of $A_y$ is the same at
any $p_T$; i.e., assuming a factorized form $A_y(p,p_T) = C(p_T)/p$,
the ratio of $A_y(p,p_T)$ to its known value $A_y(p_0,p_T)$ at a
reference momentum $p_0$ is given by $p_0/p$, \emph{regardless of
  $C(p_T)$}. While the observed shape of the $p_T$ dependence 
of $A_y$ is approximately momentum-independent for the three
higher $Q^2$ points, the $p_T$ dependence of $A_y$ at $Q^2 = 3.5$
GeV$^2$ is slightly different, with a
larger maximum value than suggested by a linear extrapolation from the
higher-$Q^2$ data and a faster falloff at large $p_T$. A plausible,
but unproven explanation for the difference
in behavior is that the thicker 100 cm analyzer used for the three
highest-$Q^2$ measurements smears out the $p_T$ distribution of both
the efficiency and the analyzing power of the FPP relative to the thinner
58 cm analyzer used for the measurement at $Q^2 = 3.5$ GeV$^2$. This
observation does not, however, invalidate the $p_0/p$ scaling of
$A_y$ in the analysis, because the data from the three higher-$Q^2$
points, as well as data from other experiments~\cite{Punjabi05,Ay_pCH2}, show
that the $1/p$ scaling is respected for any given FPP configuration,
though the details of $A_y(p_T)$ may differ slightly between different
configurations. In any case, the value of $A_y$ assigned in the analysis is never changed by
more than $\pm 5\%$ for any individual event, so the actual
effect of this prescription on the
relative weighting of events is rather small.

The description of
$A_y(p,\vartheta)$ in the present reanalysis differs slightly from that of
the original analysis. In this reanalysis,
$A_y(p,\vartheta)$ is assigned to each event based on the smooth
parametrization of $A_y(p_T)$ shown in the curves of Figure
\ref{fig:Ay}, which describe the data very well, and an overall $1/p$
scaling. The original analysis, on
the other hand, neglected the momentum dependence of $A_y$ and
assigned $A_y(\vartheta)$ to each event based on the calibration
results in discrete $\vartheta$ bins. Since $A_y$ cancels in the ratio
$P_t/P_\ell$, its description only matters to the extent that it
optimizes the statistical precision of the extraction. Different descriptions of
$A_y(p,\vartheta)$ correspond to different event weights in the
analysis, leading to slight differences in the results for $P_t$,
$P_\ell$ and $R$ reflecting statistical fluctuations of the data as a
function of $p$ and $\vartheta$. While these differences are always well within the
statistical uncertainty of the combined data, better descriptions
of $A_y(p,\vartheta)$ naturally lead to better overall results. 

\subsubsection{False Asymmetries}
\label{subsubsec:falseasym}
Consistent with the original analysis, no false asymmetry corrections
were applied in the present work; i.e.,
$\lambda_0 = 0$ was assumed in equations~\eqref{likelihoodfunction}
and \eqref{weightedsum}. ``Weighted sum'' estimators, as
defined in \cite{Besset}, can be
constructed for the focal plane
asymmetries $A_y^{FPP} \equiv -P_e A_y P_y^{FPP}$ and
$A_x^{FPP} \equiv P_e A_y P_x^{FPP}$, equivalent to
equation~\eqref{weightedsum} in the absence of precession effects. Including false
asymmetry terms up to $2\varphi$, it can be shown that the weighted-sum
estimators $\hat{A}_x^{FPP}$ and $\hat{A}_y^{FPP}$ for the focal plane
asymmetries are given to second order in the false and physical asymmetry
terms by 
\begin{eqnarray}
  \hat{A}_x^{FPP} &=& A_x^{FPP} \left(1-\frac{c_2}{2}\right) - A_y^{FPP} \frac{s_2}{2} \nonumber \\
  \hat{A}_y^{FPP} &=& -\frac{s_2}{2} A_x^{FPP} + A_y^{FPP}\left(1+\frac{c_2}{2}\right),
\end{eqnarray}
where $c_2$ and $s_2$ are the false asymmetries as in
equation~\eqref{fpasym}. Only
the $2\varphi$ Fourier moments of the false asymmetry contribute at
this order. The $\cos (2\varphi)$ false
asymmetry moment induces a ``diagonal'' correction to each physical asymmetry term
proportional to the asymmetry itself, while the $\sin (2\varphi)$
false asymmetry moment induces an ``off-diagonal'' correction to $A_x^{FPP}$
($A_y^{FPP}$) proportional to $A_y^{FPP}$ ($A_x^{FPP}$). 

Fourier analysis of the helicity sum distribution $n_+ + n_-$ showed that
the acceptance-averaged magnitude of $c_2$ and $s_2$ did not
exceed $2.5 \times 10^{-3}$ at any $Q^2$, and neither term exceeded
$1\%$ at any $\vartheta$ within the useful range. The possible effect of $c_2$ on the ``diagonal'' terms is therefore
at the $10^{-3}$ (relative) level, while the ``off-diagonal''
correction is at the $10^{-5}$ level
(absolute) for the small $A_y^{FPP}$ term, and even smaller for the larger $A_x^{FPP}$ term. Compared to both the size and statistical uncertainty in
the asymmetries (see Fig. \ref{fig:fpasym}), and the systematic
uncertainties in $P_t$ and $P_\ell$ resulting from the spin transport calculation, such
corrections are completely negligible. This is in contrast to the
GEp-III and GEp-2$\gamma$ analyses, in which a sizeable $\cos
(2\varphi)$ false asymmetry in the Hall C FPP induced a correction
that, while small, made a non-negligible contribution to the total systematic uncertainty.

\subsection{Background Estimation and Subtraction}
\label{subsec:background}

From Figure \ref{elasticplots}, two qualitative features of the data
are obvious. First, the non-elastic background before applying two-body
correlation cuts is substantial. Second, examination of the $\Delta x$
and $\Delta y$ spectra before and after applying the $\delta p$ cut
reveals that the $\delta p$ cut provides significant additional
background suppression power relative to $\Delta x$ and $\Delta y$ cuts
alone, with minimal reduction of the elastic peak strength,
implying that events outside the $\delta p$ cut are background-dominated, even after calorimeter cuts. 

As alluded to in sections \ref{subsec:hallccomp} and \ref{section:elasticeventselection},
the non-elastic background for the measurements using a calorimeter
for electron detection consists predominantly of two reactions, quasi-elastic Al($e,e'p$) scattering in the cryocell
entrance and exit windows, and $\pi^0$ production initiated by the flux of real Bremsstrahlung
photons radiated along the target material (photoproduction) as well
as virtual photons present
in the electron beam independent of target thickness
(electroproduction). Due to the kinematic acceptance of the experiment
and the $Q^2$
dependence of the respective cross sections, the contribution of $\pi^0 p$ electroproduction is
mostly limited to ``quasi-real'' photons; i.e., $Q^2 \approx 0$, and
is practically indistinguishable from real photoproduction. By
detecting both scattered particles in coincidence, the two-body $ep \rightarrow ep$
kinematics are overdetermined, providing for a clean selection of
elastic events and a direct determination of the remaining
background from the data, with no external
inputs, using the sideband-fitting method described in section
\ref{subsubsec:sideband} below. The main disadvantage of this approach
to background estimation is that it makes no reference to the underlying physics of the
signal and background. For this reason, a Monte Carlo simulation of
the experiment was carried out to confirm the conclusions regarding
backgrounds obtained directly from the data. However, the results of
the simulation were not used in any way as input to the final analysis. 

\subsubsection{Monte Carlo Simulation}

\label{subsubsec:MC}
The
simulation code is the same as that used in the data analysis of
\cite{Qattan05}, which already includes a realistic model of the
HRSL. Modifications of the code used in the analysis of
Ref.~\cite{Qattan05} to reproduce non-Gaussian tails of the HRS
resolution, caused by multiple scattering and other effects, were not included here. The only significant addition to
the code was a description of the acceptance and resolution of the
GEp-II calorimeter. Because the $15
\times 15$ cm$^2$ cell
size of the GEp-II calorimeter is large compared to the Moli\'{e}re
radius of lead-glass, coordinate reconstruction essentially consists of assigning
the shower coordinates to the center of the cell with maximum energy
deposition. Furthermore, the discriminator threshold applied to form
the timing signal was roughly 20\% of the elastically scattered electron
energy, meaning that signals below this amplitude would be rejected in
software by the timing cut. The electron energy and coordinates were
thus defined by the signal in a single block in the overwhelming
majority ($\gtrsim 90\%$) of elastic events. Physics
ingredients of the simulation include cross section models for
$^1$H$(e,e'p)$, Al$(e,e'p)$ and
$^1$H$(\gamma,\pi^0 p)$ reactions, a realistic calculation of the
Bremsstrahlung flux for $\pi^0$ photoproduction, and event-by-event radiative
corrections to the $(e,e'p)$ cross
sections following the approach of \cite{EntRadCorr}, providing for a
rigorous deconvolution of the signal and background contributions
to the $\Delta x$, $\Delta y$, and $\delta p$ distributions for
arbitrary cuts. Another
reaction that can contribute to the background is Real Compton
Scattering $\gamma p \rightarrow \gamma p$ (RCS), whose end-point
kinematics are identical to $ep \rightarrow ep$. However, the cross
section for this reaction is generally much smaller than for $\pi^0$
photoproduction \cite{Danagoulian2007,Shupe1979}, and was neglected. 

\begin{figure}
  \begin{center}
    \includegraphics[width=.45\textwidth]{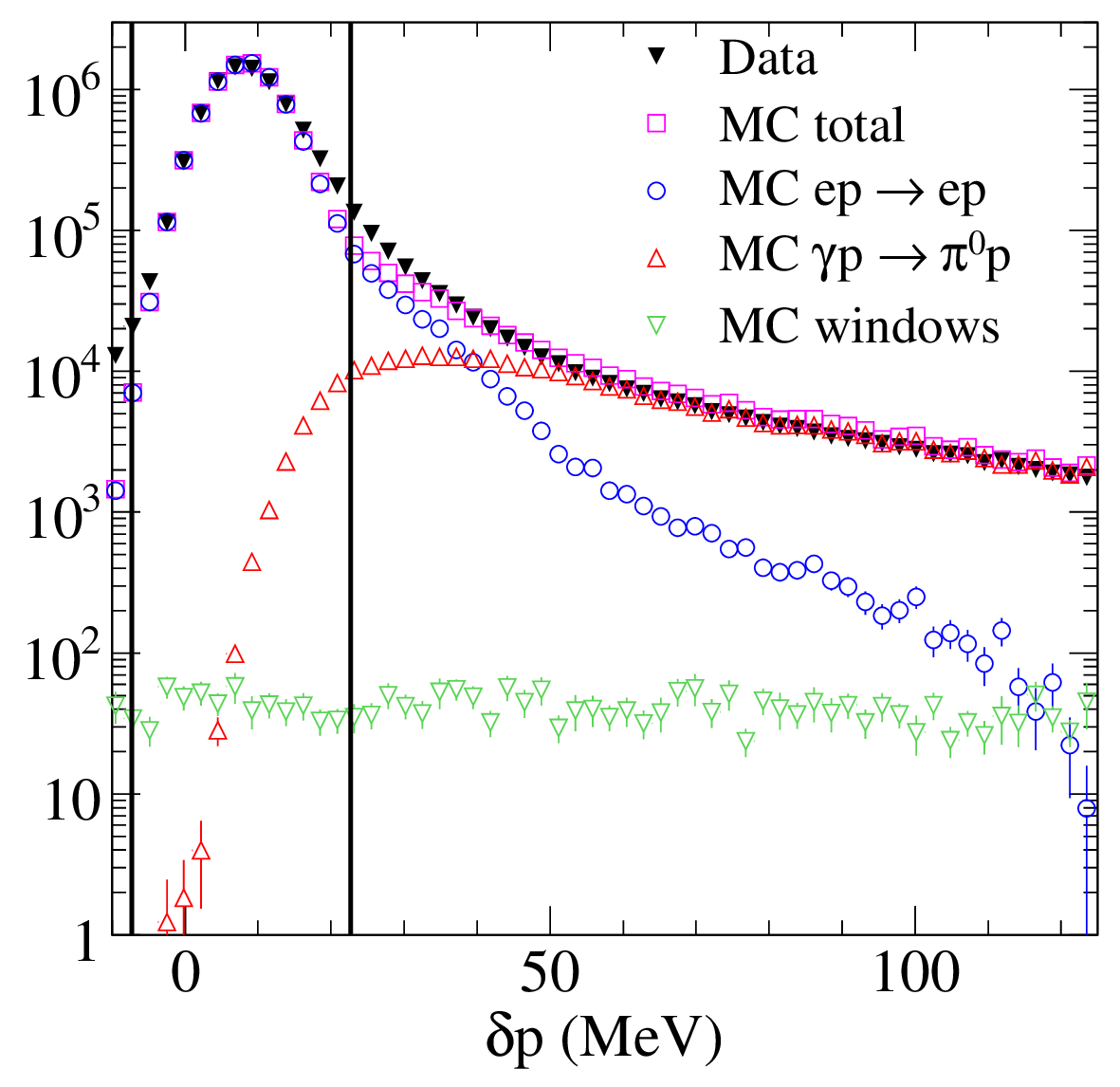}
  \end{center}
  \caption{\label{simc_data_comp} (color online) Contributions to the $\delta p$
    distribution at $Q^2 = 4.8$ GeV$^2$ estimated from the Monte Carlo
    simulation. Monte Carlo distributions are shown for elastic $ep$
    (empty circles), $\gamma p \rightarrow \pi^0 p$ (empty triangles),
    and quasi-elastic $(e,e'p)$ in the target windows (empty inverted
    triangles). The sum of all Monte Carlo contributions (empty squares)
    is compared to the data of Figure
    \ref{elasticplots}(c) (solid inverted triangles). Monte Carlo and data
    distributions are obtained after applying
    $\Delta x$ and $\Delta y$ (calorimeter) cuts. Black vertical lines
    show the $\delta p$ cut region of the final
    analysis. Uncertainties shown are statistical only. See text for details.
  }
\end{figure}
Figure \ref{simc_data_comp} shows the simulated $\delta p$
distribution in the vicinity of the elastic peak for each reaction considered, after applying $\Delta
x$ and $\Delta y$ cuts. As described below, the
simulated target window yield was normalized to match the window yield
obtained from the data in the
super-elastic ($\delta p < 0$) region. Then, the overall normalization constants for
$\pi^0 p$ and elastic $ep$ events were fitted simultaneously to minimize the
statistics-weighted sum of
squared differences between the data and the sum of Monte Carlo yields. The agreement between data and Monte Carlo is good, but not perfect,
primarily because non-Gaussian tails are not included in the
simulated $\delta p$ resolution. Nonetheless, the $\delta p$
distribution after cuts is described to within $\sim 20\%$ in
the relevant $\delta
p$ range, with the exception of disagreements of up to $\sim 40\%$ in
the $\delta p$ region from 20-40 MeV just above the elastic peak,
which is rather sensitive to non-Gaussian tails and the details of the Bremsstrahlung spectrum and
the $\pi^0$ production cross section near
end-point. 
Since
the purpose of the simulation
was to provide a qualitative illustration of the physics of the signal and the
background, and since the background contamination and its
polarization were determined directly from the data for the final
analysis, no additional fine-tuning of the simulation was attempted.

Two key features of the simulation results deserve special
emphasis. First, the contribution of the $ep$ radiative tail in the
inelastic region falls off too quickly to describe the observed tail
of the data. This is a consequence of the $\Delta
x$ cut, with $\Delta x$ calculated using the GEp-II method
\cite{GayouThesis}\footnote{The $\Delta x$ cut suppresses the $ep$
  radiative tail even more strongly when $\Delta x$ is calculated
  using the GEp-III method.}. The background fraction exceeds 80\%
above 50 MeV and 90\% above 75 MeV. The $ep$ yield falls below the $\pi^0p$
yield at $\sim$40 MeV and becomes negligible above $\sim 120$ MeV, confirming the conclusion
that the inelastic region of the $\delta p$ distribution is dominated by
the $\pi^0 p$ background rather than the $ep$ radiative tail. Second, the target window contribution is vanishingly small compared to the elastic and
$\pi^0p$ contributions in the entire $\delta p$ range of interest. More
specifically, in the region below $\pi^0$ threshold, the window
contribution is the dominant component of the background, but is too small relative to the elastic
yield to affect the measured asymmetry, while in the region where the
contamination is sufficiently large to affect the asymmetry, the
$\pi^0$ contribution is dominant. Moreover, the proton recoil
polarization in quasi-elastic Al$(\vec{e},e'\vec{p})$ scattering at
high $Q^2$ should be similar, in principle, to that in elastic $\vec{e}p \rightarrow e\vec{p}$, since the
former process is simply the latter process embedded in a nucleus,
whereas the spin structure of $\vec{\gamma} p \rightarrow \pi^0
\vec{p}$ can be (and is) dramatically different. 

The only kinematically allowed
reactions producing protons in the super-elastic region are
quasi-elastic Al$(e,e'p)$ and other reactions occurring on the
Al nuclei in the cryocell windows, in which
the initial Fermi motion of the struck proton can lead to
proton knockout with $p_p > p_p(\theta_p)$. However, a significant
fraction of the yield in the super-elastic region actually comes from
hydrogen, because the combined thickness of the entrance and exit windows
of the Hall A cryotarget \cite{HallANIM} in g cm$^{-2}$ is only about
4\% of the liquid hydrogen thickness, and the non-Gaussian tails of the
$\delta p$ resolution smear a fraction of hydrogen events into the
unphysical $\delta p$ region. The reconstructed vertex distribution in this region exhibits narrow
peaks at the window locations and a smooth hydrogen
background extending over the full target length. To estimate the
yield from the target windows, the vertex $z$ distribution was plotted
as a function of $\delta p$ in the super-elastic region for events \emph{failing}
the $\Delta x$ and $\Delta y$ cuts, in order to enhance the very small
window ``signal'' relative to the large hydrogen elastic ``background''. For each of six $\delta
p$ bins in $-180 \le \delta p\mbox{ (MeV)} \le 0$, a polynomial fit to the smooth hydrogen
background was subtracted from the vertex $z$ distribution, leaving
only the window peaks. For each window, the simulated $\delta p$
distribution with identical cuts applied was
normalized to match the background-subtracted window yield obtained from the
data. The resulting normalization
factor was then
applied to the simulated $\delta p$ distribution of window events
\emph{passing} the $\Delta x$ and $\Delta y$ cuts, leading to the contribution shown in Fig.~\ref{simc_data_comp}.

Given the vertex
resolution of the HRS, a vertex cut chosen to exclude the windows at
the $3\sigma$ level can
further suppress the very small window
background, at
the expense of a $\sim$20\%
reduction in elastic $ep$ statistics. However, the aforementioned analysis of the
window yield suggests that even when the
full target length is included, the fraction of the total yield from
the windows is negligible after all cuts are applied, making additional vertex cuts unnecessary. This conclusion is
further supported by comparing the
$\delta p$ distributions with and without
such a vertex cut, and by comparing the $\delta
p$ spectra for the $Q^2 \ge 4.0$ GeV$^2$ settings to the $\delta p$
spectrum of the $Q^2 = 3.5$ GeV$^2$ setting, for which the precise measurement of the
electron kinematics with a magnetic spectrometer provides an essentially background-free selection of
elastic events, as discussed in Section \ref{subsubsec:e35}. Based on
these considerations, the window contamination was deemed negligible,
and the study of the background contamination focused mainly on the
inelastic ($\delta p > 0$) region. 

The background subtraction procedure used for
the final analysis is
agnostic regarding the reaction mechanism responsible for the
contamination, with the caveat that the conclusion of negligible
window contamination is used to justify the assumption of constant
background polarization, which reduces the statistical uncertainty in
the background correction. In summary, the simulation provides a qualitative
description of the data that supports the conclusions of this analysis regarding
backgrounds. Averaged over the final
$\delta p$ cut region, the fractional background contamination obtained from the
simulation agrees with that obtained directly from the
data at a level similar to its systematic uncertainty, which is
determined by the data.

\subsubsection{Sideband subtraction}
\label{subsubsec:sideband}
For the final analysis, the fractional background contamination in the sample of elastic $ep$
events selected by a given set of cuts was estimated by fitting the tails of the
$\Delta x$ and $\Delta y$ distributions on either side of the
elastic peak and extrapolating into the
peak region, as shown in Figures~\ref{elasticplots}(a) and \ref{elasticplots}(b). This approach to
background estimation implies
two assumptions. First, the contribution of elastic scattering to
the tails of the $\Delta x$ and
$\Delta y$ distributions is assumed to be negligible for values of $\Delta x$ and
$\Delta y$ sufficiently far away from the elastic peak. Second, the
background is assumed to have a smooth distribution under the elastic
peak, so that joining the tails with a smooth interpolating function
is a good approximation to the true background shape.
The first assumption can in
principle be violated by
the $ep$ radiative tail and by non-Gaussian smearing effects in the
HRS angle and momentum reconstruction. Radiation redistributes
elastic $ep$ events away from the elastic peak toward negative $\Delta
x$ values, but does
not markedly affect the $\Delta y$ distribution of elastic events, since
$\Delta y$ reflects the extent to
which the two detected particles are non-coplanar, and the coplanarity
of outgoing particles is not strongly affected by radiation. Furthermore, the
$\delta p$ cut suppresses the radiative tail of the $\Delta x$
distribution. Non-Gaussian
smearing effects do not contribute a significant
fraction of events in the tails except when the background
contribution is very small. The second assumption (smooth background
distribution) was confirmed by inspecting the correlations between
$\Delta x$ and $\Delta y$; i.e., by plotting $\Delta x$ ($\Delta y$)
for $\Delta y$ ($\Delta x$) well outside the elastic peak. This
assumption was also supported by the simulations described in section
\ref{subsubsec:MC}. Although the simulation does not include the
contribution of random coincidences, the contamination of the data by
random coincidences is negligible after timing and kinematic cuts.

In the following discussion, the fractional background contamination
$f$ is defined as $f \equiv B/(S+B)$, where $B$ is the number of
background events and $S$ is the number of signal events; i.e., $f$ is
the ratio of the background yield to the \emph{total} yield.
The value of $f$ and its systematic uncertainty $\Delta f$ were estimated using a
conservative approach involving a total of twelve different fits. The
tails of the $\Delta x$ and $\Delta y$ distributions, obtained after applying all other cuts, were each fitted with Gaussian and polynomial
background shapes, for three different sizes of the elastic peak
region excluded from the fit (two spectra $\times$ two parametrizations $\times$ three
sideband ranges $=$ twelve fits). The average fit result was taken as
the value of $f$, while the \emph{rms} deviation of the fit result
from the mean was taken as the systematic
uncertainty $\Delta f$. The variations among the different fit results
reflect the level of agreement (or disagreement) among the different
spectra, assumed background lineshapes, and regions excluded from the
fit.

A central conclusion of the present reanalysis is that the background
was underestimated in the
original analysis. Using the polynomial
sideband fitting method, the estimated average values of $f$
for the cuts of the original analysis, in which no $\delta p$ cut was
applied, are 1.6\%, 2.8\% and 5.3\% for $Q^2 = 4.0$, 4.8 and 5.6
GeV$^2$, respectively. Compared to the estimates reported in
\cite{GayouThesis} for the original analysis, these estimates are higher by factors of 2.3, 7.0
and 3.8, respectively. Even at the few percent level, neglected or
underestimated inelastic contamination can have a non-negligible effect on the measured
asymmetries if the polarization of the background differs strongly
enough from that of the signal, as in this case. 

With the addition of the $\delta p$ cut, the present analysis
maximally exploits the two-body kinematic
correlations of both detected particles. 
In the inelastic region, $\pi^0$
production dominates. In terms of $\delta p$, the $\pi^0$
production ``threshold'' is very close to the elastic peak. When
reconstructed assuming elastic scattering, protons from $\gamma p \rightarrow
\pi^0 p$ at the Bremsstrahlung end-point have $\delta p = 7.4$, 8.1 and 8.8 MeV for $Q^2 = $ 4.0, 4.8 and
5.6 GeV$^2$, respectively. When compared to the $\delta p$ resolution of
$\sim 5$ MeV, there is clearly substantial overlap of the $\pi^0 p$
kinematic phase space with the elastic peak, as in the example of
Figure~\ref{simc_data_comp}. As $Q^2$ increases at a
given beam energy, the $\pi^0 p$ cross section becomes large compared
to the $ep$ cross section. 

\begin{figure}
  \includegraphics[width=.48\textwidth]{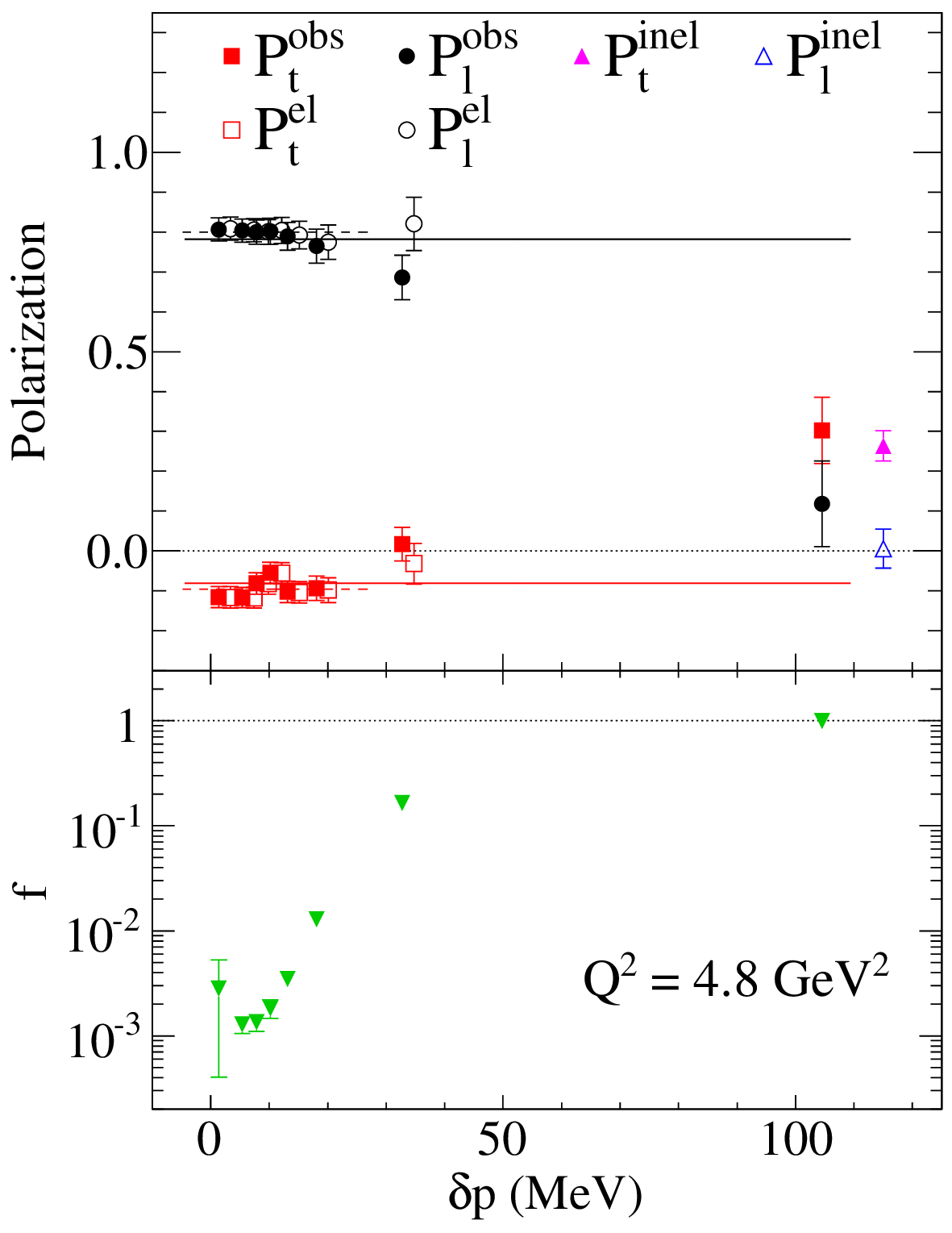}
  \caption{\label{ptplpmiss} (color online) Top panel: $\delta p$
    dependence of $P_t$ and
    $P_\ell$ for events selected by the $\Delta x$ and $\Delta y$ cuts
    of Figure \ref{elasticplots}. Data are binned in $\delta p$ as
    described in the text and plotted at the average $\delta p$
    value in each bin. Raw polarizations $P_t^{obs}$ (filled
    squares) and $P_\ell^{obs}$ (filled circles) approach the
    background polarizations $P_t^{inel}$ (filled triangle) and
    $P_\ell^{inel}$ (empty triangle) at large $\delta p$. Corrected values
    $P_t^{el}$ (empty squares) and $P_\ell^{el}$ (empty circles) are offset in $\delta p$
    for clarity. Dashed and solid horizontal lines are weighted averages
    of the corrected and raw data, respectively. Bottom panel: $\delta
    p$ dependence of the fractional background contamination $f$. Uncertainties in $f$
    are systematics-dominated, while the uncertainties in the
    polarization components are statistics-dominated. See text for details.}
\end{figure}

The effect of underestimating the $\pi^0$ background on the form factor ratio
extraction is illustrated in Figure
\ref{ptplpmiss}, which shows $P_t$, $P_\ell$ and $f$ as a function of $\delta p$, for events identified as
elastic in the original analysis, at $Q^2 = 4.8$ GeV$^2$. 
The data were divided into eight $\delta p$ bins, including six equal-statistics bins inside the cut
region of Figure \ref{elasticplots}(c), where $f$ is very small ($-7.3
\le \delta p \le 22.7$ MeV), a seventh bin with a significant
fraction of both signal and background ($22.7 \le \delta p \le 60$
MeV), and an eighth bin dominated by background ($\delta p > 60$
MeV). Because the $\Delta x$ and $\Delta y$
distributions in the last $\delta p$ bin showed no obvious signature
of an elastic peak, $f=1$ was assumed for this bin, consistent with the simulation
results shown in Figure \ref{simc_data_comp}. Meaningful background
estimation and subtraction were not possible for
this bin. As $\delta p$ increases, the raw transferred polarization components $P_t^{obs}$
and $P_\ell^{obs}$ evolve from their roughly constant values in the signal-dominated
region to values that are consistent with the background polarization components
$P_t^{inel}$ and $P_\ell^{inel}$. The $\delta p$-integrated
results for the background polarization, extracted from events
rejected by the cuts of Figure \ref{elasticplots}, are plotted at an
arbitrary $\delta p=115$ MeV for comparison. 

The background polarization components were obtained by applying
anti-cuts twice as wide as the final elastic event selection cuts;
i.e., $\Delta x$($\Delta y$) was required to be at
least 24(32) cm away from the midpoint
between half-maxima of the peak. Events selected by this
anticut are background-dominated and have negligible elastic
contamination. In order to study the $\delta p$ dependence of $P_i^{inel}$, no cut was applied to
$\delta p$ in the extraction of the background polarization. No
statistically significant $\delta p$ dependence of the background
polarization was observed, consistent with dominance of the
background by $\pi^0 p$ events. Therefore,
$P_i^{inel}$ was assumed constant in the background
subtraction procedure.

In Figure \ref{ptplpmiss}, the signal
polarization $P_i^{el} (i=t,\ell)$ was obtained from $P_i^{obs}$ in
the first seven bins using the subtraction 
\begin{eqnarray}
  P_i^{el} = \frac{P_i^{obs} - fP_i^{inel}}{1-f}. \label{bgsubtract}
\end{eqnarray}
By comparing the weighted average of all uncorrected data in
Figure~\ref{ptplpmiss} to the weighted average of the six corrected
data points inside the cut region, it is found that the background
contamination of the sample with no $\delta p$ cut induces
relative systematic shifts of $\left|\Delta P_t/P_t\right| = 15.8\%$ and
$\left|\Delta P_\ell/P_\ell\right| = 2.4\%$. From Figure
\ref{ptplpmiss}, it is clear that the tails of the $\delta p$ distribution
outside the cut region of Figure \ref{elasticplots}(c)
contribute very little to the statistical precision of the measurement
of $P_t/P_\ell$ while causing a large systematic effect. For the final analysis, rather
than correcting the results bin-by-bin in $\delta p$ using
equation \eqref{bgsubtract}, as in Figure
\ref{ptplpmiss}, the background fraction $f$ and polarization
$P_i^{inel}$ were included at the individual event level in equation \eqref{weightedsum} by making the following
replacements:
\begin{eqnarray}
  \lambda_{t,\ell}^{(i)} &\rightarrow& \lambda_{t,\ell}^{(i)}
  (1-f_{i} ) \nonumber \\
  \left(1-\lambda_0^{(i)}\right) &\rightarrow& \left(1 -
    \lambda_0^{(i)} - \lambda_{inel}^{(i)}\right),
\end{eqnarray}
where $f_{i}$ is the background contamination as a function of $\delta
p^{(i)}$ and $\lambda_{inel}^{(i)}$, representing the background asymmetry,
is given by
\begin{eqnarray}
  \lambda_{inel}^{(i)} &\equiv& f_i h_i P_e
  A_y^{(i)}\left[\left(S_{xt}^{(i)}\cos \varphi_i - S_{yt}^{(i)} \sin
      \varphi_i \right)P_t^{inel} + \right. \nonumber \\ 
  & & \left. \left(S_{x\ell}^{(i)} \cos \varphi_i - S_{y\ell}^{(i)}
      \sin \varphi_i \right)P_\ell^{inel} \right].
\end{eqnarray}
This method is functionally equivalent to correcting the results
``after the fact'' using equation \eqref{bgsubtract}. It also simplifies
the evaluation of systematic uncertainties associated with the
background correction, which were obtained by varying $f$,
$P_t^{inel}$ and $P_\ell^{inel}$ within their uncertainties and
observing the shift in $R$.

\subsection{Systematic Uncertainties}
\label{sec:systematics}
As a result of the cancellation of the beam polarization and analyzing
power in the ratio $P_t/P_\ell$ and the cancellation of the FPP
instrumental asymmetry by the beam helicity reversal, there are few
significant sources of systematic uncertainty in the results of this
experiment (as is also the case in the GEp-I, GEp-III and GEp-2$\gamma$
experiments). The dominant source of systematic uncertainty is the
spin transport calculation. Since the procedure for the evaluation of systematic
uncertainties associated with this calculation is documented at length
in Refs. \cite{QuadAlign,PentchevSpinHRS,Punjabi05,GayouThesis}, only
a brief summary of the studies and the conclusions is given here. 

The range of non-dispersive plane trajectory bend
angles $\phi_{bend}$ accepted by the HRS is roughly $\pm 60$ mrad,
independent of momentum. The maximum accepted range of the non-dispersive
plane precession angle $\chi_\phi = \gamma \kappa_p \phi_{bend}$ is
roughly $\pm 30^{\circ}$ at the highest $Q^2$ of 5.6 GeV$^2$. To first order in $\chi_\phi$, the ratio
$P_t/P_\ell$ is given in terms of the focal plane ratio
$P_y^{FPP}/P_x^{FPP}$ by $P_t/P_\ell \approx \chi_\phi - \sin \chi
P_y^{FPP}/P_x^{FPP}$. Because the non-dispersive plane precession
mixes $P_t$ and $P_\ell$, the ratio is highly sensitive to
uncertainties in $\phi_{bend}$. To first order, an uncertainty $\Delta
\phi_{bend}$ leads to an uncertainty $\Delta R \approx
\left(\mu_p\sqrt{\tau(1+\epsilon)/2\epsilon}\right)\gamma \kappa_p \Delta
\phi_{bend}$ in the extracted form factor ratio. The error
magnification factor multiplying $\Delta \phi_{bend}$ grows as large
as 33 at $Q^2 = 5.6$ GeV$^2$. To manage the systematic uncertainty due
to the precession calculation, $\phi_{bend}$ must be known to very
high accuracy. On the other hand, since $\theta_{bend}$ only enters
$P_t/P_\ell$ through the factor of $\sin \chi$ multiplying
$P_y^{FPP}/P_x^{FPP}$, and since the reconstruction of $\theta_{bend}$
involves relatively small deviations about the 45$^\circ$ central bend angle, the
accuracy of $P_t/P_\ell$ is far less sensitive to systematic errors in
$\theta_{bend}$ and $p_p$. 

The major sources of uncertainty in $\phi_{bend}$ are horizontal misalignments and rotations of the
three quadrupoles relative to the HRSL optical axis defined by the dipole
magnet. In order to control the uncertainty in $\phi_{bend}$ to the highest
possible accuracy, dedicated studies of the optical properties of HRSL
in the non-dispersive plane were performed. Electrons were scattered
from a thin carbon foil aligned with the HRSL optical axis, and a
special ``sieve-slit'' collimator was installed in front of the
entrance to HRSL before the first quadrupole magnet. The sieve-slit
collimator, part of the standard equipment of the HRSs, consists of a
5 mm thick stainless steel sheet with a
pattern of 49 holes ($7 \times 7$), spaced 25 mm apart vertically and
12.5 mm apart horizontally, used for optics calibrations
\cite{HallANIM}. In the studies described here, electrons passing through the
central sieve hole aligned with the HRS optical axis were selected. For a series
of deliberate mistunings of the HRS quadrupoles relative to the
nominal tune, the displacements in
both position and angle of
the image of the central sieve hole at the focal plane were
observed. Combined with the known first-order HRS optics coefficients
describing the effects of quadrupole misalignments and rotations, the
information gained from these studies placed a much more stringent constraint on the
misalignments than the nominal accuracy of the quadrupole
positions. By reducing the uncertainty $\Delta \phi_{bend}$ to $\pm$0.3
mrad, the optical studies reduced the systematic uncertainty in $R$ at
$Q^2 = 5.6$ GeV$^2$, where the result is most sensitive to
$\phi_{bend}$, to a level comparable with other contributions.

Additional model uncertainties in the precession calculation due to the
field layout in COSY are more difficult to quantify, but are typically
smaller than the errors associated with the accuracy of the inputs to
the calculation; i.e., the reconstructed
proton kinematics. The
COSY model uncertainties were estimated by performing the calculation in several
different ways. For the final analysis, the proton trajectory angles, momentum and vertex
coordinates, calculated using the standard HRS optics matrix tuned to
calibration data as described in \cite{HallANIM}, were used to calculate the forward spin transport matrix, as
described in section \ref{subsubsec:precession}. To estimate
systematic uncertainties, the
calculation was also performed using the same forward spin transport
matrix, but the kinematics were reconstructed using an alternate set of optics matrix
elements calculated by COSY. Finally, COSY was used to
calculate the expansion of the reverse spin transport matrix, which
was then inverted to obtain the forward matrix elements that enter the
likelihood function of equation \eqref{likelihoodfunction}. A model
systematic uncertainty was assigned based on the variations
in the results among the different methods, as described in
\cite{GayouThesis}. 

Apart from the uncertainties associated with the
non-elastic background, which were underestimated by the original
analysis, the main additional source of uncertainty is the accuracy of
the scattering angle
reconstruction in the FPP. Uncertainties associated with FPP
reconstruction were minimized by
a software alignment procedure using ``straight-through'' data
obtained with the CH$_2$ analyzers removed. The systematic errors in
$R$ due to the absolute accuracy in the determination of the
beam energy ($\delta E/E \sim 2 \times 10^{-4}$) and the proton
momentum ($\delta p/p \sim 4 \times 10^{-4}$) \cite{HallANIM}, which
mainly enter the ratio $R$ through the kinematic factor $\mu_p
\sqrt{\tau(1+\epsilon)/2\epsilon}$ of equation \eqref{recoilformulas},
are negligible compared to the precession-related uncertainties. 

The updated systematic uncertainties associated with the background estimation
and subtraction procedure are very small as a result of the added
$\delta p$ cut, and are generally
at the $10^{-3}$ level. The ``Bckgr.'' uncertainty in Table
\ref{systtable} was obtained by varying $f$, $P_t^{inel}$ and
$P_\ell^{inel}$ within their uncertainties, which are
systematics-dominated for $f$ and statistics-dominated for
$P_i^{inel}$, and observing the shift in $R$. The contributions from
$f$ and $P_t^{inel}$ are comparable, while the contribution from
$P_\ell^{inel}$ is much smaller. 

The present analysis also examined the sensitivity of $R$ to
variations in elastic event selection cuts. The analysis was performed
for various $\Delta x$, $\Delta y$ and $\delta p$ cut widths, using
both the GEp-II and GEp-III definitions of $\Delta x$ and $\Delta y$
(see section \ref{section:elasticeventselection}). The analysis was
also performed using the original polygon cut, supplemented by the new
$\delta p$ cut. For consistency of background corrections,
the contamination was estimated separately for each case. The rms
variation of $R$ due to cut variations is quoted as the ``Cuts'' uncertainty of
Table \ref{systtable}. It is generally larger than the ``Bckgr.''
uncertainty calculated using the final cuts, and reflects fluctuations
among slightly different selections of
events, not necessarily related to the background. It is, however,
much smaller than the statistical uncertainty at each $Q^2$.

The present reanalysis of the GEp-II data is identical to the original
analysis in event reconstruction, spin transport calculations, and all
cuts other than $\Delta x$, $\Delta y$ and $\delta p$ used to select
elastic events. The only other meaningful difference between the present
reanalysis and the original analysis is the improved description of
the analyzing power discussed in section \ref{subsubsec:Ay},
which only affects the results through slight modification of the $p$
and $\vartheta$-dependent weighting of events. Therefore, aspects of
systematic uncertainty analysis other than elastic
event selection and background subtraction were not revisited. These
aspects of the analysis are documented at length in \cite{GayouThesis}. 
\begin{table}
  \caption{\label{systtable} Total systematic uncertainty in $R$ and
    its contributions. See text for details.}
  \begin{center}
    \begin{ruledtabular}
      \begin{tabular}{ccccc}
        $Q^2$, GeV$^2$ & 3.5 & 4.0 & 4.8 & 5.6  \\ \hline 
        $\vartheta_{FPP}$ & $1.4 \times 10^{-3}$ & $0.8 \times 10^{-3}$
        & $1.4 \times 10^{-3}$ & $0.7 \times 10^{-3}$ \\ 
        $\varphi_{FPP}$ & $5.1 \times 10^{-3}$ & $6.3 \times 10^{-3}$ &
        $6.1 \times 10^{-3}$ & $2.9 \times 10^{-3}$ \\ 
        $\theta_{bend}$ & $4.6 \times 10^{-3}$ & $0.1 \times 10^{-3}$ &
        $2.6 \times 10^{-3}$ & $4.3 \times 10^{-3}$ \\ 
        $\phi_{bend}$ & $1.3 \times 10^{-3}$ & $1.1 \times 10^{-3}$ &
        $6.1 \times 10^{-3}$ & $12.3 \times 10^{-3}$ \\ 
        COSY & $0.4 \times 10^{-3}$ & $0.4 \times 10^{-3}$ & $1.2 \times
        10^{-3}$ & $12.7 \times 10^{-3}$ \\ 
        Bckgr. & N. A. & $0.9 \times 10^{-3}$ & $1.1 \times 10^{-3}$ &
        $1.2 \times 10^{-3}$ \\ 
        Cuts & N. A. & $5.4 \times 10^{-3}$ & $7.2 \times 10^{-3}$ &
        $3.9 \times 10^{-3}$ \\ \hline  
        Total & $7.0 \times 10^{-3}$ & $8.5 \times 10^{-3}$ & $11.7
        \times 10^{-3}$ & $18.9 \times 10^{-3}$
      \end{tabular}
    \end{ruledtabular}
  \end{center}
\end{table}

Table \ref{systtable} shows all known contributions to the systematic
uncertainty in $R$ at each $Q^2$, including the
polar ($\vartheta_{FPP}$) and azimuthal ($\varphi_{FPP}$) angle
reconstruction in the FPP, the dispersive ($\theta_{bend}$) and
non-dispersive ($\phi_{bend}$) trajectory bend angles, the COSY model
uncertainty (COSY), the non-elastic background contribution
(Bckgr.) and the cut sensitivity (Cuts). All contributions are added in quadrature to obtain the
total systematic uncertainty. Uncertainties due to FPP instrumental
asymmetries are negligible as discussed in section
\ref{section:polextract}. In the final analysis of the GEp-II experiment, the total
accuracy of the results is statistics-limited, with systematic
uncertainties at a much lower level. 

As in the original publication~\cite{Gayou02}, no radiative corrections have been applied to the data presented here. Standard,
model-independent radiative corrections to $R$ were calculated in \cite{AfanasevRadCorr} for kinematics very close to those of the GEp-II
experiment and found to be less than 1\% (relative) for all four $Q^2$
values. Though even 1\% relative corrections are much smaller than the
statistical uncertainties in the data, the calculations in
\cite{AfanasevRadCorr} were performed assuming a much wider
``inelasticity'' cut than that effected by the combination of cuts
applied in the present analysis, such that in reality, the standard radiative
corrections to the GEp-II data are even smaller, which justifies
neglecting them here. 

\section{Results}
\label{sec:results}
\subsection{Discussion of the Data}
\label{subsec:data}
\begin{figure}
  \includegraphics[width=.47\textwidth]{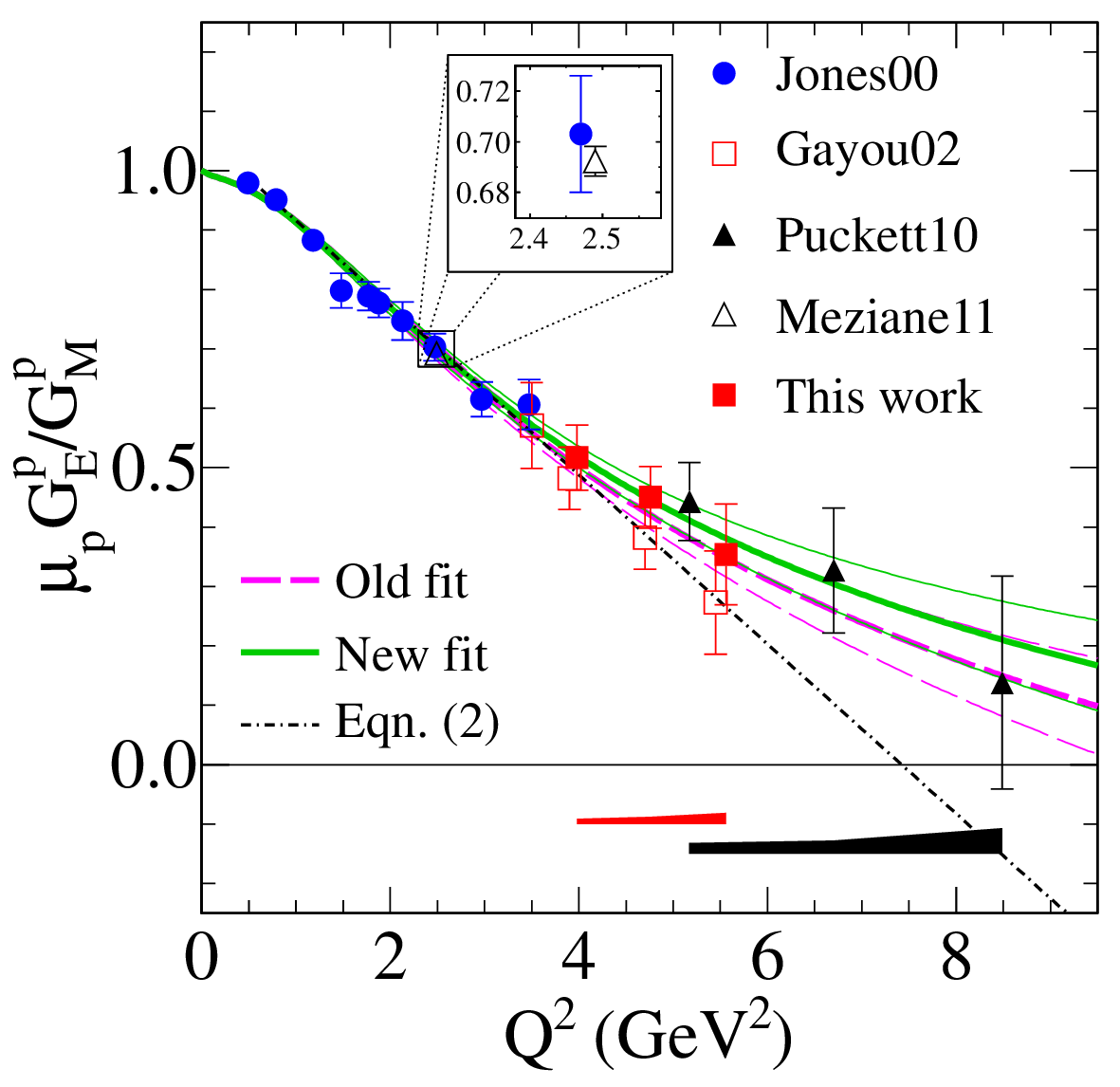}
  \caption{\label{ratiofig} (color) Polarization transfer data for
    $G_E^p/G_M^p$ from \cite{Punjabi05} (Jones00), \cite{Gayou02} (Gayou02), \cite{PuckettPRL10} 
    (Puckett10), \cite{MezianePRL11} (Meziane11) and the present work. Error bars are
    statistical. The data of \cite{Gayou02} are offset
    slightly in $Q^2$ for clarity. Systematic uncertainties for the
    present work and \cite{PuckettPRL10} are shown as bands
    below the data. The inset shows a zoomed view of the data near
    $Q^2 = 2.5$ GeV$^2$, demonstrating the excellent agreement between
    high-precision data from Hall A~\cite{Punjabi05} and Hall
    C~\cite{MezianePRL11} at this $Q^2$. Curves are global proton form
    factor fits using the originally published GEp-II data
    \cite{Gayou02} (Old fit) and the present work (New fit), with
    standard $1\sigma$ point-wise uncertainty bands. Both fits include
    the GEp-III data. The linear fit of equation \eqref{gayoufit} is
    shown for comparison. See text for details.}
\end{figure}

\begin{table*}
  \caption{\label{resultstable} Final results of the GEp-II
    experiment. $\left< Q^2 \right>$ is the acceptance-averaged $Q^2$,
    while $\Delta Q^2$ is the half-width of the total $Q^2$ interval,
    which is centered at the nominal $Q^2$. The raw
    ($P_i^{obs}$), background ($P_i^{inel}$) and corrected
    ($P_i^{el}$) polarization transfer components and the raw form
    factor ratio $R$ are presented with
    statistical uncertainties only. The background fraction
    $\left<f\right>$ averaged over the final cut region is quoted with its
    \emph{systematic} uncertainty $\Delta f$. The final results for $R =
    \mu_p G_E^p/G_M^p$ are quoted with statistical and systematic
    uncertainties. The data at $Q^2 = 3.5$ GeV$^2$ were not
    reanalyzed, and the quoted result is identical to that of the
    original publication \cite{Gayou02}. The originally published
    results~\cite{Gayou02} are quoted on the bottom line for comparison.}
  \begin{ruledtabular}
    \begin{tabular}{lccccc|}
      Nominal $Q^2$ (GeV$^2$) & 3.5 & 4.0 & 4.8 & 5.6 \\ \hline 
      $\left<Q^2\right> \pm \Delta Q^2$ (GeV$^2$) & $3.50 \pm 0.23$ & $3.98 \pm 0.26$ &
      $4.76 \pm 0.30$ & $5.56 \pm 0.34$ \\ 
      $P_t^{obs} \pm \Delta P_t^{obs}$ & N. A. & $-0.108 \pm 0.011$ & $-0.094
      \pm 0.011$ & $-0.070 \pm 0.017$ \\ 
      $P_\ell^{obs} \pm \Delta P_\ell^{obs}$ & N. A. & $0.683 \pm 0.012$ &
      $0.795 \pm 0.013$ & $0.886 \pm 0.030$ \\ 
      $R \pm \Delta R$ (raw) & N. A. & $0.514 \pm 0.055$ & $0.445 \pm 0.052$
      & $0.350 \pm 0.085$ \\ \hline 
      $\left<f\right> \pm \Delta f$ & N. A. & $(0.30 \pm 0.04)\%$ &
      $(0.38 \pm 0.06)\%$ & $(0.47 \pm 0.07)\%$ \\ 
      $P_t^{inel} \pm \Delta P_t^{inel}$ & N. A. & $0.116 \pm 0.051$ & $0.264
      \pm 0.038$ & $0.128 \pm 0.034$ \\
      $P_\ell^{inel} \pm \Delta P_\ell^{inel}$ & N. A. & $0.224 \pm 0.053$ &
      $0.006 \pm 0.049$ & $0.278 \pm 0.072$ \\ \hline 
      $P_t^{el} \pm \Delta P_t^{el}$ & $-0.118 \pm 0.015$ & $-0.109 \pm 0.011$ & $-0.096
      \pm 0.011$ & $-0.071 \pm 0.017$ \\
      $P_\ell^{el} \pm \Delta P_\ell^{el}$ & $0.616 \pm 0.017$ & $0.685 \pm 0.012$ &
      $0.799 \pm 0.013$ & $0.890 \pm 0.030$ \\
      $R \pm \Delta R_{stat} \pm \Delta R_{syst}$ (final) & $0.571 \pm
      0.072 \pm 0.007$ & $0.517 \pm
      0.055 \pm 0.009$ & $0.450 \pm 0.052 \pm 0.012$ & $0.354
      \pm 0.085 \pm 0.019$ \\ \hline 
      $P_t \pm \Delta P_t$ (equation~\eqref{recoilformulas})\protect\footnotemark[1] & $-0.118
      \pm 0.014$ & $-0.109 \pm 0.011$ & $-0.096 \pm 0.011$ & $-0.071
      \pm 0.017$ \\
      $P_\ell \pm \Delta P_\ell$ (equation~\eqref{recoilformulas})\protect\footnotemark[1] &
      $0.616 \pm 0.005$ & $0.685 \pm 0.003$ & $0.799 \pm 0.002$ &
      $0.890 \pm 0.002$ \\ \hline
      $R \pm \Delta R_{stat} \pm \Delta R_{syst}$
      (\cite{Gayou02}) & $0.571 \pm 0.072 \pm 0.007$ & $0.482 \pm
      0.052 \pm 0.008$ & $0.382 \pm 0.053 \pm 0.011$ & $0.273 \pm
      0.087 \pm 0.028$ 
    \end{tabular}
  \end{ruledtabular}
  \footnotetext[1]{These are the values of $P_t$ and $P_\ell$ calculated
    from equation~\eqref{recoilformulas}, with uncertainties due
    solely to the uncertainty in $R$.}
\end{table*}
The final results of the GEp-II experiment are reported in Table
\ref{resultstable} and presented in Figure
\ref{ratiofig}.
The values and statistical uncertainties of $P_t^{el}$ and $P_\ell^{el}$
presented in Table \ref{resultstable} (and Figure
\ref{ptplpmiss}) are obtained from
equation~\eqref{weightedsum}. Because the analyzing power is calibrated using equation
\eqref{recoilformulas}, the extracted $P_t^{el}$ and $P_\ell^{el}$ values are,
by definition, equal to those of equations \eqref{recoilformulas},
which depend only on $R$ and kinematic factors,
\emph{regardless} of the value of $P_e$ assumed in the
analysis. For reference, the values of $P_e$ used in the analysis
at each $Q^2$ are shown in Table \ref{kintable}. These values are based on the average of all beam polarization
measurements at a given setting. Because $P_e$ was
stable at the few percent level throughout the duration of each kinematic setting, a
single $P_e$ value was assigned to all data taken at a given $Q^2$. As
presented, the statistical uncertainties in $P_t^{el}$ and $P_\ell^{el}$ correspond to the uncertainties in the raw asymmetries measured by the FPP,
which are large compared to the corresponding systematic
uncertainties. 

$P_t$ and $P_\ell$ can also be calculated from
$R$ and kinematic factors using equation~\eqref{recoilformulas}. Neglecting the very small covariance of $P_t$ and
$P_\ell$ and the uncertainty in the kinematic factors involved, the uncertainty in $R$ is given by $\left(\Delta
  R/R\right )^2 = \left(\Delta P_t/P_t \right )^2 + \left(\Delta
  P_\ell / P_\ell \right)^2$. While the uncertainties in $P_t$ and
$P_\ell$ obtained from equation~\eqref{weightedsum} are similar,
and $\Delta P_\ell^{el}$ is generally larger than $\Delta P_t^{el}$ due
to the unfavorable precession angle,
the uncertainty in $R$ is nevertheless dominated by the uncertainty in $P_t$, since
$P_\ell$ is generally large compared to $P_t$. Due to the weak
sensitivity of $P_\ell$ to $R$, the uncertainty in $P_\ell$ calculated
from equation~\eqref{recoilformulas} is
much smaller than the uncertainty in $P_\ell$ extracted from the FPP
asymmetry. On the other hand, since $P_t$ is proportional to $R$ and
the relative uncertainties in $P_t$ and $R$ are similar, the
uncertainty in $P_t$ calculated from equation~\eqref{recoilformulas}
is very similar to the uncertainty in its extraction from the measured asymmetry.

Figure \ref{ratiofig} shows the final results with the GEp-I data~\cite{Punjabi05}, the originally published
GEp-II data \cite{Gayou02}, the GEp-III data \cite{PuckettPRL10}, and
the combined GEp-2$\gamma$ result \cite{MezianePRL11}, consisting of a
weighted average of measurements of $R$ at three $\epsilon$ values for
a fixed $Q^2$ of 2.5 GeV$^2$. The curves illustrate the effect of the revised data on a
global fit using the Kelly parametrization
\cite{KellyFit} of $G_E^p$ and $G_M^p$ to elastic $ep$ cross
section and polarization data, including the GEp-III data \cite{PuckettPRL10}. The
data selection and fit method are detailed in~\cite{Puckett:2010kn}. The dashed ``Old fit'' curve uses
the original GEp-II results, while the solid ``New fit'' curve replaces the three highest-$Q^2$
points from GEp-II with the results of the present
reanalysis. The combined contribution of the six highest-$Q^2$ data
points to
the $\chi^2$ of the ``Old'' global fit is 2.68. In
the ``New'' fit, the same
$\chi^2$ contribution\footnote{In this context, the $\chi^2$
  contribution of the six data points is defined as $\sum_{i=1}^{6} \left(R_i
    - R_{fit}(Q^2_i)\right)^2/(\Delta R_i)^2$; i.e., it is not normalized by
  the number of data points.} drops to 0.55,
indicating a significant improvement in
the consistency of the data at high $Q^2$.

The noticeable systematic increase in
the results for $R$ in the improved data analysis is mostly
attributable to the systematic effect of the background,
underestimated by the original analysis. Indeed,
the added $\delta p$ cut of the present work suppresses the background
contamination to well below 1\%, minimizing the associated correction
and its uncertainty. The most significant difference between the final
and original results not attributable to
background or changes in elastic event selection cuts is
caused by the improved description of the FPP analyzing power in the
present analysis. In
the original analysis, the
momentum dependence of the analyzing power was neglected, and the data
were divided into discrete bins in the FPP polar scattering angle
$\vartheta$. In each bin, the analyzing power was extracted from the
measured asymmetries using equation~\eqref{recoilformulas} as
described in section~\ref{subsubsec:Ay}. Then, the analyzing
power, which enters equation~\eqref{likelihoodfunction}
as a weight, was assigned to each event in a given $\vartheta$ bin according
to the extracted $A_y$ result in that bin. This method is approximately
equivalent to analyzing the data in bins of $\vartheta$ assuming $A_y
= 1$, and then combining the results of
all $\vartheta$ bins in a weighted average to obtain the final result. 

In the present work, the final results were obtained from a completely
unbinned maximum-likelihood analysis in which $A_y(p,p_T)$ was described
using the smooth parametrization of the $p_T$ dependence presented in section
\ref{subsubsec:Ay} and a global momentum scaling $A_y \propto 1/p$,
leading to a slightly different relative weighting of events as a
function of $p$ and $\vartheta$. At $Q^2 = 5.6$ GeV$^2$, where the
statistical uncertainty is large, roughly half the
difference between the originally published result and the final
result is attributable to the different description of $A_y$ (with
the other half coming from the background), while at $Q^2 = 4.0$ and
$4.8$ GeV$^2$, the effect of the $A_y$ description is small and the
difference is dominated by the background effects. This observation
can be understood by examining the
$\vartheta$ dependence of $R$ in
Figure~\ref{fig:rthfpp} and the $p_T$-dependence of $A_y$ in
Figure~\ref{fig:Ay} at $Q^2 = 5.6$ GeV$^2$. A negative fluctuation of $R$ in the
$\vartheta$ bin near 11$^\circ$ coincides with a positive fluctuation
of $A_y$ in the $p_T$ bin near 0.7 GeV. Assigning this value of $A_y$ to all events
in this bin artificially overweights the corresponding negative
statistical fluctuation in $R$, inducing a slight negative bias to the
result. Because this particular fluctuation is relatively large,
the effect of using a smooth parametrization of the analyzing power
instead of a discretely binned description is noticeable. This is in contrast to
the two lower-$Q^2$ points, for which no large $\vartheta$-dependent statistical fluctuations of $A_y$ or
$R$ are observed, making the combined result rather insensitive to the
description of $A_y$. It cannot be too strongly emphasized that the
dependence of the result on the description of $A_y$ derives
\emph{only} from $p$ and $\vartheta$-dependent statistical fluctuations of
the data, since $A_y$ cancels in the ratio $P_t/P_\ell$ (see Figure
\ref{fig:rthfpp} and the discussion in section
\ref{subsubsec:Ay}). Therefore, the sensitivity of the results
to the description of
$A_y$ is properly regarded as part of the statistical
uncertainty, and no additional systematic uncertainty
contribution is assigned. 

Despite discarding up
to $10\%$ of the events included in the original analysis, the
statistical error of the final result for $R$ is actually slightly
reduced at $Q^2 = 4.8$ and 5.6 GeV$^2$ relative to the original
publication. The improvement reflects an increase in the effective $A_y$
of the final sample of events due to the improved suppression of the
background, which tends to dilute the measured asymmetry. On the other hand, the statistical error
at $Q^2 = 4.0$ GeV$^2$ has slightly increased relative to the original
publication, since at this $Q^2$ the
loss of statistics slightly outweighs the increase in $A_y$ from
improved background suppression. Nonetheless, the quality of the
result is improved by the removal of a previously underestimated
systematic error. 


Compared to the situation before the GEp-III experiment, the emerging
picture of the large-$Q^2$ behavior of $G_E^p/G_M^p$ is considerably
modified. Before GEp-III, the GEp-I and GEp-II data suggested a strong
linear decrease of $R$ continuing to high $Q^2$. The linear trend of
the data suggested a zero crossing of $G_E^p/G_M^p$ before 8
GeV$^2$. The GEp-III data showed that the linear decrease probably does
not continue to higher $Q^2$, at least not at the slope suggested by
the GEp-I and original GEp-II results. Although the lower-$Q^2$ data from
GEp-2$\gamma$ appeared to rule out any neglected systematic error in
the GEp-III data, the fact that all three data points from GEp-III
were systematically above the trendline of the previous data raised 
concern about the consistency between different experiments and the
reproducibility of the polarization transfer method. Moreover, while there
was no \emph{a priori} reason to expect the linear decrease to
continue, and the apparent $\sim 1.8\sigma$ disagreement between GEp-II and GEp-III did
not rise to the level of statistical significance, the lessons learned
from the GEp-III analysis, particularly with respect to backgrounds, motivated a reanalysis of the GEp-II data, leading to the results
presented in this article. With improved analysis, the data from
Halls A and C \cite{Punjabi05,Gayou02,PuckettPRL10,MezianePRL11} are
now in excellent agreement over a wide $Q^2$ range, bringing added
clarity to the experimental situation
regarding $G_E^p/G_M^p$. 

In a simple global analysis using the Kelly
parametrization \cite{Puckett:2010kn}, the data before GEp-III
implied a zero crossing at $Q^2 = 9$ GeV$^2$, with an uncertainty
range of $7.7 \mbox{ GeV}^2 \le Q^2 \le 12.5$ GeV$^2$, based
on the pointwise $1\sigma$ error bands of the fit result. After adding
the GEp-III data and replacing the GEp-II data with the final analysis
results, the zero crossing is shifted to 15 GeV$^2$, with an uncertainty
range of roughly $12 \mbox{ GeV}^2 \le Q^2 \le 29$ GeV$^2$. Although the size of the
error band in $G_E^p$ shrinks by a factor of two at large $Q^2$
when the GEp-III data are added, the reduced slope of $G_E^p$ increases the uncertainty in the location of the
potential zero
crossing. The Kelly parametrization, despite having the correct static limit
and sensible pQCD-based asymptotic behavior at high $Q^2$, does not
describe the actual physics involved in the transition between low and high-$Q^2$
asymptotic behavior. Therefore, its extrapolation beyond the
range of the existing data necessarily understates the true uncertainty in the
behavior of $G_E^p$ at large $Q^2$. Only future measurements at higher $Q^2$ with higher precision \cite{GEP5} can definitively reveal the behavior of
$G_E^p$ in the region where the predictions of leading models of
the nucleon diverge, as discussed in the following section. 

\subsection{Physics Interpretation}
\label{subsec:theory}
\subsubsection{Perturbative QCD}
Perturbative QCD (pQCD) makes rigorous predictions for the $Q^2$
dependence of the nucleon
form factors when $Q^2$ is sufficiently large that the scattering
amplitude can be factorized as the convolution of a baryon
distribution amplitude with a perturbatively calculable hard
scattering kernel~\cite{Lepage:1980fj}. At leading order in 1/$Q^2$, the Dirac form factor is proportional to 
$\alpha_s/Q^4$ times slowly varying logarithmic terms, because the large
momentum transfer absorbed by the struck quark must be shared among
the two spectator quarks via two hard gluon exchanges in order for the
nucleon to recoil as a whole. The Pauli form factor is suppressed by
an extra power of $Q^2$ at leading order due to helicity conservation
\cite{BrodskyFarrar1975}, implying that $Q^2 F_2/F_1$ (and therefore $G_E/G_M$) becomes constant
at very high $Q^2$. While pQCD predicts the asymptotic $Q^2$
dependence of the nucleon form factors, it does not predict
the value of $Q^2$ at which the hard
scattering mechanism becomes dominant. Isgur and Llewellyn Smith
\cite{Isgur:1988iw,Isgur:1989cy} have argued that pQCD
is not applicable to observables of exclusive reactions such as the
nucleon form factors in the experimentally
accessible $Q^2$ region. Ralston and Jain~\cite{RalstonJain}, inspired
by the results of the GEp-I and GEp-II experiments, revisited
the leading power behavior in $1/Q$ of $F_2/F_1$ in the pQCD
hard-scattering picture by considering the
violation of hadron helicity conservation that ensues when quark wavefunction
components with nonzero orbital angular momentum are included, and
found that $F_2/F_1 \propto 1/Q$. 

Belitsky, Ji and Yuan \cite{BelitskyJiYuan2003}, like Ralston and Jain
\cite{RalstonJain}, argued that quark
orbital angular momentum is the dominant mechanism for nucleon helicity
flip at large $Q^2$ in pQCD, owing to the very small mass of the current
quarks involved in the hard scattering. They performed a pQCD analysis of the
proton's Pauli form factor $F_2^p$ including the subleading-twist
contribution to the proton's light-cone wavefunction. The
leading-order pQCD contribution to $F_2^p$ involves intial and final-state light-cone
wavefunctions differing by one unit of quark orbital angular
momentum, with zero orbital angular momentum in either the initial or
final state. In this calculation, logarithmic singularities in the
convolution integrals lead to the modified scaling $Q^2 F_2^p/F_1^p \propto
\ln^2 \left(Q^2 / \Lambda^2 \right)$, where $\Lambda$ is an infrared cutoff
parameter related to the size of the nucleon. 

\begin{figure}
  \begin{center}
    \includegraphics[width=.48\textwidth]{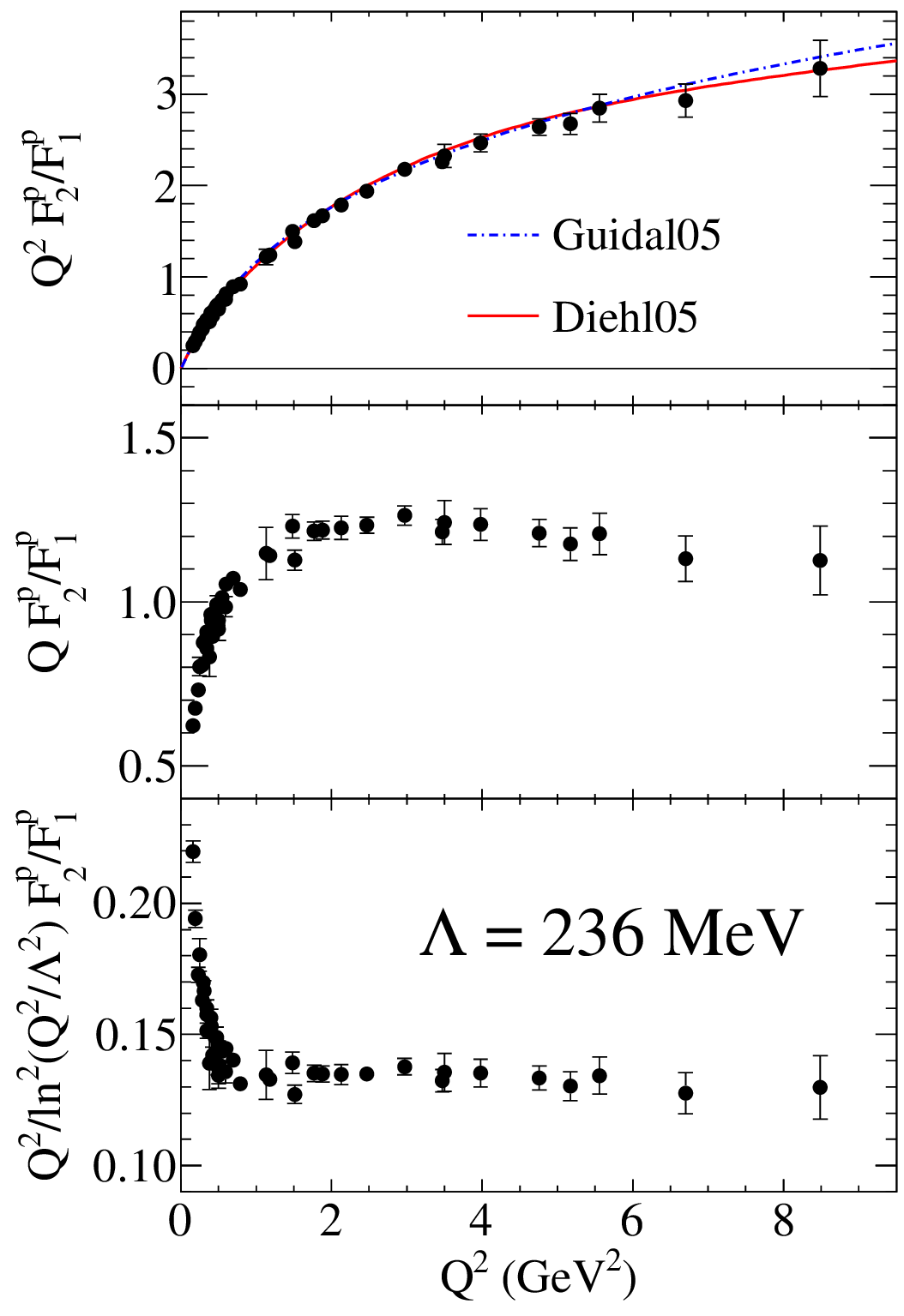}
  \end{center}
  \caption{\label{pQCDfig} (color online) Data for the ratio $F_2^p/F_1^p$ from
    selected polarization experiments including Refs.~\cite{Crawford07,Punjabi05,PuckettPRL10,Jones06,Maclachlan06,Milbrath98,Ron11,Zhan201159}
    and the final GEp-II data from Table \ref{resultstable} of this
    work. From top to bottom: $Q^2 F_2^p/F_1^p$, $QF_2^p/F_1^p$ and
    $Q^2/\ln^2\left(Q^2/\Lambda^2\right) F_2^p/F_1^p$, where $\Lambda
    = 236$ MeV was fitted to the data for $Q^2 \ge 1$ GeV$^2$. The
    curves in the top panel are GPD model fits from \cite{GuidalGPD2005} (Guidal05) and \cite{DiehlGPDtomography2005} (Diehl05).}
\end{figure}
Figure \ref{pQCDfig}
shows the experimental data for $F_2^p/F_1^p$ plotted as $Q^2
F_2^p/F_1^p$, $QF_2^p/F_1^p$ and $Q^2 / \ln^2
\left(Q^2/\Lambda^2\right)F_2^p/F_1^p$. Clearly, the leading-twist,
leading-order pQCD scaling behavior is not respected by the data in
the presently accessible $Q^2$ region, although the slope of
$Q^2F_2^p/F_1^p$ does appear to be trending toward a flat behavior at
the highest-$Q^2$ values measured so far. The scaling of $QF_2/F_1$
predicted by \cite{RalstonJain} is approximately satisfied up to 8.5
GeV$^2$, although there is a hint that $F_2$ may start to fall faster
than $F_1/Q$ for higher $Q$. The
logarithmic scaling of~\cite{BelitskyJiYuan2003} is satisfied for
$Q^2 \gtrsim 1$ GeV$^2$ at a value of the cutoff parameter $\Lambda =
236$ MeV ($\hbar c / \Lambda =0.835$ fm) determined by fitting the
data for $Q^2 \ge 1$ GeV$^2$. 

While the ``precocious'' scaling of $F_2^p/F_1^p$ is interesting, it
is probably largely accidental, perhaps a consequence of delicate
cancellations of higher-order effects in the ratio \cite{BelitskyJiYuan2003}. The scaling of
$F_2^p/F_1^p$ is a necessary but insufficient condition for the onset
of the perturbative regime. pQCD-based form
factor predictions based on light-cone sum rules
\cite{LCQCD2006,LCQCD2008} have yet to reach the level of accuracy
achieved by the phenomenological models discussed below in describing all four nucleon form
factors. In the GPD model fits shown in Figure \ref{pQCDfig}, the
'Feynman' mechanism corresponding to the overlap of soft wavefunctions
dominates the form factor behavior. The neutron data for $F_2^n/F_1^n$ do not scale in the currently measured
$Q^2$ region up to 3.4 GeV$^2$ for values of
$\Lambda$ similar to that which describes the proton data
\cite{Riordan}. Moreover, combining the proton and neutron data
to separate the up and down-quark contributions to the nucleon form
factors \cite{FlavorDecomposition} reveals that the ratios
$F_2^u/F_1^u$ and $F_2^d/F_1^d$ become
approximately constant above 1 GeV$^2$, at odds with the asymptotic pQCD picture,
while the ratios $F_1^d/F_1^u$
and $F_2^d/F_2^u$ decrease at high $Q^2$, a behavior that can be
explained in terms of diquark degrees of freedom \cite{Cloet09}. Based
on these and other considerations, it is
generally believed that the nucleon form factors are dominated by
non-perturbative physics in the 1-10 GeV$^2$
region addressed by present experiments.   

\subsubsection{Generalized Parton Distributions}
The generalized parton distributions (GPDs) are universal non-perturbative
matrix elements involved in the QCD factorization of hard exclusive
processes such as deeply virtual Compton scattering
(DVCS)~\cite{RadyushkinGPD,JiGPDPRL1997,JiGPDPRD1997,HardExclusiveReview2001}. The
GPDs are functions of the longitudinal momentum
fraction $x$, the momentum fraction asymmetry or
``skewness'' $\xi$ and the squared momentum transfer to the nucleon
$t$ (not to be confused with the photon virtuality $Q^2$). GPDs play a
crucial role in the synthesis of seemingly disparate nucleon structure
information obtained from inclusive and exclusive reactions. The Dirac
and Pauli form factors $F_1$ and $F_2$ equal the first $x$ moments
of the vector ($H(x,t)$) and tensor ($E(x,t)$) GPDs, respectively.
In the forward ($t \rightarrow 0$) limit, $H(x,t=0)$ is the valence quark
density. Precise measurements of the Pauli form factor $F_2$ at large
$Q^2$ constrain the behavior of $E(x,t)$, yielding new information on nucleon
structure that is inaccessible in inclusive deep inelastic
scattering (DIS). With increasing $Q^2$, the strength in the GPD
integrals corresponding to the form factors is
increasingly concentrated in the high-$x$ region. Therefore, the $x
\rightarrow 1$ behavior of $H(x,t)$ and $E(x,t)$
can be constrained by fitting the high-$Q^2$ nucleon form factors. 

While systematic studies of the observables of DVCS and other hard exclusive
reactions promise an eventual direct extraction of GPDs from global
analysis (for recent examples, see \cite{GuidalGPDfitter,Kumericki:2009uq,Goldstein:2010gu,Moutarde:2009fg,Kumericki:2011rz}),
the experimental mapping of these observables is still at an early
stage. Meanwhile, constraints from the elastic form factors
and the forward parton distributions
measured in DIS have been explored using physically motivated GPD
parametrizations based on Regge phenomenology
\cite{DiehlGPDtomography2005,GuidalGPD2005}. In both models, the
high-$x$ behavior of $E$ was determined by the high $Q^2$ behavior of
$F_2^p$ measured by the GEp-I and GEp-II experiments, enabling an evaluation
of Ji's sum rule~\cite{JiGPDPRL1997,JiGPDPRD1997} for the total angular
momentum carried by the up ($J^u$) and down ($J^d$) quarks in the
nucleon. The calculations of \cite{GuidalGPD2005} found $2J^d =
-0.06$ and  $2J^u = +0.58$, in qualitative agreement with lattice QCD
calculations available at the time~\cite{Gockeler:2003jfa}, as well as
more recent calculations~\cite{LHPC08,ETMC11}. The predictions
of the GPD models of~\cite{GuidalGPD2005} and
\cite{DiehlGPDtomography2005} are compared to
the data for $Q^2 F_2^p/F_1^p$ in Figure~\ref{pQCDfig}. 


The two-dimensional Fourier transform of the $t$-dependence of the
GPDs at $\xi = 0$ yields a three-dimensional impact parameter
representation $\rho(x,\mathbf{b}_\perp)$ in
two transverse spatial dimensions and one longitudinal momentum
dimension \cite{BurkardtIJMPA}. By forming the charge-squared weighted
sum over quark flavors and
integrating over all $x$, Miller~\cite{MillerGPDchargedensity2007}
derived the model-independent infinite momentum frame transverse charge
density $\rho_{ch}(b)$ as the two-dimensional Fourier transform of the Dirac form
factor $F_1$. The Pauli form factor $F_2$ can also be related to the
transverse anomalous magnetic moment
density $\rho_{m}(\mathbf{b})$~\cite{MillerTransverseDensityReview2010}. Miller \emph{et
  al.}~\cite{MillerProtonTransverseDensity2011} performed the first analysis
of the uncertainties in the transverse charge and magnetization
densities of the proton due to the uncertainties and incomplete $Q^2$
coverage of the form factor
data. Measurements of $G_E^p$ at yet higher $Q^2$ are needed to reduce the
uncertainty in $\rho_m(b)$ at small $b$. 

Since an exact solution to QCD in the
non-perturbative regime is not yet possible, predicting nucleon form
factors in the domain of strong coupling and confinement is rather
difficult. Consequently, many phenomenological
models have aimed to unravel the complicated internal structure of the
nucleon in this domain. The following discussion provides an overview
of a wide range of models.

\subsubsection{Vector Meson Dominance}
The global features of the nucleon form factors were explained by
early models based on vector meson dominance (VMD)
\cite{Iachello1973}. VMD models 
are a special case of dispersion relation analyses, which provide a
model-independent, non-perturbative framework to interpret the
electromagnetic structure of the nucleon in both the space-like 
and time-like regions. Early VMD model calculations
included the $\rho$ and its excited states
for the isovector form factors, and the  $\omega$  and $\phi$ for the
isoscalar form factors. The number of
mesons included and the coupling constants and masses can be varied to
fit the data. In practice, many parameters are fixed or strongly
constrained by experimental
data, including but not limited to nucleon FF data, reducing the number of free parameters and increasing the
predictive power of the approach. More recent calculations have used 
the pQCD scaling relations to constrain the large $Q^2$ behavior of
the fits. An example 
is Lomon's fit~\cite{Lomon2002}, which uses $\rho(770)$,
$\omega(782)$, $\phi(1020)$, $\rho^\prime(1450)$ and
$\omega^\prime(1420)$ mesons and has a total of 12 variable
parameters~\cite{Lomon2002,Lomon2006}. Bijker and
Iachello~\cite{Iachello2004} updated the 1973 model of
Iachello, Jackson and Land\'e \cite{Iachello1973}, performing a new fit including the $\rho(770)$,
$\omega(782)$, and $\phi(1020)$ mesons, and a phenomenological
``direct coupling'' term attributed to an 
intrinsic three-quark structure of $rms$ radius $\sim 0.34$~fm. 

Despite the relatively good fits obtained by VMD models, the approach 
is at odds with general constraints from unitarity. This difficulty can
be overcome using dispersion relations. 
H\"ohler's dispersion relation analysis~\cite{Hohler} was extended in the mid-nineties by 
Mergell, Meissner, and Drechsel~\cite{MergellMeissnerDrechsel}
to include nucleon form factor data in the time-like region~\cite{Hammer1996}. 
The analysis of \cite{MergellMeissnerDrechsel} has been further 
improved by Belushkin {\it et al.}~\cite{Belushkin2007}. 
In addition to the $2 \pi$ continuum present in the isovector 
spectral functions, the $\rho \pi$ and $K \bar K$ continua were 
included in the isoscalar spectral functions. 
In~\cite{Belushkin2007}, the $2 \pi$ continuum was reevaluated using 
the latest experimental data for the pion form factors in the time-like 
region. A simultaneous fit to the world data 
for all four form factors in both the space-like and time-like 
regions was performed. The results are in very good agreement with the
data available at the time. Dubnicka {\it et al.} developed
a unitary and analytic ten-resonance model including the $2 \pi$
continuum \cite{Dubnicka:2003wz,Adamuscin:2005aq}, which fits all
nucleon form factors in both the space-like and time-like regions. 

\begin{figure}
  \begin{center}
    \includegraphics[width=.48\textwidth]{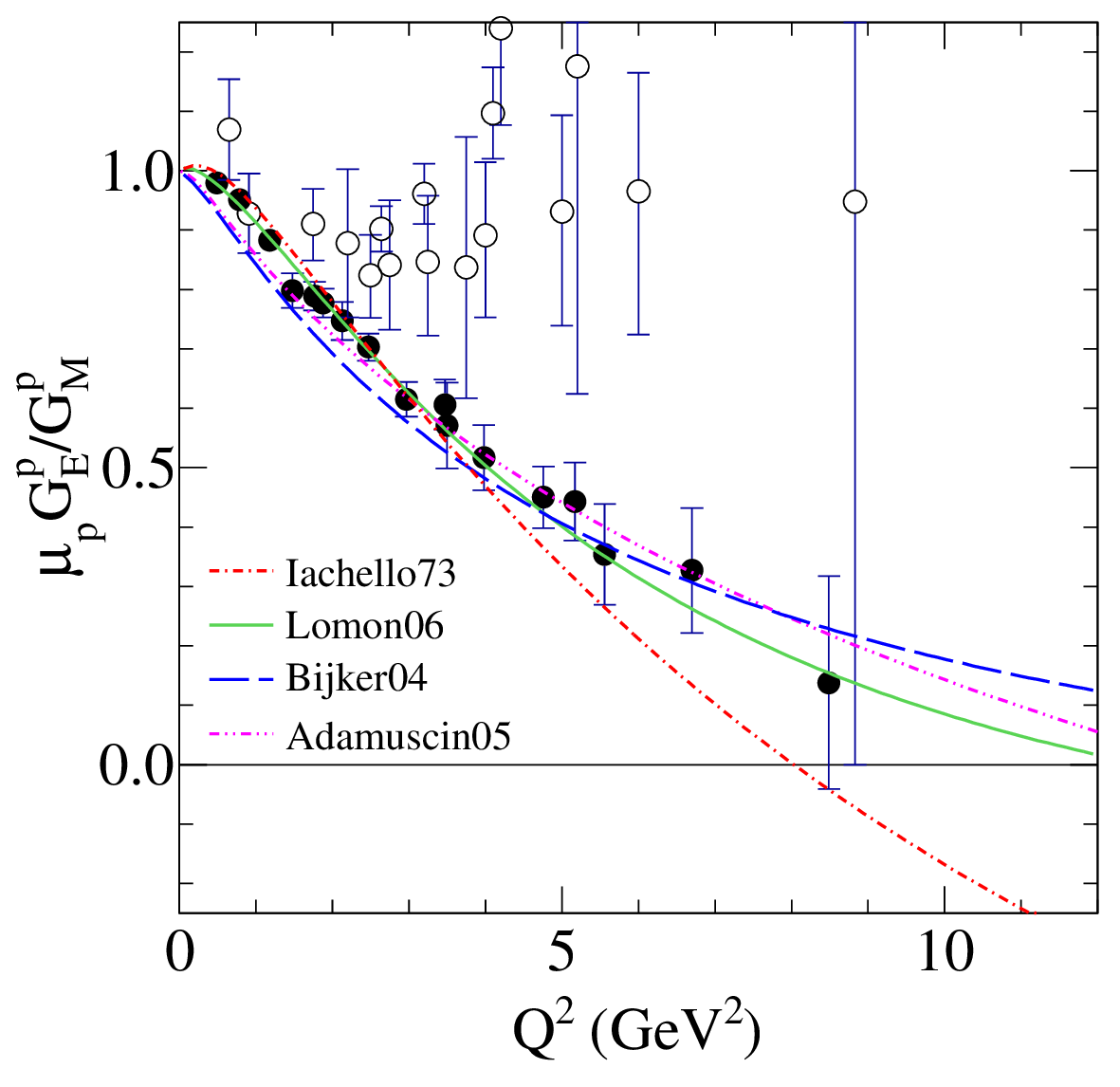}
  \end{center}
  \caption{\label{VMDfig} (color online) Comparison of selected VMD model predictions from \cite{Iachello1973} (Iachello73), \cite{Lomon2006} (Lomon06), \cite{Iachello2004} (Bijker04), and \cite{Adamuscin:2005aq} (Adamuscin05)
    to selected $\mu_p G_E^p/G_M^p$ data. Data are from cross section \cite{Andi94,Christy04,Qattan05}
    (empty circles) and polarization
    \cite{Punjabi05,Gayou02,PuckettPRL10}
    (filled circles) measurements, where the results of \cite{Gayou02}
    are replaced by the results of the present work. See text for details.}
\end{figure}
Figure \ref{VMDfig} compares the predictions of selected VMD-based
models to the experimental data for $\mu_p G_E^p/G_M^p$. Of the models shown,
the latest version of Lomon's fit \cite{Lomon2006} with twelve
adjustable parameters achieves the best
overall agreement with the data for all four form factors at spacelike
$Q^2$, emphasizing a smooth evolution from VMD
behavior at low $Q^2$ to pQCD scaling at asymptotically high
$Q^2$. Apart from fitting to a more complete data set, the main added
feature of the model of Bijker and Iachello
\cite{Iachello2004} relative to the 1973 model of Iachello, Jackson
and Land\'e is the inclusion of a ``direct'' coupling term in the
isoscalar Pauli form factor which improves the large-$Q^2$ behavior of
$G_E^p$ and $G_E^n$. This model achieves a rather good fit to all four FFs
using just six adjustable parameters (compared to five in the 1973 model). 

\subsubsection{Lattice QCD}
Lattice QCD calculations provide {\it ab initio} evaluations of static
and dynamic hadron properties,
including the nucleon electromagnetic form factors, from numerical
solutions of QCD on a finite-volume lattice of discrete space-time
points. At present, the lattice calculations
are done using unphysically large quark masses which are given in terms of 
the pion mass, $m_{\pi}$. Moreover, most recent calculations focus on
the isovector form factors, for which the contributions from
disconnected diagrams are reduced. Calculations are performed for various
$m_{\pi}$ values and lattice spacings $a$ and then extrapolated to the physical
pion mass and the continuum limit $a \rightarrow 0$. Recently the
QCDSF/UKQCD Collaboration has performed calculations \cite{Collins:2011mk} at $m_{\pi} = 180$~MeV with different lattice spacings and volume sizes, but  the upper $Q^2$ range is limited to 3~GeV$^2$.
Lattice QCD form factor calculations in the $Q^2$ region measured by
the GEp-II and GEp-III experiments are difficult
due to large statistical and systematic errors. Calculations by Lin {\it et al.} employ 
a novel technique to extend the reliable $Q^2$ range of the
calculations to $Q^2 = 6$~GeV$^2$ at $m_{\pi} > 450 $~MeV for quenched and dynamical
ensembles \cite{Lin:2010fv}. Nonetheless, calculations at such high
$Q^2$ must ultimately be performed with a finer lattice spacing to reduce the systematic error.

\subsubsection{Constituent Quark Models}
In the constituent quark model (CQM), the nucleon consists of three
constituent quarks, which can be thought of as valence quarks that
become much heavier than the
elementary quarks appearing in the QCD Lagrangian when dressed by
gluons and quark-antiquark pairs. The dressing effects are absorbed into
the masses of these quasiparticle effective degrees of freedom. The early non-relativistic
constituent quark model achieved considerable success in describing
the spectrum of baryons and
mesons with correct masses~\cite{isgurkarl}. In order to describe
dynamical quantities such as form factors in terms of constituent quarks, a relativistic 
description (rCQM) is mandatory because the $Q^2$ values involved in modern
experiments have reached as high as ten times the nucleon mass
squared and $\sim 10^6$ times the ``bare'' quark mass squared. 

Frank \emph{et al.}~\cite{FrankJenningsMiller} calculated
$G_E^p$ and $G_M^p$ in the light-front CQM using
the light-front nucleon wave function of Schlumpf~\cite{schlumpfCQM}, and
predicted that $G_E^p$ might change sign near 5.6 GeV$^2$, a behavior
inconsistent with current data, though qualitatively correct. In this
model, constructing a Poincar\'{e}-invariant
nucleon wavefunction that is also an eigenstate of spin
leads to the substantial violation of hadron helicity
conservation~\cite{MillerFrank} responsible for the observed scaling of
$QF_2/F_1$ in the $Q^2$ range of present experiments. This feature is
a consequence of the unitary Melosh rotation~\cite{MeloshRotation}
which mixes quark spin states in the process of boosting the nucleon
spin-flavor wavefunction from the
rest frame to the light front. Miller extended this model to include
pion-cloud effects \cite{MillerLFCBM}, important to the understanding
of the low-$Q^2$ behavior generally and $G_E^n$ in particular. 


Gross \emph{et al.}~\cite{gross06,gross08} modeled the nucleon as a
bound state of three dressed valence
constituent quarks in the covariant spectator formalism, in which the
virtual photon is absorbed by an off-shell constituent quark, and the
two spectator quarks always propagate as an on-shell diquark. In this
model, the constituent quarks have internal structure described by
form factors which become pointlike at large $Q^2$ as required by
pQCD and exhibit VMD-like behavior at low $Q^2$. The model nucleon
wavefunction of~\cite{gross08} obeys the Dirac equation, includes only \emph{s}-wave
components, and its spin-isospin structure
reduces to that of the SU(2) $\times$ SU(2) quark model in the
non-relativistic limit. 

Cardarelli {\it et al.}~\cite{Cardarelli:2000tk} calculated the ratio
using light-front dynamics and 
investigated the effects of SU(6) symmetry breaking. As in
\cite{FrankJenningsMiller}, they showed that
the decrease of $R$ with increasing $Q^2$ is caused by the
relativistic effect of the Melosh rotations of the constituent
quark spins.
De Sanctis {\it et al.} calculated the nucleon form
factors in the relativistic 
hypercentral constituent quark model (hCQM)~\cite{Sanctis:2007zz}. A good fit
to all the nucleon form factors was obtained
using a linear combination of monopole and dipole constituent quark form factors. The
calculation was recently extended to $Q^2 =
12$~GeV$^2$~\cite{Santopinto2010}. The same group also performed
calculations within the relativistic interacting quark-diquark model
\cite{DeSanctis2011}, which does not achieve the same level of
agreement with the data as the hCQM.

De Melo {\it et al.}~\cite{deMelo09} examined the non-valence components of the
nucleon state in light-front dynamics, achieving a good description of
all spacelike and timelike nucleon FF data with the inclusion of
the Z-diagram involving $q\bar{q}$ pair creation in addition to the
triangle (valence) diagram. The chiral
constituent quark model based on Goldstone-boson-exchange dynamics was
used by Boffi {\it et al.}~\cite{boffi:2001zb} to describe the elastic
electromagnetic and weak form factors in a covariant framework using
the point-form approach to relativistic
quantum mechanics.

\begin{figure}
  \begin{center}
    \includegraphics[width=.48\textwidth]{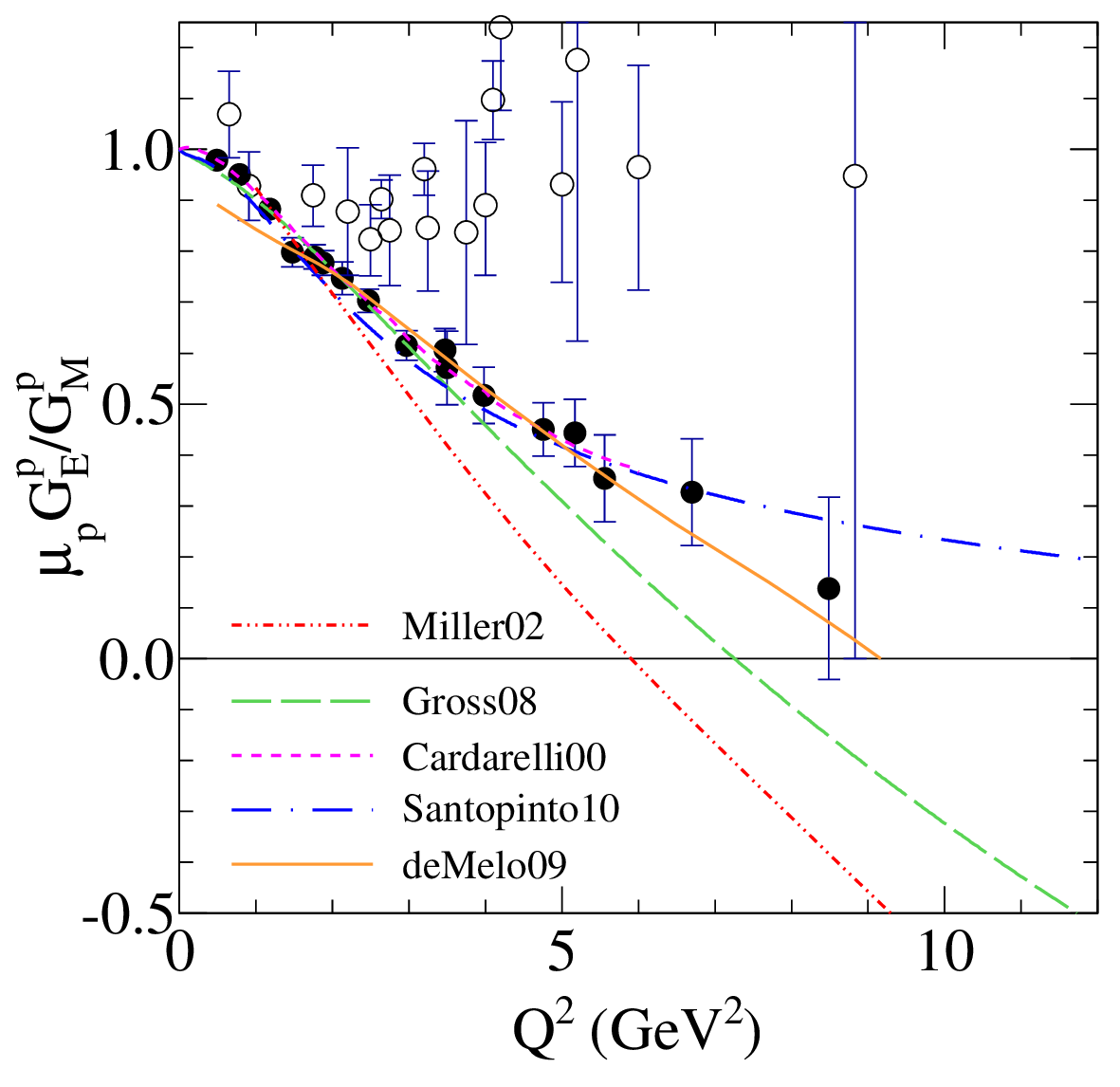}
  \end{center}
  \caption{\label{CQMfig} (color online) Selected rCQM predictions from \cite{MillerLFCBM} (Miller02), \cite{gross08} (Gross08), \cite{Cardarelli:2000tk} (Cardarelli00), \cite{Santopinto2010} (Santopinto10) and \cite{deMelo09} (deMelo09) for $R = \mu_p
    G_E^p/G_M^p$, compared to selected data from cross
    section~\cite{Andi94,Christy04,Qattan05} (empty circles) and
    polarization~\cite{Punjabi05,Gayou02,PuckettPRL10} (filled
    circles) experiments,
    where the results of~\cite{Gayou02} are replaced by the results of
    the present work. See text for details.}
\end{figure}
Figure \ref{CQMfig} compares the predictions of selected rCQM
calculations to selected data for $R$. Of the calculations shown, those
in which the constituent quarks have internal structure represented by
CQ form factors (\cite{Cardarelli:2000tk, gross08, Santopinto2010}) and/or significant
VMD-related contributions to the photon-nucleon vertex
(\cite{deMelo09}) describe the data better than those in which the
constituent quarks are pointlike (\cite{MillerLFCBM}) and have only
direct coupling to the photon. Although this
may be related to the greater number of adjustable parameters in
models with CQFFs, it is apparently physically meaningful that
most of the models require structure of the constituent quarks
and/or significant nonvalence ($q\bar{q}$ pair creation) contributions
to achieve a good description of the data.


\subsubsection{Dyson-Schwinger Equations}
A different theoretical approach to the prediction of nucleon form
factors is based on QCD's Dyson-Schwinger equations (DSEs). The DSEs
are an infinite tower of coupled integral equations for QCD's Green
functions that provide access to emergent phenomena of
non-perturbative QCD, such as dynamical chiral symmetry breaking and
confinement~\cite{Roberts1994477}. The DSEs admit a
symmetry-preserving truncation scheme that enables a
unified description of meson and baryon properties. The approach has already
achieved considerable success
in the pseudoscalar meson sector~\cite{CD200850}. The prediction of
nucleon form factors in the DSE approach involves the solution of a
Poincar\'{e}-covariant Faddeev equation. In the calculations of~\cite{Cloet09}, dressed quarks form
the elementary degrees of freedom and correlations between them are
expressed via scalar and axial vector diquarks. The
only variable parameters in this approach are the diquark masses,
fixed to reproduce the nucleon and $\Delta$ masses, and a diquark
charge radius $r_1^+$ embodying the electromagnetic structure of the
diquark correlations. A different approach to DSE-based
form factor calculations effects binding of the nucleon through a single
dressed gluon exchange between any two
quarks~\cite{EichmannFaddeev} without explicit diquark degrees of freedom. In this calculation, the only
parameters are a scale fixed to reproduce the pion decay
constant and a dimensionless width parameter $\eta$ describing the
infrared behavior of the effective coupling strength of the
quark-quark interaction. 

\begin{figure}
  \begin{center}
    \includegraphics[width=.48\textwidth]{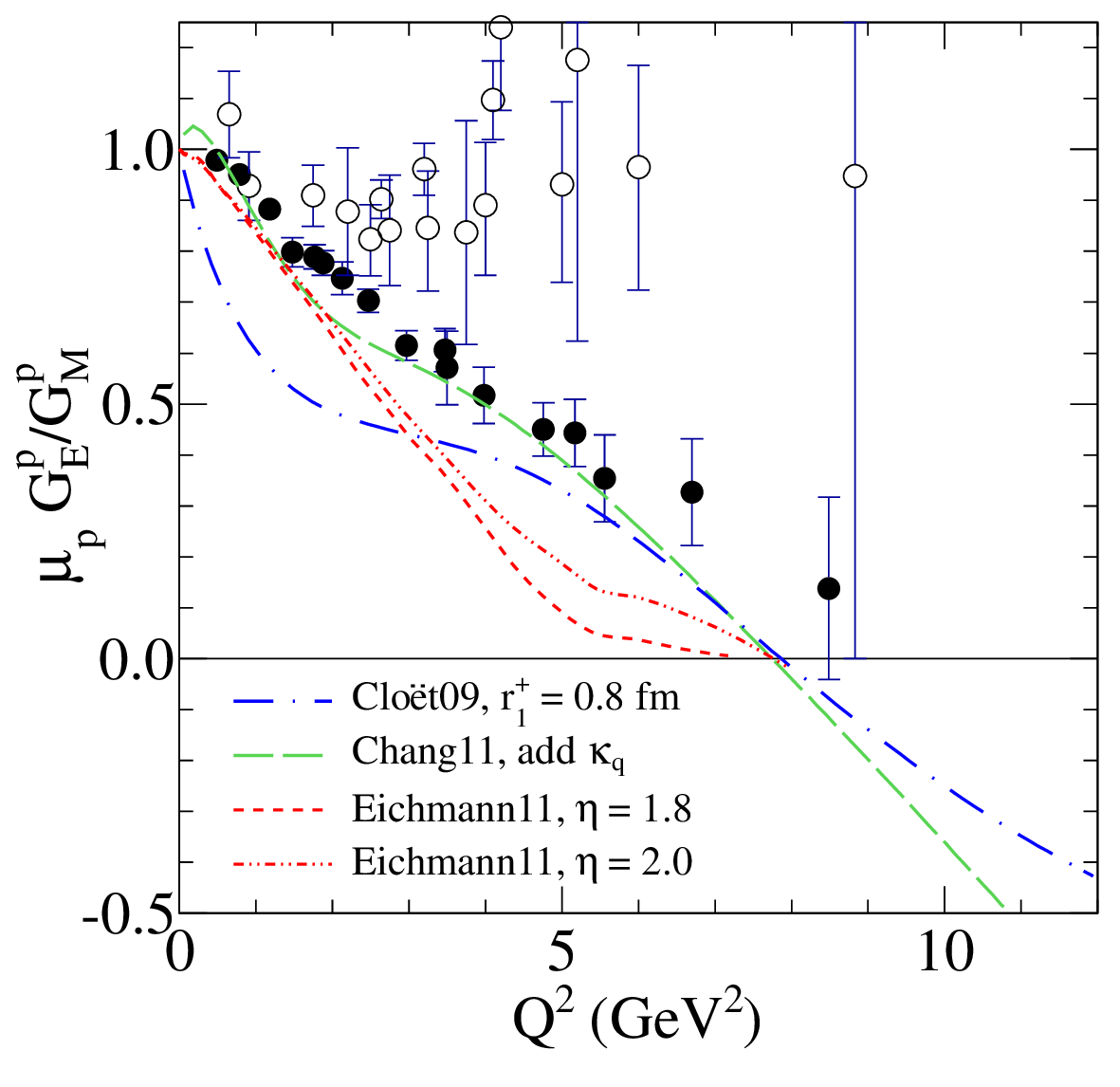}
  \end{center}
  \caption{\label{DSEfig} (color online) Predictions of DSE-based calculations for
    $R = \mu_p G_E^p/G_M^p$ compared to experimental data from cross
    section~\cite{Andi94,Christy04,Qattan05} (empty circles) and
    polarization~\cite{Punjabi05,Gayou02,PuckettPRL10} (filled
    circles) experiments,
    where the results of \cite{Gayou02} are replaced by those of the
    present work. The results of \cite{Cloet09} (Clo\"{e}t09) are shown for a
    particular choice of the diquark charge radius. The curve from \cite{Chang:2011tx} (Chang11) is that of \cite{Cloet09} 
    with the addition of dressed quark anomalous magnetic moments. The
    results of~\cite{EichmannFaddeev} (Eichmann11) are shown for two values of
    $\eta$, showing the weak sensitivity of the form factor results to this parameter.}
\end{figure}

The predictions of several DSE-based calculations for the
proton Sachs form factor ratio $R = \mu_p G_E^p/G_M^p$ are shown in Figure
\ref{DSEfig}. The quark-diquark model calculation~\cite{Cloet09}
underpredicts the data at low $Q^2$ but agrees reasonably well at higher
$Q^2$. The disagreement at low $Q^2$ is attributed to the omission of
meson cloud effects. The addition of dynamically
generated, momentum-dependent dressed-quark anomalous magnetic
moments~\cite{Chang:2011tx} that become large at infrared momenta improves the description of $R$ at
low $Q^2$. The three-quark model calculation~\cite{EichmannFaddeev} agrees
with the data at low $Q^2$, but underpredicts the data at higher $Q^2$,
becoming numerically unreliable for $Q^2 \gtrsim 7$ GeV$^2$. 

The deficiencies of the DSE approach, including the approximation schemes
required to make the calculations analytically tractable and the
omission of meson-cloud effects, are evident in the disagreement
between the predicted form factors and the experimental data, which is
more severe than in the various models described above, which have
more adjustable parameters. The advantage of the approach is that it
provides a systematically
improvable framework for the \emph{ab initio}
evaluation of hadron properties in continuum non-perturbative
QCD, that is complementary to discretized lattice simulations. As
fundamental measurable properties of nucleon structure, the
electromagnetic form factors are essential to the feedback between
theory and experiment required to make further progress in this direction.  
\begin{figure*}
  \begin{center}
    \includegraphics[width=.96\textwidth]{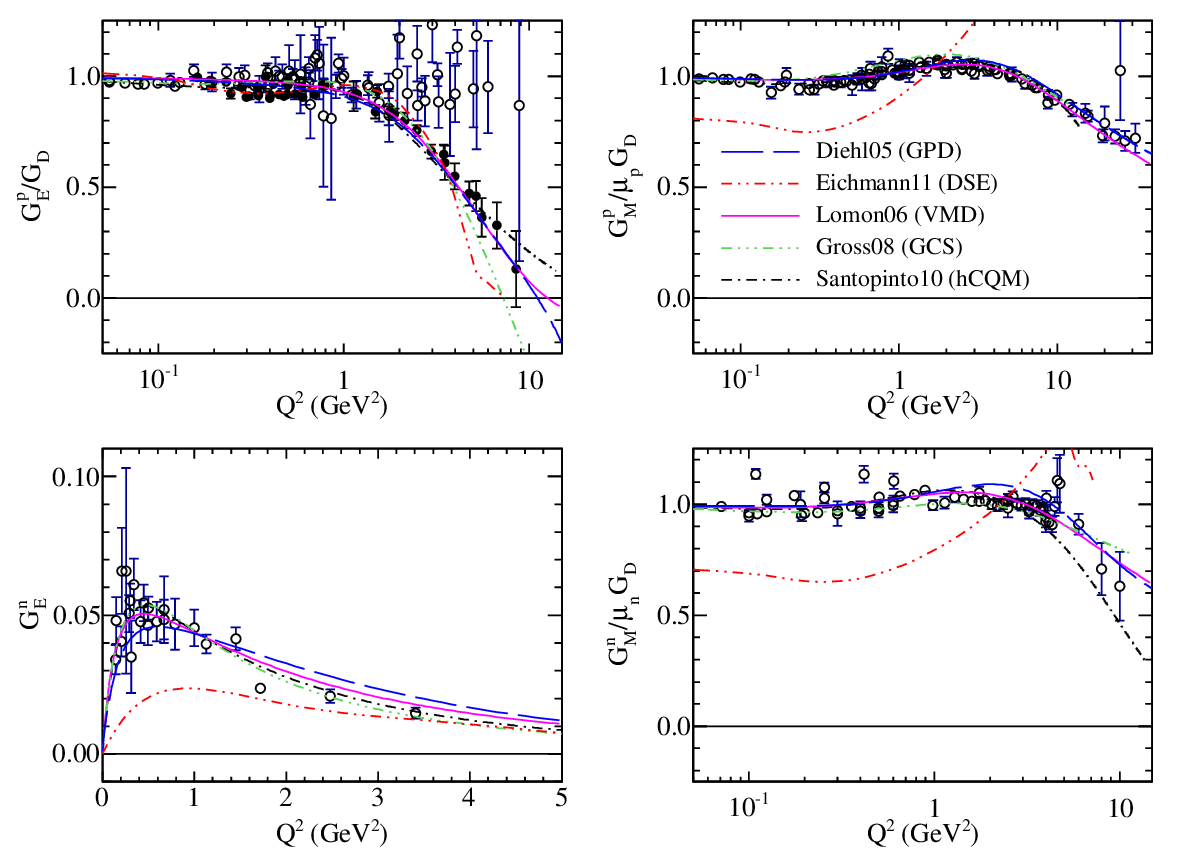}
  \end{center}
  \caption{\label{FourFFsFig} (color online) Comparison of selected theoretical
    predictions to data for all four nucleon FFs at space-like
    $Q^2$. Theory curves are \cite{DiehlGPDtomography2005} (Diehl05), \cite{EichmannFaddeev} (Eichmann11), \cite{Lomon2006} (Lomon06), \cite{gross08} (Gross08) and \cite{Santopinto2010} (Santopinto10). $G_E^p$ data are from \cite{Andi94,Bork75,Christy04,Janssens66,Price71,Qattan05,Simon80,Walker94,Berger71}
    (cross section data, empty circles) and
    \cite{Crawford07,Punjabi05,Gayou02,PuckettPRL10,Jones06,Maclachlan06,Milbrath98,Ron07,Zhan201159}
    (polarization data, filled circles), where the results of
    \cite{Gayou02} have been replaced by the results of the present
    work (Table \ref{resultstable}). $G_M^p$ data are from \cite{Bartel73,Berger71,Andi94,Bork75,Christy04,Kirk1973,Janssens66,Price71,Qattan05,Sill93,Walker94}. $G_E^n$
    data are from \cite{Bermuth03,Eden94,Geis08,Glazier05,Herberg99,Meyerhoff93,Ostrick99,Passchier99,Plaster06,Riordan,Rohe99,Warren04,Zhu01}. $G_M^n$
    data are from \cite{Anklin94,Anklin98,Lachniet,Bruins95,Gao94,JLab07,Kubon02,Lung93,Markowitz93,Rock82,Xu00,Xu03}. $G_D = \left(1+Q^2/\Lambda^2 \right)^{-2}$,
    with $\Lambda^2 = 0.71$ GeV$^2$, is the standard dipole form factor.}
\end{figure*}

\subsubsection{AdS/QCD}
In the past decade, theoretical activity has flourished in modeling QCD
from the conjecture of the anti-de Sitter space/conformal field theory (AdS/CFT)
correspondence~\cite{Maldacena:1997re,Witten:1998qj,Gubser:1998bc}, a
mapping between weakly coupled gravitational theories in curved
five-dimensional space-time and strongly coupled gauge theories in
flat four-dimensional space-time. Since QCD is not a conformal field theory, the symmetry of the anti-de Sitter space
is broken by applying a boundary condition. 
Brodsky and de Teramond~\cite{Brodsky:2008pg} have calculated $F_1$ for the proton and neutron
and emphasized the agreement of the predicted $Q^2 F_1$ dependence with the data.
Abidin and Carlson~\cite{Abidin:2009hr} have calculated both proton and neutron
$F_1$ and $F_2$ along with the tensor form factors using both hard and soft wall 
boundary conditions. This model predicts the same asymptotic $Q^2$
dependence as the dimensional scaling of pQCD, but does not reproduce
the detailed features of the data in the presently measured $Q^2$ region.

\subsubsection{World nucleon form factor data compared to theory}
Figure \ref{FourFFsFig} summarizes the theoretical interpretation of
the nucleon electromagnetic form factors, with representative examples from each of
the classes of models discussed compared to the world data for all
four nucleon electromagnetic form factors. Published results for $R = \mu_p G_E^p/G_M^p$
were converted to $G_E^p$ values using the global fit of $G_E^p$ and $G_M^p$
from~\cite{Puckett:2010kn}, updated to use the $R$ values of
the present work, a change that does not noticeably affect $G_M^p$. Except at
very low $Q^2$, the contribution of the uncertainty in $G_M^p$ to the
resulting uncertainty in $G_E^p$ is negligible. At
this juncture, it is worth recalling that the $G_E^p$ results extracted
from cross section data are believed to be unreliable at high
$Q^2$ due to incompletely understood TPEX corrections, which have not
been applied to the data shown in
Figures~\ref{VMDfig}-\ref{FourFFsFig}. Except for the DSE
calculation of~\cite{EichmannFaddeev}, all of the models shown describe
existing data very well, which is to be expected given
that the parameters of the models are fitted to reproduce the
data. However, their predictions tend to diverge when extrapolated
outside the $Q^2$ range of the data. That the DSE-based calculation
of~\cite{EichmannFaddeev} fails to describe the data as well as the
other calculations is not surprising, since it represents a more
fundamental \emph{ab initio} approach with virtually no adjustable
parameters, but requires approximations that
are not yet well-controlled. Significant progress in the quality of
the predictions is nonetheless evident, as the data expose the
weaknesses of different approximation schemes. Since the hard scattering
mechanism leading to the asymptotic pQCD scaling relations is not
expected to dominate the form factor behavior at presently accessible
$Q^2$ values, phenomenological models and the ambitious ongoing efforts in
lattice QCD and DSE calculations are of paramount importance to understanding
the internal structure and dynamics of the nucleon. Planned measurements at higher
$Q^2$ following the 12 GeV upgrade of JLab promise to be of 
continuing interest and relevance owing to their power to discriminate
among the various models and to guide the improvement of the more
fundamental calculational approaches.

\section{Conclusion}
\label{sec:conclusion}
This work has presented an expanded description and an improved final
data analysis of the GEp-II
experiment, originally published in
\cite{Gayou02}, which measured the proton electromagnetic form factor
ratio for $3.5 \mbox{ GeV}^2 \le Q^2 \le 5.6$ GeV$^2$ in Jefferson Lab's Hall A
using the polarization transfer method. The improved data analysis
finds a systematic increase in the
results for $R = \mu_p G_E^p/G_M^p$ that improves the agreement
between the GEp-II and GEp-III \cite{PuckettPRL10} data. This increase
mainly reflects the underestimated impact of the $\pi^0$ production background in
the original analysis of GEp-II. Section \ref{sec:expsetup} presented the
details of the experimental apparatus and described the differences
between the GEp-II and GEp-III experiments. Section
\ref{sec:analysis} presented the full details of the data analysis,
including the selection of elastic events in section
\ref{section:elasticeventselection}, the extraction of polarization
observables in section \ref{section:polextract}, and the treatment of
the background in section \ref{subsec:background}. The analysis of
systematic uncertainties was presented in section
\ref{sec:systematics}. In section \ref{subsec:data}, the features of
the data and the sources of the increase in the results relative to
the original analysis were discussed at length. An overview of
recent progress in the theoretical understanding of nucleon form
factors was given in section \ref{subsec:theory}. In conclusion, this work represents
the final results of the GEp-II experiment. The revised data presented
here and the results of the GEp-III experiment \cite{PuckettPRL10}
have considerably improved the experimental knowledge of $G_E^p$ at
large $Q^2$.

\begin{acknowledgments}
The collaboration thanks the Hall A technical staff and the Jefferson
Lab Accelerator Division for their outstanding support during the
experiment. This work was supported by the U. S. Department of Energy,
the U. S. National Science Foundation, the Italian Istituto Nazionale
di Fisica Nucleare (INFN), the French Commissariat \`{a} l'Energie
Atomique (CEA) and Centre National de la Recherche Scientifique
(CNRS-IN2P3), the Natural Sciences and Engineering Research Council of
Canada (NSERC), the EEC Grants No. INTAS 99-00125 for the Kharkov
Institute of Physics and Technology, and No. CRDF UP2-2271, the
Swedish Natural Science Research Council, and the Los Alamos National
Laboratory LDRD program. The Southeastern Universities
Research Association (SURA) operates the Thomas
Jefferson National Accelerator Facility for the U. S. Department of Energy under Contract No. DEAC05-84ER40150.
\end{acknowledgments}

\bibliography{main}
\end{document}